\documentclass[lettersize,onecolumn]{IEEEtran}
\usepackage[utf8]{inputenc}
\usepackage{amsmath,amsfonts}
\usepackage{mathrsfs}
\usepackage{algorithmic}
\usepackage{array}
\usepackage[caption=false,font=normalsize,labelfont=sf,textfont=sf]{subfig}
\usepackage{textcomp}
\usepackage{stfloats}
\usepackage{url}
\usepackage{verbatim}
\usepackage{graphicx}
\usepackage{cite}
\usepackage[normalem]{ulem}
\usepackage{amsmath,amsfonts,amssymb,amscd,amsthm,xspace}
\theoremstyle{plain}

\newtheorem{theorem}{Theorem}
\newtheorem{corollary}[theorem]{Corollary}
\newtheorem{lemma}[theorem]{Lemma}

\newtheorem{assumption}[theorem]{Assumption}
\theoremstyle{definition}

\theoremstyle{remark}
\newtheorem{remark}[theorem]{Remark}

\newcommand{\cref}[1]{Chapter~\ref{#1}}

\usepackage[linesnumbered,ruled,vlined]{algorithm2e}
\usepackage{tikz}
\usetikzlibrary{shapes.geometric}
\usetikzlibrary{arrows.meta, angles, quotes, calc}
\usepackage{pgfplots}     
\pgfplotsset{compat=1.18}
\usepackage{float}
\usepackage{booktabs}

\newcommand{\dtv}{d_{\mathrm{TV}}}

\newcommand{\I}{\mathbb{I}}
\newcommand{\Po}{\mathrm{Po}}
\newcommand{\N}{\mathrm{N}}
\newcommand{\IE}{\mathbb{E}}

\newcommand{\IP}{\mathbb{P}}
\newcommand{\NN}{\mathcal{N}}
\newcommand{\ZZ}{\mathcal{Z}}
\newcommand{\BB}{\mathcal{B}}
\newcommand{\HH}{\mathcal{H}}
\newcommand{\M}{\mathcal{M}}
\newcommand{\E}{\mathcal{E}}

\newcommand{\IR}{\mathbb{R}}
\newcommand{\C}{\mathbb{C}}
\newcommand{\R}{\mathcal{R}}
\newcommand{\X}{\mathcal{X}}
\newcommand{\PP}{\mathcal{P}}

\newcommand{\bigo}{\mathrm{O}}
\newcommand{\lito}{\mathrm{o}}
\newcommand{\Var}{\operatorname{Var}}
\newcommand{\Cov}{\operatorname{Cov}}

\newcommand{\law}{\mathscr{L}}
\newcommand{\dx}{\mathrm{d}}
\newcommand{\Dmixd}[1]{\mathcal{D}_d #1}
\newcommand{\dequal}{\overset{d}{=}}
\newcommand{\convergeP}{\overset{\IP}{\rightarrow}}

\newcommand{\til}{\tilde}

\newcommand{\lb}{\left(}
\newcommand{\rb}{\right)}
\newcommand{\lsb}{\left[}
\newcommand{\rsb}{\right]}
\newcommand{\lcb}{\left\{}
\newcommand{\rcb}{\right\}}
\newcommand{\given}{\;\middle|\;}

\newcommand{\posRealLine}{\IR^\circ_+}
\newcommand{\toinf}{\to\infty}
\newcommand{\sigmatoinf}{{\sigma\toinf}}
\newcommand{\tilx}{\tilde{X}}
\newcommand{\sig}{{(\sigma)}}
\newcommand{\BallD}{B_{\bf0}(\R)}
\newcommand{\vtj}{\Theta^{(j)}}
\newcommand{\tilvtj}{\til{\Theta}^{(j)}}
\newcommand{\vpj}{\Psi^{(j)}}
\newcommand{\vtk}{\Theta^{(k)}}
\newcommand{\vpk}{\Psi^{(k)}}
\newcommand{\vtzero}{\Theta^{(d-1)}}
\newcommand{\tilvtzero}{\til{\Theta}^{(d-1)}}
\newcommand{\vpzero}{\Psi^{(d-1)}}
\newcommand{\hatr}{\hat{R}}

\newcommand{\Efirst}{\mathcal{E}(R_1)}
\newcommand{\eps}{\varepsilon}

\begin{document}

\title{Performance Analysis for Wireless Localization with Random Sensor Network}

\author{Mengqi~Ma and Aihua~Xia
\thanks{M.~Ma is with the School of Mathematics and Statistics, The University of Melbourne, Parkville, VIC 3010, Australia (e-mail: mamm1@student.unimelb.edu.au).}
\thanks{A.~Xia is with the School of Mathematics and Statistics, The University of Melbourne, Parkville, VIC 3010, Australia (e-mail: aihuaxia@unimelb.edu.au).}
\thanks{Manuscript received January 9, 2026; revised May 15, 2026.}}

\markboth{Journal of \LaTeX\ Class Files,~Vol.~1, No.~2, Month~Year}%
{Shell \MakeLowercase{\textit{et al.}}: A Sample Article Using IEEEtran.cls for IEEE Journals}


\maketitle

\begin{abstract}
Accurate wireless localization underpins applications from autonomous systems to smart infrastructure. We study the mean-squared error (MSE) and the conditional MSE (CMSE), given the number of sensors, of practical fusion-based estimators---both the simple average and the weighted estimators---in $d$-dimensional, stationary isotropic (translation- and rotation-invariant) random sensor networks, where a central processor combines received-signal-strength (RSS) and angle-of-arrival (AOA) measurements collected from the sensors to infer a target’s position. Our contributions are twofold. First, we establish an approximation theorem: when the signal propagation effect is sufficiently strong (e.g., strong shadowing or multipath environments), the joint distribution of RSS and AOA observations collected from a broad class of stationary isotropic deployments becomes, in distribution, indistinguishable from that induced by a homogeneous Poisson point process (HPPP). Second, leveraging this equivalence, we analyze the performance of the fusion-based estimators in an HPPP sensor network. For such a deployment within a finite observation region, we derive tractable analytical upper bounds for both the MSE and CMSE, establishing explicit scaling laws with respect to sensor density, observation radius, and noise variance. The approximation theorem then certifies these HPPP-based bounds as reasonable proxies for non-HPPP deployments in noisy regimes. Overall, the results translate deployment and sensing parameters into achievable accuracy targets and provide robust, cost-aware guidance for the design of next-generation location-aware wireless networks.
\end{abstract}


\section{Introduction}\label{Section: Intro}
\IEEEPARstart{A}{ccurate} wireless localization is a fundamental enabler for a wide variety of emerging applications, ranging from autonomous vehicles and asset tracking to industrial automation, smart cities, and defense operations. While satellite-based systems such as GPS have transformed global positioning capabilities, their performance deteriorates in obstructed or indoor environments where signal blockage and multipath propagation become severe. To address these challenges, network-based wireless localization techniques have gained increasing attention.

In a typical wireless localization setup, multiple spatially distributed sensor nodes collect measurements of received signal waveforms from an unknown target source. The signal may either be directly emitted by the target or transmitted by the sensor and reflected by the target. Commonly used features include time-of-arrival (TOA), angle-of-arrival (AOA), and received signal strength (RSS). The TOA is obtained by measuring the signal propagation time between the target and the sensor; ideally, the distance estimate is the product of the known propagation speed and the measured travel time. The AOA represents the angle at which a signal arrives at the sensor, and it can be measured either directly using a directional antenna or indirectly by using an antenna array to compare the relative TOA across elements. The RSS refers to the power of the received signal, which can be used to estimate the distance to the target, although its accuracy is often limited by the difficulty of precisely modeling the relationship between signal strength and propagation distance \cite[Section~1]{WinEtAl2010a}. These measurements are then transmitted to a central processor or \emph{fusion center} that jointly estimates the target’s position.

\begin{figure*}[!t]
\centering
\subfloat[]{\includegraphics[width=2.3in]{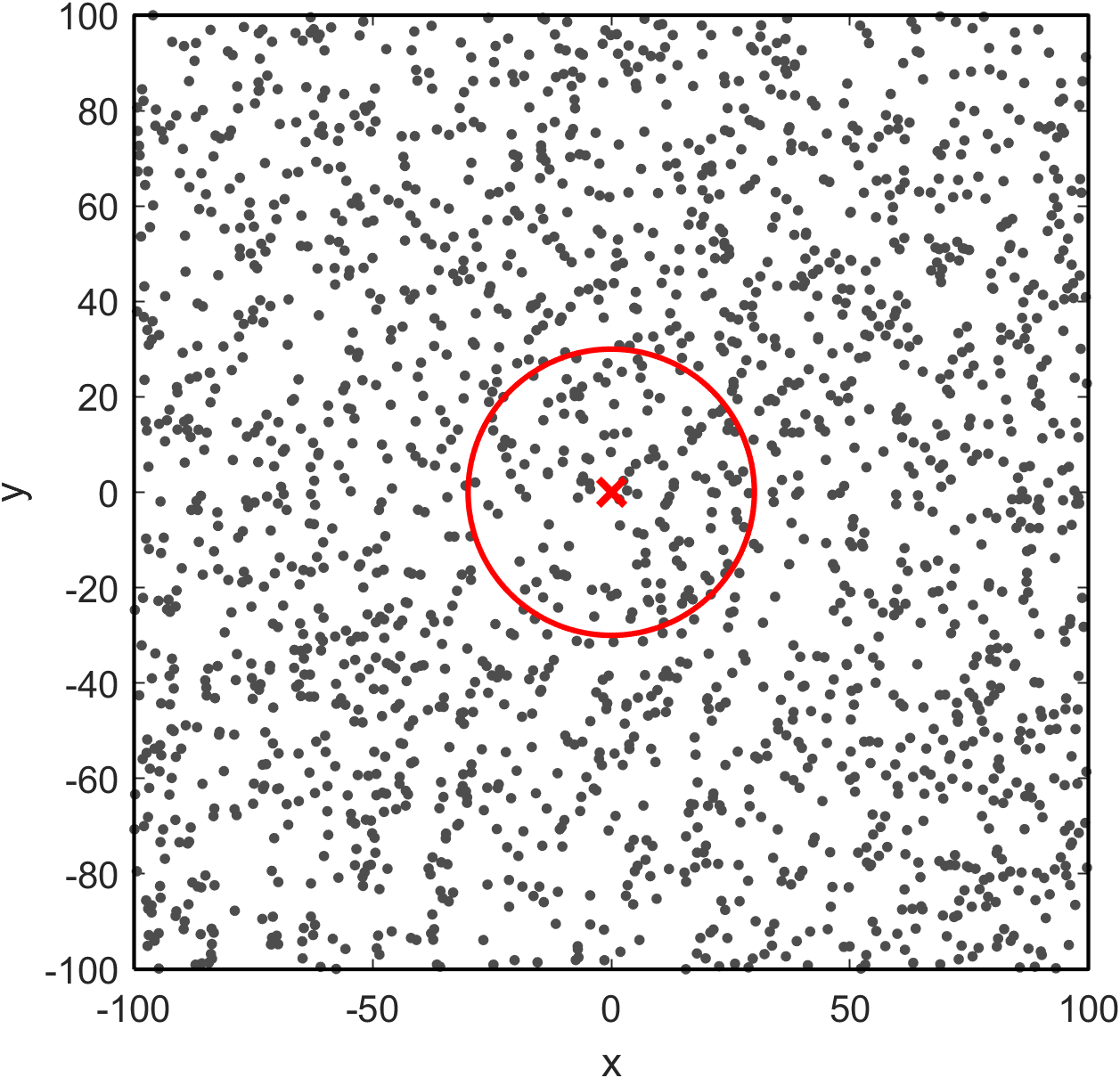}%
\label{fig:intuition_a}}
\hfil
\subfloat[]{\includegraphics[width=2.3in]{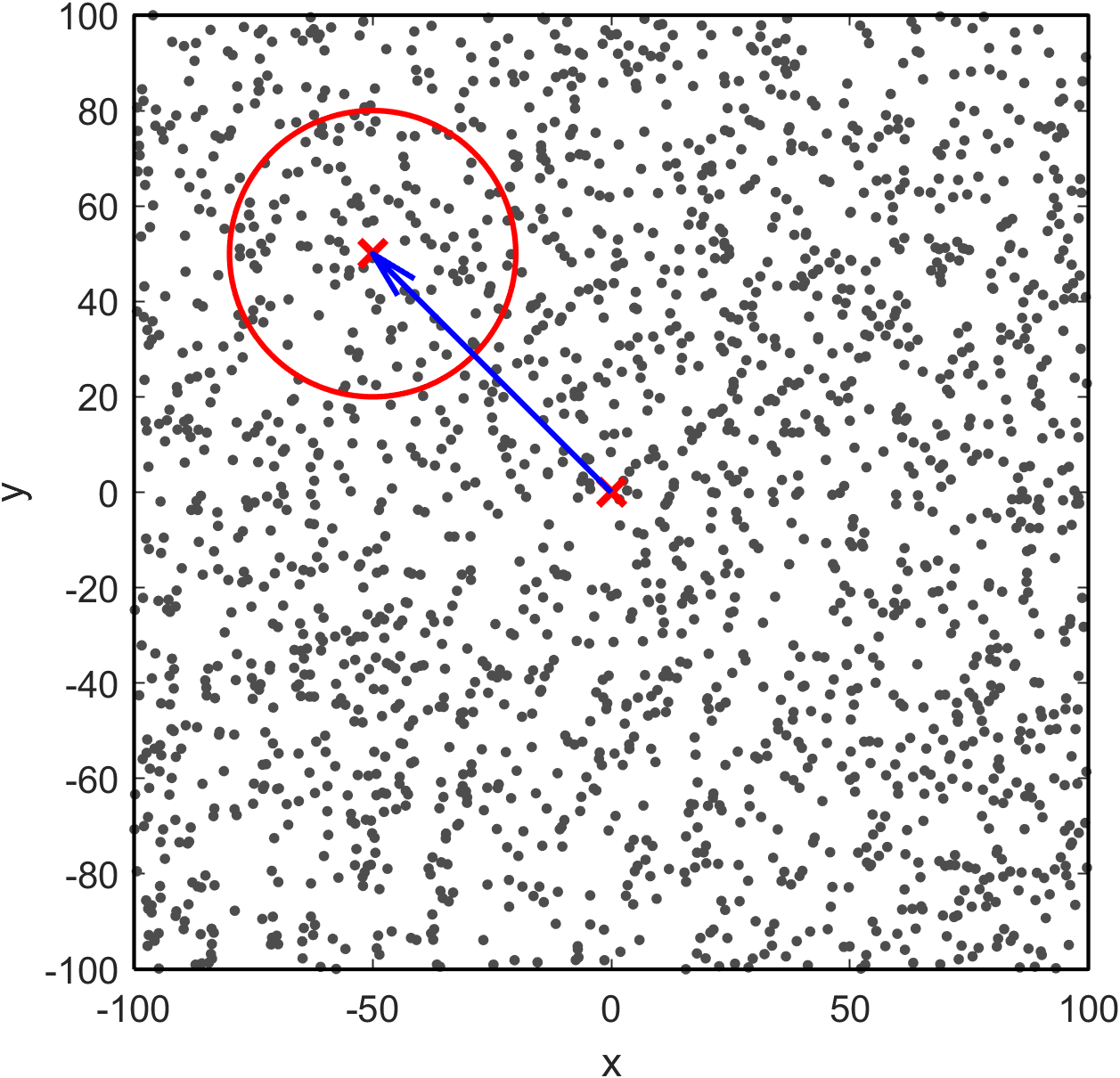}%
\label{fig:intuition_b}}
\hfil
\subfloat[]{\includegraphics[width=2.3in]{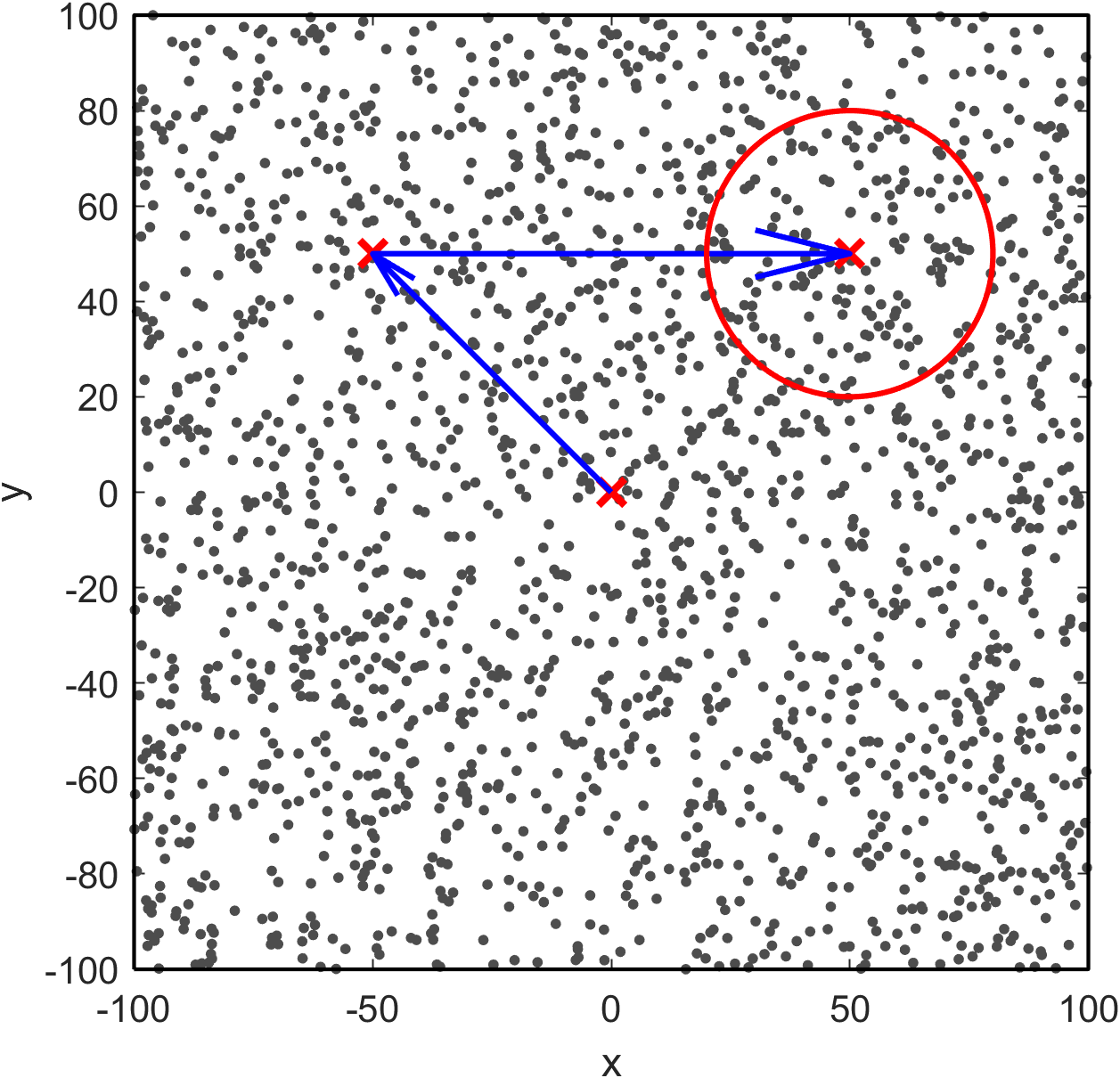}%
\label{fig:intuition_c}}

\subfloat[]{\includegraphics[width=2.3in]{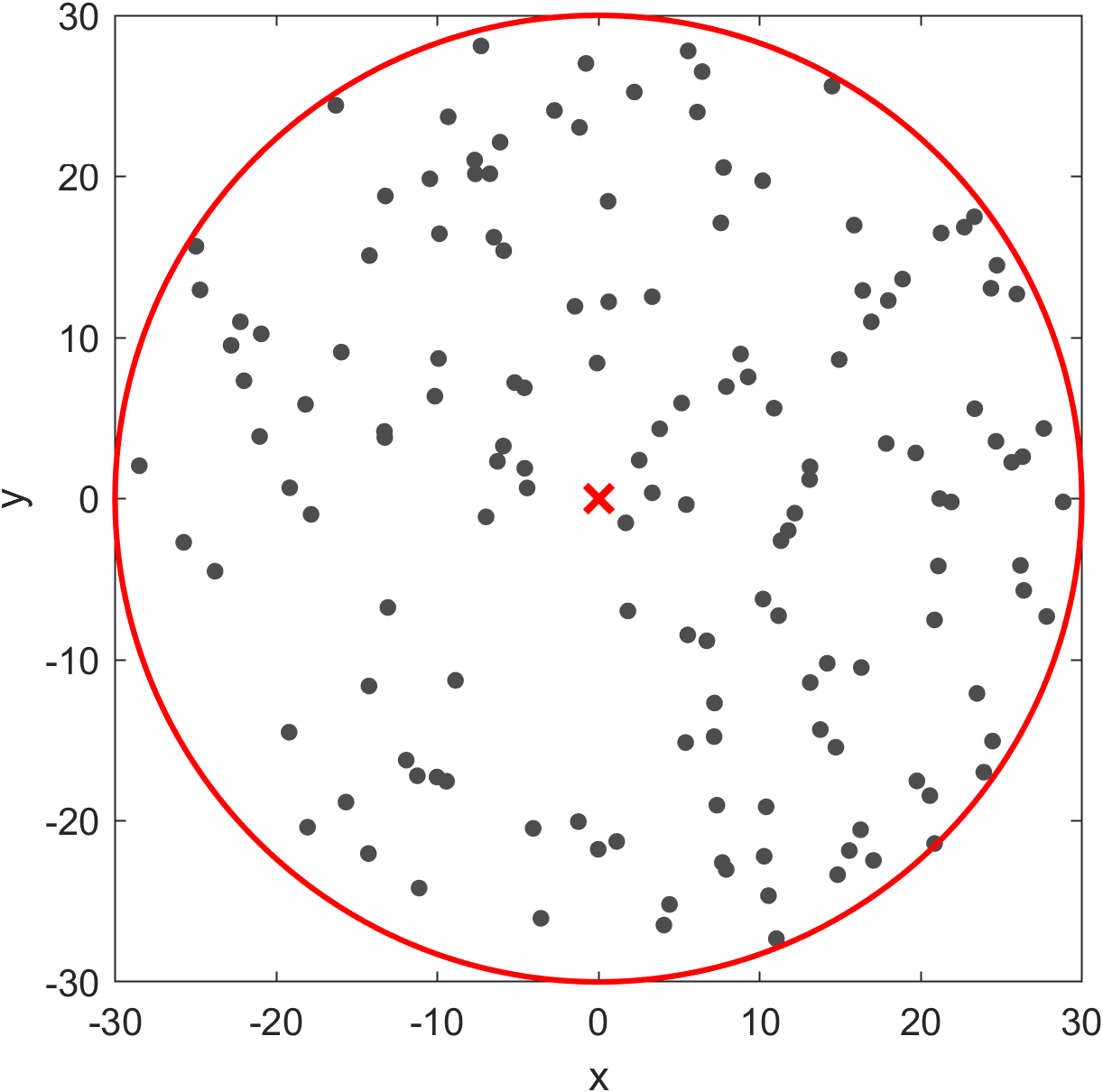}%
\label{fig:intuition_d}}
\hfil
\subfloat[]{\includegraphics[width=2.3in]{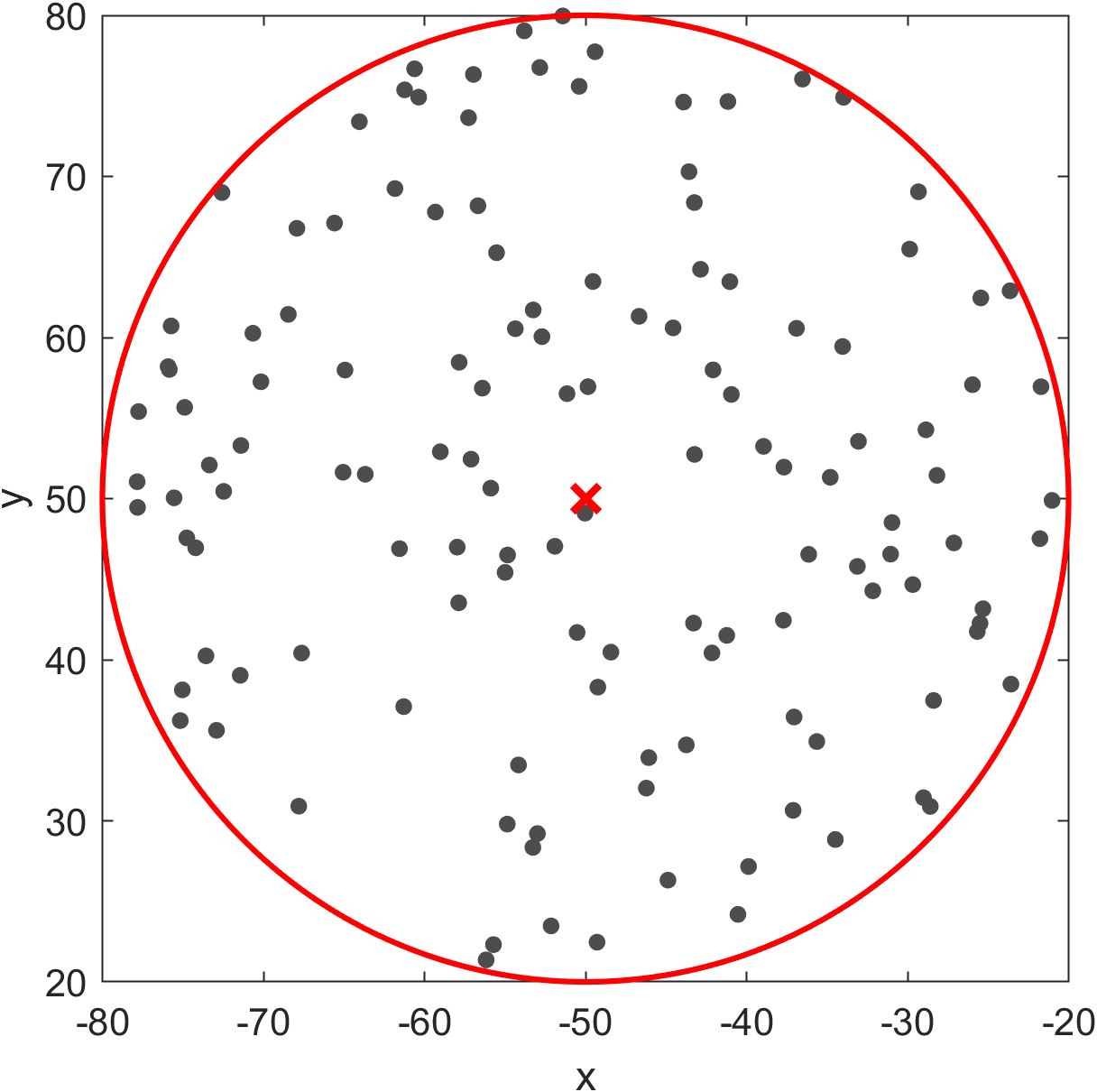}%
\label{fig:intuition_e}}
\hfil
\subfloat[]{\includegraphics[width=2.3in]{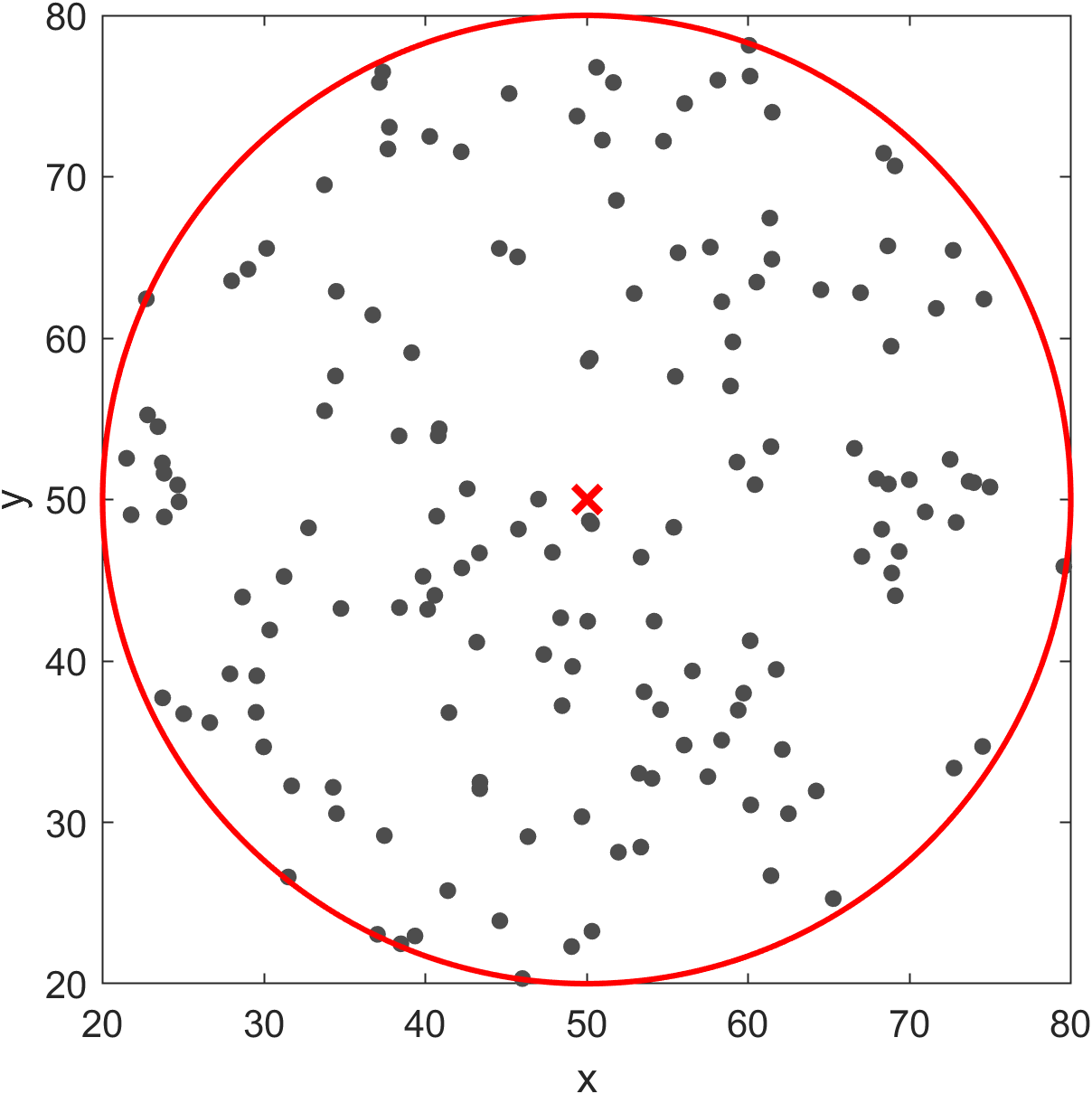}%
\label{fig:intuition_f}}

\caption{Illustration of the modeling intuition. Panels (a)–(c) show a moving target (red cross) and the nearby sensors used for localization (those inside the red circle). Panels (d)–(f) present the corresponding zoomed-in views, highlighting the local sensor configuration around the target. As the target moves, the set of nearby sensors changes and appears as a random point pattern.}
\label{fig:intuition_combined}
\end{figure*}

In this work, we focus on the problem of localizing a moving target in wireless sensor networks under realistic signal propagation effects. The sensor nodes may be deterministically placed or randomly deployed, subject to applications. In practice, the localization algorithm typically uses only the sensors that are sufficiently close to the target, since distant sensors provide measurements that are too weak or too noisy to be informative. As the target moves through space, the subset of nearby sensors continually changes. This evolving set of “usable” sensors, illustrated in Figures~\ref{fig:intuition_a}–\ref{fig:intuition_f}, appears from the target’s perspective as a random point pattern—even if the underlying sensor deployment is fixed and deterministic. Moreover, localization performance depends only on the relative spatial configuration between the target and the surrounding sensors, rather than their absolute geographical coordinates. These observations motivate our modeling approach: instead of treating the sensor field as fixed and the target as moving, we adopt an equivalent representation in which the target is modeled as an emitting source fixed at the origin, while the sensor nodes are modeled as points drawn from a point process (PP). By treating the effective neighboring sensors as realizations of a random PP, we preserve the inherent spatial uncertainty encountered by the moving target and obtain a mathematically tractable framework for characterizing average localization performance. This stochastic-geometry-based modeling paradigm—widely used in the literature \cite{KeelerRossXia2018,7745970,BlaszczyszynKarrayKeeler2013,BlaszczyszynKarrayKeeler2015,HaenggiEtAl2009}—enables rigorous analysis of the expected localization performance under varying network densities and measurement uncertainties.

Building on these insights, this work advances the performance analysis of wireless localization in \(d\)-dimensional networks where sensor nodes are modeled by a broad class of stationary and isotropic PPs (i.e., distributions invariant in law under translations and rotations; see \cite[Chapters~12 and~15]{DaleyVereJones2008}). Recent studies~\cite{WSNPPP1,BergelNoam2018} have primarily focused on information-theoretic characterizations---most notably the Cramér--Rao bounds (CRBs), which establish lower bounds on the mean squared error (MSE) of any unbiased estimator without specifying an explicit estimation procedure; see also~\cite{ReviewCRLB}. Among these, the approach of~\cite{BergelNoam2018} is most closely related to ours, as it assumes that sensor nodes follow a homogeneous Poisson point process (HPPP), which constitutes a special case of the broader class of stationary isotropic PPs considered in this work. In contrast, we adopt a complementary perspective by directly quantifying the MSE and conditional MSE (CMSE), given the number of sensors, of a concrete and practically implementable localization strategy (see Algorithm~\ref{Algo1}). Specifically, we consider fusion-based estimators---both the simple-average and the weighted estimators---wherein a central processor aggregates RSS and AOA measurements from spatially distributed sensor nodes to jointly infer the target position. Analytical and approximate upper bounds are then derived for the MSE and CMSE across a general class of stationary isotropic sensor networks.

As the exact MSE and CMSE for stationary isotropic sensor networks are generally intractable, we begin our analysis by establishing an approximation theorem. Specifically, when the signal propagation effect is sufficiently strong, the joint distribution of RSS and AOA observations collected from a broad class of stationary isotropic sensor networks satisfying the short-range dependency (SRD) condition \eqref{equ: SRD} becomes, in distribution, indistinguishable from that of a homogeneous Poisson point process (HPPP) network. Leveraging this equivalence, we analyze the performance of the fusion-based estimators under an HPPP model. For such a deployment within a finite observation region, we derive tractable analytical upper bounds for both the MSE and CMSE of the simple average estimator, and establish explicit scaling laws with respect to sensor density, observation radius, and noise variance. The approximation theorem then justifies the use of these HPPP-based bounds as reasonable approximations for stationary isotropic deployments satisfying the SRD condition \eqref{equ: SRD} in noisy regimes. We further investigate weighted fusion estimators and show that, as the sensor network becomes denser, these estimators are generally asymptotically inferior to the simple average estimator due to the inefficient use of information from distant sensors. Extensive simulation studies are conducted to evaluate the tightness of the proposed upper bounds in HPPP, repulsive, and clustered sensor networks under realistic propagation environments. The performance of the simple average estimator and weighted estimators is also compared numerically.

The remainder of this paper is organized as follows. Section~\ref{Section: Convergence} introduces the system model, the target estimator, and presents the convergence result, which establishes the asymptotic equivalence between observations collected from a broad class of stationary isotropic sensor networks and those from an HPPP network. Examples based on repulsive and clustered PPs are provided to illustrate this convergence behavior. Section~\ref{Section: Poisson model} studies the HPPP model and provides analytical upper bounds on the MSE and CMSE of the simple average estimator. Weighted fusion estimators are also considered and compared with the simple average estimator. In Section~\ref{Section: Simulation}, simulation studies are conducted to support and validate the analytical findings. Section~\ref{Section: Discussion} concludes with a summary of our findings and outlines potential directions for future research, while Section~\ref{Section: proof} contains the proofs of the main theoretical results.

\section{Model Definition And Convergence Result}\label{Section: Convergence}
In this section, we present the system model, state our convergence results in greater detail, and discuss their applicability through examples.

\subsection{Model Definition and Target Estimator}
We investigate the performance of localizing a moving target in wireless sensor networks. The target may either emit signals with constant power or reflect signals transmitted by the sensors, where the sensors themselves transmit with the same constant power. In the latter case, the only difference is that the propagation distance is doubled. Without loss of generality, we assume the target is an emitting source located at the origin $\bf0$ with constant transmission power. The sensor locations, denoted by $\Xi$, are distributed according to a simple (no multiple points at the same location) stationary and isotropic PP on $\mathbb{R}^d\backslash\{\bf0\}$ with intensity $\lambda > 0$. Write $\Xi =\sum_{i \in \mathcal{I}^\Xi} \delta_{X_i}$, where $\mathcal{I}^\Xi$ is some index set and $\delta_x$ is the Dirac measure at $x$. The Cartesian coordinate $X_i:=(X_{i,1},...,X_{i,d})$ of each sensor location can be represented using the hyperspherical coordinates $\lb  R_i, \boldsymbol{\Psi}_{i} \rb $ with $\boldsymbol{\Psi}_{i} := \lcb  \Psi_{i}^{(j)}\rcb_{j=1}^{d-1}$ \cite{Blumenson1960}, where
\begin{equation*}
\begin{cases}
R_i :=|X_i|:= \sqrt{\sum_{j=1}^d X_{i,j}^2},\\
\vpj_i := \arctan\!\lb  \sqrt{\sum_{k=j+1}^d X_{i,k}^2}\Big/{X_{i,j}}  \rb , 1\leq j\leq d-2,\\
\vpzero_i := \arctan\!\lb  X_{i,d}/X_{i,d-1} \rb ,
\end{cases}
\end{equation*}
where $ R_i>0$ is the radial coordinate, $0\leq\vpj_i \leq \pi$ for $1\leq j\leq d-2$ and $0\leq \vpzero_i<2\pi$ are the angular coordinates. The variable $\vpzero_i$ is called the azimuthal angle, and $\lcb \vpj_i\rcb_{j=1}^{d-2}$ are referred to as the elevation angles. Conversely, the Cartesian coordinates can be recovered via
\begin{equation*}
    X_i = R_i\eta(\boldsymbol{\Psi}_{i}).
\end{equation*}
where $\eta(\boldsymbol{\Psi}_{i})$ denotes the unit direction vector associated with $\boldsymbol{\Psi}_{i}$, given by

\begin{equation*}
\begin{cases}
\eta(\boldsymbol{\Psi}_{i})_j =  \cos\vpk_i  \prod_{k=1}^{j-1} \sin \vpk_i  , 1\leq j\leq d-1, \\[6pt]
\eta(\boldsymbol{\Psi}_{i})_{d} =  \sin\vpzero_i  \prod_{k=1}^{d-2} \sin \vpk_i.
\end{cases}
\end{equation*}
For example, when $d=2$, the Cartesian coordinate $(X_{i,1},X_{i,2})$ can be represented using the polar coordinate
\[
\lb  R_i,\Psi_i^{(1)} \rb  = \lb \sqrt{X_{i,1}^2+X_{i,2}^2},\arctan(X_{i,2}/X_{i,1})\rb ,
\]
or equivalently,
\[
(X_{i,1},X_{i,2}) = \lb  R_i \cos \Psi_i^{(1)},R_i \sin \Psi_i^{(1)} \rb .
\]
When $d=3$, the Cartesian coordinate $(X_{i,1},X_{i,2},X_{i,3})$ can be represented using the spherical coordinate 
\[
\lb  R_i,\Psi_i^{(1)},\Psi_i^{(2)}\rb  = \lb \sqrt{X_{i,1}^2+X_{i,2}^2+X_{i,3}^2},\arctan\!\lb  \sqrt{X_{i,2}^2+X_{i,3}^2}\Big/X_{i,1}\rb ,\arctan(X_{i,3}/X_{i,2})\rb ,
\]
or equivalently
\[
(X_{i,1},X_{i,2},X_{i,3}) = \lb  R_i \cos\Psi_i^{(1)}, R_i \cos \Psi_i^{(2)}  \sin\Psi_i^{(1)},R_i \sin \Psi_i^{(2)}  \sin\Psi_i^{(1)} \rb .
\]

Each sensor $X_i$ receives signals from the emitting source and measures the RSS and AOA of the incoming signals. We denote by $P_i$ the RSS and by $\boldsymbol{\Theta}_{i}:=  \lcb \vtj_i\rcb_{j=1}^{d-1}$ the received AOAs, all measured relative to the sensor location $X_i$. These measurements are then transmitted to a fusion center, where they are used to determine the target's position.

We first define our choice of the target estimator before stating our model for the RSS and AOAs. In practice, localization algorithms typically rely only on sensors that are sufficiently close to the target, since sensors located far away produce signal measurements that are too weak or too noisy to be informative. Accordingly, we restrict attention to sensors lying inside the ball \(\BallD\) of radius \(\mathcal{R} > 0\) centered at the origin $\bf 0$ (i.e., at the target), where $\R$ is referred to as the observation radius, chosen by the fusion center. We define the restriction of $\Xi$ to the ball $\BallD$ of radius $\R$ centered at the origin by
\[
\Xi_\R := \sum_{i \in \mathcal{I}^\Xi} \delta_{X_i}\, \I\!\lb X_i \in \BallD\rb = \sum_{i \in \mathcal{I}^{\Xi_\R}} \delta_{X_i},
\]
where $\mathcal{I}^{\Xi_\R} \subset \mathcal{I}^\Xi$ denotes the index set of points belonging to $\Xi_\R$. We denote by $|\Xi_\R|$ the number of sensors contained in the observation region.

Each sensor at $X_i,i \in \mathcal{I}^{\Xi_\R}$, provides an individual estimate $\tilx_i$ of the target's location based solely on the RSS and AOA measurements. The RSS $P_i$ can be used to estimate the range $R_i$, and we denote the corresponding estimator by $\hatr_i$. In this work, unless otherwise stated, we only assume $\hatr_i$ is a non-increasing function of $P_i$, but do not commit to any specific form for the estimator. Later in Lemma~\ref{lemma: unbiasedEstimator}, we will show that the target estimator is unbiased regardless of the choice of the distance estimator $\hatr_i$. The individual estimated target location $\tilx_i := \lb \tilde{X}_{i,1},...,\tilde{X}_{i,d}\rb $ is given by projecting the estimated distance $\hatr_i$ along the measured AOA direction vector $\eta(\boldsymbol{\Theta}_{i})$ from the sensor located at $X_i$. That is,
\begin{equation}\label{equ: individual_estimate_AOA_vec}
    \tilde{X}_{i} = X_i+\hatr_i \eta(\boldsymbol{\Theta}_{i}).
\end{equation}
Figure~\ref{fig: prediction in the two-dimensional system} shows the estimated target location provided by individual sensors in the two-dimensional noisy system.

\begin{figure}[!t]
\centering
\begin{tikzpicture}[scale=1.2]

  \draw[->] (-1,0) -- (5,0) node[right] {$x$};
  \draw[->] (0,-1) -- (0,4) node[above] {$y$};

  \coordinate (O) at (0,0);
  \coordinate (P) at (3,2); 
  \coordinate (Xaxis) at (1,0); 
  \coordinate (Xaxispoint) at (4,2);
  \coordinate (predict) at (1,-0.5);

  \node at (O) [star, star points=5, star point ratio=2.25,
                fill=black, minimum size=10pt, inner sep=0pt] {};
    \node at (predict) {$\times$};
    \node at (predict) [right] {$(\tilde{X}_{i,1}, \tilde{X}_{i,2})$};
    
  \draw[dotted, thick] (O) -- (P) node[midway, above left] {$R_i$};
  \draw[dotted, thick] (predict) -- (P) node[midway, below right] {$\hatr_i$};

  \fill (P) circle (2pt) node[above right] {$(X_{i,1}, X_{i,2})$};

  \draw[dotted] (P) -- ++(1,0);
  \draw[dotted] (P) -- ++(0,1);

  \draw pic["$\Psi_i^{(1)}$", draw=black, angle eccentricity=1.4, angle radius=20]
    {angle = Xaxis--O--P};

  \draw pic["$\Theta_i^{(1)}$", draw=black, angle eccentricity=1.4, angle radius=20]
    {angle = predict--P--Xaxispoint};

\end{tikzpicture}
\caption{Estimated target location provided by individual sensors in the two-dimensional noisy system.}
\label{fig: prediction in the two-dimensional system}

\end{figure}

The final estimate of the target location, denoted by $\X := (\X_{1},...,\X_{d})$, is defined as the (simple) average of the individual estimates:
\begin{equation}\label{equ: targetEstimator}
    \X :=\frac{1}{|\Xi_\R| } \sum_{i \in \mathcal{I}^{\Xi_\R}} \tilde{X}_{i}.
\end{equation}
The localization procedure is summarized in Algorithm \ref{Algo1}. In Sections \ref{Section: weighted function result} and \ref{Section: weighted fusion}, we also consider a class of weighted average estimators \eqref{equ: weighted_estimator_order} and compare their performance with that of the simple average estimator.

\begin{algorithm}[H]
\caption{Location Estimation using RSS and AOA.}\label{Algo1}
\begin{algorithmic}
\STATE 
\STATE {\textsc{INPUT}} Sensor data $\{ P_i, \boldsymbol{\Theta}_{i} \}_{i \in \mathcal{I}^{\Xi_\R}}$ and sensor locations $\{X_i\}_{i \in \mathcal{I}^{\Xi_\R}}$
\STATE {\textsc{OUTPUT}} Location estimate $\X$
\STATE 
\FOR{each sensor $i \in \mathcal{I}^{\Xi_\R}$}
    \STATE \hspace{0.5cm} estimate the distance $\hatr_i$ from RSS $P_i$
    \STATE \hspace{0.5cm} compute individual estimate $\tilde{X}_{i} = X_i+\hatr_i \eta(\boldsymbol{\Theta}_{i})$
\ENDFOR
\STATE 
\STATE \hspace{0.5cm} average to get an overall estimate $\X = \dfrac{1}{|\Xi_\R|} \sum_{i \in \mathcal{I}^{\Xi_\R}} \tilde{X}_i$
\STATE \hspace{0.5cm} \textbf{return} $\X$
\end{algorithmic}
\end{algorithm}

A standard assumption in the literature \cite{BlaszczyszynKarrayKeeler2013,BlaszczyszynKarrayKeeler2015,KeelerRossXia2018,7745970}, adopted throughout this work, is that in the absence of signal propagation effects, the RSS $P_i$ depends deterministically on the distance $R_i$ to the target through a non-increasing path-loss function $\ell : \posRealLine \to \posRealLine$, where $\posRealLine:=(0,\infty)$. That is, $P_i = \ell(R_i)$. A common choice for $\ell$ is given by $\ell(r) =(Cr)^{-\beta}$ for some $\beta , C > 0$. $\beta$ is called the path-loss exponent, which is a system parameter that is typically estimated from measurement data. The constant $C$ is determined by the received power measured at a prescribed reference distance for a chosen transmission power; see Section~\ref{section: 2d_system_model} for an example. The azimuthal AOA is given by $\vtzero_i = \lb \vpzero_i - \pi\rb \bmod 2\pi\in[0,2\pi)$, and the elevation AOAs are given by $\vtj_i = \pi-\vpj_i\in[0,\pi]$, for $1\leq j\leq d-2$.

In practice, random propagation effects influence both the strength and direction of the received signal due to phenomena such as multipath fading (caused by signals taking multiple paths and interfering with each other) and shadow fading (due to signals colliding with obstacles). Consequently, the relationship between the RSS $P_i$ and the distance $R_i$ is no longer deterministic, and the measured AOAs deviate from the underlying true AOAs in the noiseless case.

To capture this, we assume that the RSS is further affected by random propagation effects modeled by independent and identically distributed (i.i.d.) positive random variables (RVs) $S(\sigma), S_1(\sigma), S_2(\sigma), \ldots$, parameterized by a non-negative noise parameter $\sigma$. For example, one may take 
\begin{equation}\label{equ: log-normal shadowing}
    S(\sigma) = \exp\{\sigma B-\sigma^2/\beta\}
\end{equation}
for $B$ a standard normal RV
and $\beta$ is the pass loss exponent \cite{BlaszczyszynKarrayKeeler2015,BlaszczyszynKarrayKeeler2013}, which is referred to as the log-normal shadowing. The case $\sigma=0$ represents an ideal, noiseless propagation environment. Increasing $\sigma$ corresponds to higher noise levels. Accordingly, we assume the RSS is given by
\begin{equation}\label{equ: relation between P and R}
    P_i :=P_i\sig := \ell(R_i) S_i(\sigma).
\end{equation}

Furthermore, the measured AOAs generally deviate from the exact AOAs (in the noiseless case), with variability increasing in $R_i$. 
To model this, we define the observed AOAs by
\begin{equation}\label{equ: wrapped normal AOA}
    \vtj_i \;=\; \tilvtj_i \bmod 2\pi, 
    \qquad 0 \leq \vtj_i < 2\pi,
\end{equation}
where the latent variables $\tilvtj_i$ are assumed to follow Gaussian distributions centered at the true angles, with variances increasing with the distance $R_i$. Specifically, conditioned on $\lb R_i,\vpzero_i \rb$, the azimuthal AOA $\tilvtzero_i$ satisfies
\begin{equation}\label{equ: unwrapped normal AOA 1}
\lb \tilvtzero_i \given \vpzero_i, R_i \rb 
\;\dequal\;
\N\!\lb \vpzero_i - \pi,\, \E(R_i) \rb ,
\end{equation}
where $\E:\posRealLine\to\posRealLine$ is non-decreasing. 
Similarly, for $1\leq j\leq d-2$, conditioned on $\lb R_i,\vpj_i \rb$, the elevation AOA $\tilvtj_i$ satisfies
\begin{equation}\label{equ: unwrapped normal AOA 2}
\lb \tilvtj_i \given \vpj_i, R_i \rb  
\;\dequal\;
\N\!\lb \pi - \vpj_i,\, \E(R_i) \rb .
\end{equation}

Under this construction, $\vtj_i$ follows a \emph{wrapped normal distribution} on $[0,2\pi)$ (see \cite[Section~2.2.6]{Jammalamadaka2001} for background). Moreover, as $\E(R_i)$ increases, the wrapped normal distribution approaches the uniform distribution on $[0,2\pi)$ (this follows directly from the alternative representation of its density in \cite[(2.2.15)]{Jammalamadaka2001}), corresponding to highly noisy angular observations.

In practice, physical sensors observe an azimuthal AOA in $[0,2\pi)$ and elevation AOAs in $[0,\pi]$. 
While one could enforce these ranges by wrapping the azimuthal and elevation components separately, we instead allow all angular components to take values in $[0,2\pi)$ for analytical convenience. 
Although this representation is not consistent with the standard hyperspherical coordinate system, we will show that it has no effect on the localization procedure considered in this work.

Recall from \eqref{equ: individual_estimate_AOA_vec} that each individual estimate $\tilde{X}_i$ is obtained by projecting the estimated distance $\hatr_i$ along the direction specified by the measured AOAs. 
Hence, the estimator depends on the angular measurements only through their induced Cartesian direction vector, rather than through a particular hyperspherical parametrization. The standard hyperspherical coordinate system represents directions using elevation angles in $[0,\pi]$ and an azimuthal angle in $[0,2\pi)$, yielding a unique representation of Cartesian direction vectors. 
Extending the elevation angles to $[0,2\pi)$ introduces a non-unique parametrization, meaning that different angle tuples may correspond to the same direction in $\mathbb{R}^d$. 
However, this non-uniqueness does not affect the localization procedure considered here. For example, in three dimensions, the angle tuples $\lb \Psi_i^{(1)},\Psi_i^{(2)} \rb = (0.75\pi,\,1.25\pi)$ and $\lb \Psi_i^{(1)},\Psi_i^{(2)} \rb = (1.25\pi,\,0.25\pi)$ correspond to different angular representations but generate the same direction vector in $\mathbb{R}^3$. Consequently, they yield identical projected Cartesian estimates through \eqref{equ: individual_estimate_AOA_vec}. Therefore, the projection-based localization algorithm is invariant under such reparametrizations.

For this reason, we may safely allow all angular components, including the elevation angles, to take values in $[0,2\pi)$. This simplifies the analysis and enables us to adopt a unified wrapped Gaussian model for all AOAs without affecting the resulting location estimates.

Our choice of the wrapped Gaussian model for the estimated AOAs is motivated by two considerations. 
First, in classical array processing (e.g., Maximum Likelihood/Multiple Signal Classification and their variants; see \cite{MUSIC,MUSICML}), AOA estimates are \emph{asymptotically normal} at high signal-to-noise ratio (SNR) or with many snapshots, with variance approaching the CRB \cite{AoAGaussian}. 
This provides a theoretical basis for modeling the angular perturbation as Gaussian around the true angle. 
Since physical angle measurements are inherently periodic, we adopt a wrapped Gaussian model to account for this periodicity.

Second, because angles lie on a circle or sphere, it is natural to adopt circular distributions for modeling AOA measurements. 
In the literature, von Mises distributions in 2D and von Mises–Fisher distributions in 3D are commonly used to capture periodicity and directional uncertainty \cite{circularGaussian2D,circularGaussian3D,circularGaussian3Dexpermental}. 
In this work, we instead adopt a wrapped Gaussian model, which arises naturally from a Gaussian perturbation of the true angle and provides a convenient analytical representation while respecting the periodic nature of angular data. 
Moreover, it is known that for any wrapped normal distribution, one can construct a von Mises distribution with the same mean direction and first circular moment matching (see \cite[Section~2.2.6]{Jammalamadaka2001}). 
In particular, the two distributions exhibit very similar concentration and shape in the small-noise regime, and both models converge to the uniform distribution on $[0,2\pi)$ in the large-noise limit. 
Therefore, the wrapped Gaussian model captures essentially the same statistical behavior as the von Mises distribution, while being more convenient for analytical derivations in our setting.

The observables available at each sensor $X_i, i\in\mathcal{I}^\Xi$, are $\lb  P_i, \boldsymbol{\Theta}_{i}\rb $. We let $N_i:=1/P_i = \frac{g(R_i)}{S_i\sig}$ to denote the inverse RSS, where $g(r) := 1/\ell(r)$, and define the observable process
\begin{equation}\label{equ: observable_process}
    \NN^\sig_\Xi:=\sum_{i \in \mathcal{I}^\Xi} \delta_{\lb  N_i,\boldsymbol{\Theta}_i\rb },
\end{equation}
which collects the (inverse) RSS and AOA measurements from each sensor \(X_i, i\in\mathcal{I}^\Xi\). These measurements are transmitted to the fusion center, which estimates the target position based solely on these observed quantities.

Analytical properties of Algorithm~\ref{Algo1} for a general stationary and isotropic sensor network $\Xi$ are intractable, except in a few special cases, which themselves require separate, model-specific analyses. Such an approach is technically challenging and offers limited general insight. To overcome this difficulty, we adopt an alternative strategy. We first establish that, in the presence of significant noise and propagation effects, the signal measurements $\NN^\sig_\Xi$ induced by a class of stationary and isotropic networks $\Xi$ become statistically equivalent to those generated by an HPPP network (see Theorem~\ref{convergence_thm} and the accompanying discussion). Since the localization algorithm relies only on the signal measurements, we then analyze its performance under the HPPP model, which yields tractable closed-form upper bounds for the MSE and CMSE; see Theorem~\ref{thm: CMSD and MSD}.

\subsection{Main Results}

In this section, we study the distributional discrepancy between the observable processes $\NN^\sig_\Xi$ and $\NN^\sig_\Phi$, where $\Phi$ is an HPPP on $\IR^d$ with the same intensity $\lambda$. Both processes are PPs defined on $\posRealLine \times [0,2\pi)^{d-1}$ with the same mean measure
\begin{align}
    \M^\sig\!\lb  \dx t, \dx \boldsymbol{\nu} \rb 
    &:= \IE\sum_{i \in \mathcal{I}^\Xi} \I\!\lb  N_i\in \dx t, \boldsymbol{\Theta}_{i}\in \dx \boldsymbol{\nu} \rb \nonumber\\
    &= \IE\!\lsb \sum_{i \in \mathcal{I}^\Xi} \IP\!\lb  N_i\in \dx t, \boldsymbol{\Theta}_{i}\in \dx \boldsymbol{\nu} \given X_i \rb  \rsb \nonumber\\
    &=  \lambda\int_{\IR^d} \IP\!\lb \frac{g(|x|)}{S\sig}   \in \dx t\rb \IP\!\lb  \boldsymbol{\Theta}_{1}\in \dx \boldsymbol{\nu} \given X_1=x \rb \dx x, \label{equ: same_mean_measure}
\end{align}
where the second last equality is by the tower property conditioning on $\Xi$ and the last equality is by Campbell's formula (see \cite[Theorem 4.1]{Haenggi2012}).

We first observe that $\NN^\sig_\Phi$ is a Poisson point process (PPP) on $\posRealLine \times [0,2\pi)^{d-1}$ by the Poisson Mapping Theorem; see \cite[Chapter~1.3.3]{BaccelliBlaszczyszyn2009a}. Indeed, consider the independently marked PP
\[
\sum_{i \in \mathcal{I}^\Phi}\delta_{ \lb  X_i, S_i\sig, \boldsymbol{\Theta}_{i} \rb  },
\]
which forms a PPP on $\IR^d \times \posRealLine \times [0,2\pi)^{d-1}$. Since the points of $\NN^\sig_\Phi$ are obtained as measurable transformations of this marked process, it follows that $\NN^\sig_\Phi$ is also a PPP. Consequently, we are essentially approximating $\NN^\sig_\Xi$ using a PPP with the same mean measure $\M^\sig$.

To make use of existing Poisson approximation results \cite{BarbourHolstJanson1992,Xia1995OnMI,BARBOUR19929}, we first consider the observable process $\NN^\sig_\xi$ generated by a fixed snapshot $\xi$ of the random sensor network $\Xi$. Such a snapshot can be interpreted as a realization of the sensor network configuration observed by the target at a specific geographic location. 

Our goal is to approximate $\NN^\sig_\xi$ by a PPP $\ZZ^\sig_\xi$ with the same mean measure as $\NN^\sig_\xi$, which is given by
\begin{align*}
    \M_\xi^\sig\!\lb  \dx t, \dx \boldsymbol{\nu} \rb 
    &:= \IE\!\lsb\sum_{i \in \mathcal{I}^\Xi} \I\!\lb  N_i\in \dx t, \boldsymbol{\Theta}_{i}\in \dx \boldsymbol{\nu} \rb \given \Xi=\xi \rsb\\
    &= \sum_{i \in \mathcal{I}^\xi} \IP\!\lb  N_i\in \dx t, \boldsymbol{\Theta}_{i}\in \dx \boldsymbol{\nu} \given X_i=x_i \rb \\
    &= \int_{\IR^d} \IP\!\lb \frac{g(|x|)}{S\sig}  \in \dx t \rb \IP\!\lb  \boldsymbol{\Theta}_{1}\in \dx \boldsymbol{\nu} \given X_1=x \rb \xi(\dx x).
\end{align*}

We now establish a marked PPP approximation counterpart of \cite[Theorem~2.2]{KeelerRossXia2018}. For brevity, define $p_x^\sig(\tau):=\IP\!\lb \frac{g(|x|)}{S(\sigma)}\le \tau\rb $, $p_i^\sig(\tau)=p_{x_i}^\sig(\tau)$, and $\M_\xi^\sig (\tau) := \M_\xi^\sig \!\lb  (0,\tau], [0,2\pi)^{d-1} \rb $. 

For any PP $\zeta = \sum_{i\in\mathcal{I}^\zeta} \delta_{Y_i}$ on $\posRealLine\times[0,2\pi)^{d-1}$ with mean measure $\M_\zeta$, we define 
\[
\zeta|_\tau:=\sum_{i\in\mathcal{I}^\zeta} \delta_{Y_i}\I\!\lb  Y_i\in (0,\tau]\times [0,2\pi)^{d-1}\rb 
\]
as the restriction of $\zeta$ to $(0,\tau]\times [0,2\pi)^{d-1}$, and denote its mean measure by $\M_\zeta|_\tau$. We also define
\[
\zeta(\tau):=\zeta\!\lb  (0,\tau]\times [0,2\pi)^{d-1}\rb 
\]
as the number of points in $(0,\tau]\times [0,2\pi)^ {d-1}$.

Let $\HH$ be the space of all locally finite PPs on $\posRealLine\times[0,2\pi)^{d-1}$ (that is, processes assigning a finite number of points to every bounded subset of the space), equipped with the vague topology \cite[Chapter~15.7]{Kallenberg1983}, \cite[Chapter~1.5]{BaccelliBlaszczyszynKarray2024}. Let $\BB(\HH)$ denote the Borel $\sigma$-algebra generated by this topology. 

The total variation distance between two PPs $\zeta_1$ and $\zeta_2$ on $(\HH,  \BB(\HH))$ is defined as
\begin{equation}\label{equ: supremum_def_dtv}
\dtv\!\lb\law(\zeta_1), \law(\zeta_2)\rb
=
\sup_{A \in \BB(\HH)}
\left|
\IP(\zeta_1 \in A) - \IP(\zeta_2 \in A)
\right|,
\end{equation}
where $\law(\cdot)$ denotes the distribution of a random element; see \cite[Chapter~A.1]{BarbourHolstJanson1992}.

The following theorem characterizes the total variation distance between the distributions of $\NN^\sig_\xi|_\tau$ and $\ZZ^\sig_\xi|_\tau$, where $1/\tau$ can be interpreted as the smallest RSS value of interest to the fusion center of the sensor network.

\begin{theorem}\label{thm: dtv_thm}
Let $\xi$ be a locally finite and deterministic PP on $\IR^d \backslash \{\bf0\}$, and let $\NN^\sig_\xi$ be the observable process generated by $\xi$. Let $\ZZ_\xi^\sig$ be a PPP on $\posRealLine \times [0,2\pi)^{d-1}$ with the same mean measure as $\NN^\sig_\xi$, denoted by $\M_\xi^\sig$. Then
\begin{equation}\label{equ: dtv_thm}
\frac{1}{32}\min\!\lcb 1, \M_\xi^\sig (\tau)^{-1}\rcb\sum_{i\in\mathcal{I}^\xi}p_i^\sig(\tau)^2
\leq \dtv\!\lb \law\lb \NN^\sig_\xi|_\tau\rb ,\law\lb \ZZ^\sig_\xi|_\tau\rb \rb 
\le \sum_{i\in\mathcal{I}^\xi}p_i^\sig(\tau)^2
\le \M_\xi^\sig (\tau)\sup_{i\in\mathcal{I}^\xi}p_i^\sig(\tau).
\end{equation}
\end{theorem}

\begin{remark}
Theorem~\ref{thm: dtv_thm} can be viewed as a natural extension of \cite[Theorem~2.2]{KeelerRossXia2018}. In that work, the PPP approximation is established for the inverse RSS process $\sum_{i \in \mathcal{I}^\Xi}\delta_{N_i}$, also referred to as the propagation process, which captures only the signal strength. In contrast, Theorem~\ref{thm: dtv_thm} extends this result to a (marked) PPP approximation of the observable process, where each point is augmented with angular information. As a result, the approximation accounts not only for the signal strength but also for the directions of the incoming signals.
\end{remark}

Theorem~\ref{thm: dtv_thm} shows that for any fixed sensor configuration $\xi$ and any fixed $\tau>0$, the restricted observable process $\NN^\sig_\xi|_\tau$ is close to $\ZZ^\sig_\xi|_\tau$ whenever $\sup_{i\in\mathcal{I}^\xi}p_i^\sig(\tau)$ is small. In other words, $\NN^\sig_\xi|_\tau$ can be approximated by a PPP $\ZZ^\sig_\xi|_\tau$ with the same mean measure $\M^\sig_\xi|_\tau$, and the approximation error is quantified by \eqref{equ: dtv_thm}. 

For a random sensor configuration $\Xi$, the observable process $\NN^\sig_\Xi|_\tau$ induced by each realization of $\Xi$ can similarly be approximated by a PPP with the corresponding realized mean measure $\M^\sig_\Xi|_\tau$. More generally, $\NN^\sig_\Xi|_\tau$ can be approximated by a Cox PP $\ZZ^\sig_\Xi|_\tau$, that is, a PPP with a random mean measure $\M^\sig_\Xi|_\tau$ (satisfying $\IE\M^\sig_\Xi|_\tau = \M^\sig|_\tau$). The approximation error can be quantified by the following corollary of Theorem~\ref{thm: dtv_thm}.

\begin{corollary}\label{corollary: cox_approximation}
Let $\Xi$ be a locally finite PP on $\IR^d \backslash \{\bf0\}$, and let $\NN^\sig_\Xi$ be the observable process generated by $\Xi$. Let $\ZZ_\Xi^\sig$ be a Cox PP on $\posRealLine \times [0,2\pi)^{d-1}$ directed by the random mean measure $\M_\Xi^\sig$. Then
\begin{equation}\label{equ: dtv_thm_cox}
\dtv\!\lb \law\lb \NN^\sig_\Xi|_\tau\rb ,\law\lb \ZZ^\sig_\Xi|_\tau\rb \rb 
\le \lambda\int_{\IR^d} p_x^\sig(\tau)^2 \dx x.
\end{equation}
\end{corollary}

These observations suggest that $\NN^\sig_\Xi|_\tau$ can be well approximated by $\NN^\sig_\Phi|_\tau$ whenever the random mean measure $\M^\sig_\Xi|_\tau$ (of the approximating Cox PP) is approximately deterministic and close to $\M^\sig|_\tau $. To formalize this idea, we introduce some standard terminology from PP theory and stochastic geometry (see, for example, \cite[Chapter~6.4]{Haenggi2012}) before stating the key conditions under which this approximation holds.

We use $\varrho^{(2)}:\mathbb{R}^d\times\mathbb{R}^d\to\mathbb{R}_+:=[0,\infty)$ to denote the second-order product density of $\Xi$, defined as 
\[
\varrho^{(2)}(x,y):=\frac{\IE[\Xi(\dx x)\Xi(\dx y)]}{\dx x \dx y}, \quad x\ne y,
\]
assuming it exists. Intuitively, $\varrho^{(2)}(x,y)$ describes the joint intensity of observing points of the process simultaneously near $x$ and $y$, and thus captures the spatial interaction between points.

Since $\Xi$ is stationary and isotropic, $\varrho^{(2)}(x,y)$ depends only on the distance $|x-y|$ between points. Thus, there exists $\varrho^{(2)}_{\mathrm{SI}}:\mathbb{R}_+\to\mathbb{R}_+$ such that 
\[
\varrho^{(2)}(x,y)= \varrho^{(2)}_{\mathrm{SI}}(|x-y|), \quad \forall x,y\in\mathbb{R}^d.
\]
That is, the interaction between points depends only on how far apart they are, and not on their absolute locations or directions.

We define $h_{\mathrm{SI}}:\mathbb{R}_+\to\mathbb{R}_+$ as the pair correlation function of $\Xi$, given by
\[
h_{\mathrm{SI}}(r):=\frac{\varrho^{(2)}_{\mathrm{SI}}(r)}{\lambda^2}.
\]
Conceptually, $h_{\mathrm{SI}}(r)$ measures the deviation from complete spatial randomness at distance $r$. In particular, $h_{\mathrm{SI}}(r)=1$ indicates no interaction between points ($h_{\mathrm{SI}}(r)=1,\forall r>0$ in the HPPP case), while $h_{\mathrm{SI}}(r)<1$ indicates repulsion (points tend to be more regularly spaced), and $h_{\mathrm{SI}}(r)>1$ indicates clustering (points tend to form groups).

\begin{assumption}\label{assumption}
Suppose $\Xi$ is a stationary and isotropic PP on $\IR^d$ with intensity $\lambda>0$ and pair correlation function $h_{\mathrm{SI}}(r)$ satisfying the short-range dependence (SRD) condition
\begin{equation}\label{equ: SRD}
    \int_{0}^{\infty} |h_{\mathrm{SI}}(r)-1|\, r^{d-1}\, dr < \infty.
\end{equation}
Let $\Phi$ be an HPPP on $\IR^d$ with the same intensity measure.

Let $g : \posRealLine \to \posRealLine$ be a left-continuous, nondecreasing function with generalized inverse
\[
g^{-1}(y) := \inf\{ x \ge 0 : g(x) > y \},
\]
and satisfying $\lim_{t\to 0^+} g(t) = 0$.

Let $S\sig,S_1\sig,S_2\sig,...$ be a sequence of i.i.d. positive RVs indexed by a non-negative parameter $\sigma$, such that $S\sig\convergeP0$ as $\sigmatoinf$, and
\begin{equation}\label{equ: finite assumption}
    \limsup_{\sigma \to \infty} \IE\!\lsb g^{-1}(S\sig b)^d\rsb < \infty
\quad \forall b>0.
\end{equation}

For each point $X_i$, $i \in \mathcal{I}^\Xi$, let $\lb  R_i,  \lcb  \Psi_{i}^{(j)}\rcb_{j=1}^{d-1} \rb $ denote its hyperspherical coordinate representation. Assume that the AOAs $\lcb \vtj_i\rcb_{i \in \mathcal{I}^\Xi,\,1\le j\le d-1}$ defined in \eqref{equ: wrapped normal AOA} are conditionally independent given $\Xi$.
\end{assumption}

The SRD condition \eqref{equ: SRD} ensures that point interactions are short-ranged, in the sense that correlations between points decay sufficiently fast with distance and become negligible at large spatial scales. Intuitively, this means that the PP exhibits only local dependence and behaves approximately like an HPPP when observed over large spatial scales.

From a physical perspective, the regime of $\sigmatoinf$ corresponds to wireless environments in which the RSS is dominated by random propagation effects—such as shadow fading and multipath interference—rather than by its deterministic dependence on distance. For instance, under the log-normal shadowing model in \eqref{equ: log-normal shadowing}, taking logarithms on both sides of \eqref{equ: relation between P and R} yields a decomposition of the form\
\[
\ln P_i = \ln\ell(R_i) + \sigma B_i-\sigma^2/\beta,
\]
where $B_i$ are i.i.d. standard normal RVs. As $\sigma$ increases, the random component $\sigma B_i-\sigma^2/\beta$ dominates the distance-dependent term $\ln\ell(R_i)$, so that the observed RSS becomes effectively decoupled from the underlying spatial geometry of the network. Consequently, the underlying short-range spatial dependence structure of the sensor locations has negligible influence on the observable process.

In the following results, we study the weak convergence of random measures (e.g., mean measures and PPs) under the vague topology on the space of locally finite measures on $\posRealLine \times [0,2\pi)^{d-1}$; see \cite[Chapter~4.1]{Kallenberg1983}.

\begin{theorem}\label{theorem: mean measure converge}
    Under Assumption~\ref{assumption}, we have $\lim_\sigmatoinf\M^\sig_\Xi \dequal \lim_\sigmatoinf\M^\sig$ in the vague topology.
\end{theorem}

\begin{remark}
Theorem~\ref{theorem: mean measure converge} constitutes the main original contribution of this section, where we provide conditions under which the random mean measure $\M^\sig_\Xi $ (of the approximating Cox PP) converges.
\end{remark}

Theorem~\ref{theorem: mean measure converge} shows that the (random) mean measure $\M^\sig_\Xi$ becomes asymptotically deterministic and converges in distribution to $\M^\sig$ as the noise level $\sigma$ increases. This suggests that the approximating Cox PP $\ZZ^\sig_\Xi|_\tau$ converges in distribution to the PPP $\NN^\sig_\Phi|_\tau$ as $\sigma$ gets larger. In addition, the assumption $S\sig\convergeP0$ as $\sigmatoinf$ together with Corollary~\ref{corollary: cox_approximation} implies that the observable process $\NN^\sig_\Xi|_\tau$ is well approximated by $\ZZ^\sig_\Xi|_\tau$ for large $\sigma$. These observations lead to the following theorem, which extends \cite[Theorem~2.10]{KeelerRossXia2018}, itself a generalization of \cite[Theorem~3]{BlaszczyszynKarrayKeeler2013}. In contrast to these prior works, which focus on the asymptotic behavior of the propagation process $\sum_{i \in \mathcal{I}^\Xi}\delta_{N_i}$, our result incorporates additional structural information and provides a stronger characterization of the observable process. For this reason, we primarily compare our results with those in \cite{KeelerRossXia2018}.

\begin{theorem}\label{convergence_thm}
    Under Assumption~\ref{assumption}, we have $\lim_\sigmatoinf \NN^\sig_\Xi \dequal \lim_\sigmatoinf \NN^\sig_\Phi$ in the vague topology.
\end{theorem}

\begin{remark}\label{Remark_convergenc_thm}
    By Theorem~\ref{convergence_thm}, the observable processes generated by all stationary isotropic PPs obeying \eqref{equ: SRD} have the same limiting law, although the theorem is phrased to highlight their connection to that generated by the HPPP.
\end{remark}

The convergence result shows that, when the random propagation noise $\sigma$ is large, the observable process generated by an arbitrary stationary isotropic PP satisfying \eqref{equ: SRD} becomes statistically indistinguishable from that generated by an HPPP with the same intensity. This justifies the use of the HPPP network model in noisy environments, even when the underlying sensor configuration does not strictly follow an HPPP.

The proof of Theroem~\ref{convergence_thm} follows the spirit of \cite[Theorem~2.10]{KeelerRossXia2018}, where the authors proved that the propagation process $\sum_{i \in \mathcal{I}^\Xi}\delta_{N_i}$ generated by a $d$-dimensional random network, under some regularity conditions \cite[(2.6)]{KeelerRossXia2018}, converges in distribution to a PPP on $\posRealLine$. This limiting PPP can be interpreted as the propagation process induced by an underlying PPP network in $\IR^d$, which implies that, from the perspective of an observer interested only in the RSS or its related statistics, the underlying network appears indistinguishable from a PPP network when the noise is sufficiently significant.

Our proof follows a similar line of reasoning, but extends the propagation process to the observable process $\NN^\sig_\Xi$, in which we are concerned not only with the signal strength but also with the directions of the incoming signals.

It is worth mentioning that in \cite{KeelerRossXia2018}, the context of study is based on a setting where a receiver is located at the origin and receives signals from transmitters that form a random PP. In contrast, in our framework, the target is located at the origin and is modeled as an emitting source that transmits signals to sensors, which themselves form a random PP. In our setting, a fusion center collects the observable information gathered by the sensors, analogous to the receiver in \cite{KeelerRossXia2018}. Therefore, our result can be interpreted as a dual result, highlighting that the respective roles of transmitters and receivers can be interchanged; what ultimately matters is the information collected from the network.

It is also worth noting that the result in \cite{KeelerRossXia2018} applies to a broader class of random networks in $\IR^d$, including inhomogeneous models and deterministic models, whereas in our case, we restrict our analysis to stationary and isotropic random networks.

\subsection{Examples}
In this section, we apply the convergence result to several non-HPPP models from the literature \cite{maternhardcore,repulsive,Poisson_cluster,Wireless_Network_TCP,Wireless_Network_MCP_TCP,NakataMiyoshi2014}.

\subsubsection{Sensors placed according to a Mat\'{e}rn hardcore process}
The assumption that sensors are distributed according to an HPPP is widely used \cite{BergelNoam2018,WSNPPP1,WSNPPP3} because of its mathematical tractability. However, in practice, this assumption can be overly optimistic \cite{empirical_Poisson}, as real-world sensor deployments often exhibit \textit{repulsion} or a minimum separation distance between nodes. To capture such spatial interactions, it is more realistic to model the sensor locations using PPs that incorporate repulsive behavior, such as \textit{determinantal point processes (DPPs)} \cite{repulsive} and \textit{Mat\'{e}rn hardcore processes (MHPs)} \cite{maternhardcore}.

In this section, we study both the Type I and Type II MHP (see \cite[Chapter~2.1.3]{BaccelliBlaszczyszyn2009a}), which introduce repulsion through dependent thinning of an initial HPPP. Specifically, we begin by generating parent points according to an HPPP in $\IR^d$
\[
\Phi^p = \sum_{i \in \mathcal{I}^{\Phi^p}}\delta_{X_i}
\]
with intensity $\lambda_0$. The points in $\Phi^p$ are then selectively retained based on a specified thinning rule, resulting in a new point configuration.

For the Type I MHP, each point $X_i,i \in \mathcal{I}^{\Phi^p}$, in the initial configuration is retained if there are no other points within a distance of $\rho_c$ from $X_i$, where $\rho_c$ denotes the hardcore radius. If there are points within this distance, then $X_i$ and all points within $\rho_c$ of $X_i$ are removed. The remaining points constitute a Type I MHP, denoted by $\Xi_1$, which ensures that any two retained points are at least $\rho_c$ units apart. However, this thinning mechanism can be overly aggressive: in some cases, it may remove all points, resulting in an empty configuration. 

In contrast, the Type II MHP thinning rule is more conservative. For each point $X_i,i \in \mathcal{I}^{\Phi^p}$, in the initial configuration, we assign a random mark $A_i$ sampled uniformly from $[0,1]$, representing the ``age'' of the point. Then, for each point $X_i$, we examine all other points within a distance $\rho_c$. If $X_i$ has the smallest age among its neighbors within $\rho_c$, it is retained; otherwise, it is removed. The surviving points form a Type II MHP, denoted by $\Xi_2$, which again guarantees that any two points are separated by at least $\rho_c$ units, but typically retains more points than the Type I MHP.

Finally, we will use Theorem~\ref{convergence_thm} to show that the observable processes $\mathcal{N}_{\Xi_1}^\sig$ and $\mathcal{N}_{\Xi_2}^\sig$, induced by $\Xi_1$ and $\Xi_2$ respectively, converge in distribution to the same limiting process as the observable processes induced by HPPPs with the same intensity as $\Xi_1$ and $\Xi_2$, respectively.

We note that both $\Xi_1$ and $\Xi_2$ are stationary and isotropic since they are constructed from an HPPP, which is itself stationary and isotropic, and the dependent thinning rules rely solely on inter-point distances rather than directions. Therefore, it suffices to verify the SRD condition \eqref{equ: SRD}.

For Type I MHP $\Xi_1$, its intensity is given by
\[
\lambda_1=\lambda_0 e^{-\lambda_0 V_d (\rho_c)}
\]
where $V_d (\rho_c)$ is the volume of a $d$-dimensional ball of radius $\rho_c$, and $e^{-\lambda_0 V_d (\rho_c)}$ is the Palm retention probability of the “typical point” of $\Xi_1$; see \cite[Definition~3.7]{Haenggi2012}. The pair correlation function of $\Xi_1$ is given by
\[
h_{\mathrm{SI},1}(\rho)=
\begin{cases}
0, & 0<\rho\leq \rho_c,\\[2mm]
e^{\lambda_0 A_d(\rho_c,\rho)}, & \rho> \rho_c,\\[2mm]
\end{cases}
\]
where $A_d(\rho_c,\rho)$ is the volume of the intersection of two $d$-dimensional balls of radius $\rho_c$ and midpoint distance $\rho$; see \cite[Theorem~2]{MHPTypeI}. In low-dimensional cases,
\[
A_2(\rho_c,\rho)=
\begin{cases}
2\rho_c^2 \arccos\!\lb \dfrac{\rho}{2\rho_c}\rb -\dfrac{\rho}{2}\sqrt{4\rho_c^2-\rho^2}, & 0\le\rho\le 2\rho_c,\\[2mm]
0, & \rho>2\rho_c,
\end{cases}
\]
and
\[
A_3(\rho_c,\rho)=
\begin{cases}
\displaystyle \frac{4\pi \rho_c^3}{3}\lb 1-\frac{3\rho}{4\rho_c}+\frac{\rho^3}{16\rho_c^3}\rb , & 0\le\rho\le 2\rho_c,\\[2mm]
0, & \rho>2\rho_c.
\end{cases}
\]
When $\rho>2\rho_c$, $h_{\mathrm{SI},1}(\rho)=1$ since there is no overlap between two balls of radius $\rho_c$ when their midpoints are more than $2\rho_c$ apart. Therefore, the SRD condition is immediately satisfied for Type I MHP.

For Type II MHP $\Xi_2$, its intensity is given by 
\begin{equation}\label{equ:density_MHP_II}
    \lambda_2=\lambda_0 \frac{1-e^{-\lambda_0 V_d (\rho_c)}}{\lambda_0 V_d (\rho_c)}= \frac{1-e^{-\lambda_0 V_d (\rho_c)}}{ V_d (\rho_c)},
\end{equation}
where $\frac{1-e^{-\lambda_0 V_d (\rho_c)}}{\lambda_0 V_d (\rho_c)}$ is the Palm retention probability of the “typical point” of $\Xi_2$; see \cite[Definition~3.8]{Haenggi2012}. The pair correlation function of $\Xi_2$ is given by
\[
h_{\mathrm{SI},2}(\rho)=
\begin{cases}
0, & 0<\rho\leq \rho_c,\\[2mm]
\frac{2\bar{V}_d(\rho_c,\rho)\lb 1-e^{-\lambda_0 V_d (\rho_c)} \rb  -2V_d (\rho_c)\lb 1-e^{-\lambda_0 \bar{V}_d(\rho_c,\rho)} \rb }{\lambda_2^2 V_d (\rho_c) \bar{V}_d(\rho_c,\rho) \lb  \bar{V}_d(\rho_c,\rho)- V_d (\rho_c)\rb }, & \rho> \rho_c,\\[2mm]
\end{cases}
\]
where 
\[
\bar{V}_d(\rho_c,\rho) = 2V_d (\rho_c)-A_d(\rho_c,\rho)
\]
is the volume of the union of two $d$-dimensional balls of radius $\rho_c$ and midpoint distance $\rho$; see \cite[Chapter~6.5]{Statistical_analysis_and_modelling_of_spatial}. When $\rho>2\rho_c$, $\bar{V}_d(\rho_c,\rho)=2V_d (\rho_c)$, and thus
\begin{align*}
    h_{\mathrm{SI},2}(\rho)&=\frac{4V_d (\rho_c)\lb 1-e^{-\lambda_0 V_d (\rho_c)} \rb  -2V_d (\rho_c)\lb 1-e^{-2\lambda_0 V_d (\rho_c)} \rb }{2\lambda_2^2 V_d (\rho_c)^3}\\
    &=\frac{1}{\lambda_2^2}\lb \frac{1-e^{-\lambda_0 V_d (\rho_c)}}{V_d (\rho_c)}\rb ^2\\
    &=1.
\end{align*}
Therefore, the SRD condition is also satisfied for Type II MHP.

Theorem~\ref{convergence_thm} yields
\[
\lim_\sigmatoinf \NN^\sig_{\Xi_1}\dequal \lim_\sigmatoinf \NN^\sig_{\Phi_1},
\]
where $\NN^\sig_{\Phi_1}$ is the observable process induced by an HPPP $\Phi_1$ with intensity $\lambda_1$, and 
\[
\lim_\sigmatoinf \NN^\sig_{\Xi_2}\dequal \lim_\sigmatoinf \NN^\sig_{\Phi_2},
\]
where $\NN^\sig_{\Phi_2}$ is the observable process induced by an HPPP $\Phi_2$ with intensity $\lambda_2$.

\subsubsection{Sensors placed according to a stationary isotropic determinantal point process}

Another class of repulsive PPs that has been widely used to model wireless networks \cite{DeterminantalGinibre,DeterminantalGaussian,DeterminantalGinibreSensor} is the DPP, defined on a locally compact space $\mathbb{S}$ (i.e., every point has a neighborhood whose closure is compact), such as $\IR^d$ or $\mathbb{C}$.

A PP on $\mathbb{S}$ is said to be a DPP with kernel $K:\mathbb{S}\times\mathbb{S} \to \mathbb{C}$ if its $n^{\mathrm{th}}$-order product density $\varrho^{(n)}:\mathbb{S}^n \to \IR_+$ is given by
\[
\varrho^{(n)}(x_1,\ldots,x_n)
=
\det\!\lb K(x_i,x_j) \rb_{1\le i,j\le n},
\qquad (x_1,\ldots,x_n) \in \mathbb{S}^n,
\]
where $\det\!\lb K(x_i,x_j) \rb_{1\le i,j\le n}$ denotes the determinant of the $n\times n$ matrix with $(i,j)$-th entry $K(x_i,x_j)$; see \cite[Section~2]{LavancierEtAl2015} for further background. Intuitively, $\varrho^{(n)}(x_1,\ldots,x_n)$ characterizes the joint intensity of observing points of the process simultaneously in infinitesimal neighborhoods around $x_1,\ldots,x_n$, and the determinant structure captures the repulsive interaction between points.

We first consider $\Xi$ as a stationary isotropic DPP on $\mathbb{R}^d$ with a real-valued
kernel $K:\mathbb{R}^d\times\mathbb{R}^d \to \mathbb{R}$. 
When the kernel is real, the stationarity and isotropy of $\Xi$ is \emph{equivalent} to requiring
that $K$ is translation and rotation invariant, 
so that it admits the radial representation
\begin{equation}\label{equ: determinantal_condition}
    K(x,y) = K_0(|x-y|), \qquad x,y\in\mathbb{R}^d,
\end{equation}
for some continuous function $K_0:\IR_+\to\mathbb{R}$; see \cite[Section~3]{LavancierEtAl2015}.
 We further assume that $K_0$ satisfies the following two conditions:
\begin{description}

\item{(a)} For some positive constants $r_0,C,\delta$,  
\begin{equation*}
K_0(r)\le Cr^{-d/2-\delta}, \mbox{ for all }r\ge r_0.
\end{equation*}

\item {(b)} The Fourier transform $\hat K_0$ defined as 
$$\hat K_0(x):=\int K_0(t)\exp(-2\sqrt{-1}\pi  t\cdot x)\mathrm{d}t,\qquad x\in\mathbb{R}^d,$$
where $t\cdot x$ is the dot product of $t$ and $x$, satisfies $\hat K_0\in[0,1]$.  

\end{description}

Conditions~(a) and (b) together with \cite[Corollary~1 and Theorem~1]{LavancierEtAl2015} ensure the existence of the DPP $\Xi$ with the kernel $K$. For this DPP, we have $\lambda=K_0(0)$ and $ \varrho^{(2)}(x,y)=K_0(0)^2-K_0(|x-y|)^2$. Hence 
$$h_{\mathrm{SI}}(r)=1-K_0(r)^2/K_0(0)^2,$$
and condition~(a) implies the SRD condition \eqref{equ: SRD}. Theorem~\ref{convergence_thm} yields
\[
\lim_\sigmatoinf \NN^\sig_{\Xi}\dequal \lim_\sigmatoinf \NN^\sig_{\Phi},
\]
where $\NN^\sig_{\Phi}$ is the observable process induced by an HPPP $\Phi$ with intensity $K_0(0)$.

A particularly important example in this class is the Gaussian DPP on $\mathbb{R}^d$, whose kernel is given by
\[
K(x,y)=K_0(|x-y|)=\lambda \exp\!\lcb -\frac{|x-y|^2}{\alpha^2}\rcb , \qquad x,y\in\mathbb{R}^d,
\]
for some parameter $\alpha>0$ controlling the degree of repulsion; see \cite[Table~2]{LavancierEtAl2015}. Clearly, condition~(a) is immediately satisfied.

The Fourier transform of $K_0$ is given by
\[
\hat K_0(x)
=
\lambda (\sqrt{\pi}\alpha)^d
\exp\!\lcb -\pi^2\alpha^2 |x|^2\rcb ,
\qquad x\in\mathbb{R}^d;
\]
see \cite[Table~2]{LavancierEtAl2015}. Therefore, condition~(b) holds whenever
\[
\lambda (\sqrt{\pi}\alpha)^d \le 1.
\]
This condition highlights a fundamental trade-off between the achievable intensity and the strength of repulsion in a DPP: stronger repulsion (larger $\alpha$) necessarily limits the maximum admissible density $\lambda$.

While the above conditions characterize a broad class of stationary and isotropic DPPs with real-valued kernels, they do not exhaust all possible constructions satisfying our assumptions. In particular, stationary and isotropic DPPs with complex-valued kernels may also satisfy the SRD condition~\eqref{equ: SRD}.

The $\alpha$-Ginibre point process (GPP), defined on $\mathbb{C}$ (or equivalently on $\mathbb{R}^2$), is a canonical example of such class; see \cite[Section~5.6.1]{BaccelliBlaszczyszynKarray2024} for background. When the kernel is complex-valued, condition~\eqref{equ: determinantal_condition} remains sufficient (though not necessary) for stationarity and isotropy; see \cite[Section~3]{LavancierEtAl2015}. Identifying $\mathbb{R}^2$ with $\mathbb{C}$, the $\alpha$-GPP with parameters $\alpha\in(0,1]$ and intensity $\lambda>0$ has kernel \cite[Section~3.2]{GinibreTutorial}
\[
K(z,w)
=
\lambda
\exp\!\lcb 
-\frac{ \lambda\pi}{2\alpha}\lb|z|^2+|w|^2\rb
\rcb 
\exp\!\lcb 
\frac{ \lambda\pi}{\alpha} z\overline{w}
\rcb ,
\qquad z,w\in\C.
\]
A direct computation shows that the second-order product density is given by
\[
\varrho^{(2)}(x,y)
=
\lambda^2\lb 1 - e^{-\frac{\pi\lambda}{\alpha}|x-y|^2}\rb
=
\varrho^{(2)}_{\mathrm{SI}}(|x-y|).
\]
Accordingly, the stationary isotropic pair correlation function is given by
\[
h_{\mathrm{SI}}(r)
=
\frac{\varrho^{(2)}_{\mathrm{SI}}(r)}{\lambda^2}
=
1 - e^{-\frac{\pi\lambda}{\alpha}r^2},
\]
which implies that the $\alpha$-GPP satisfies the SRD condition~\eqref{equ: SRD}.

\subsubsection{Sensors placed according to a Poisson cluster process}
Wireless sensor deployments frequently exhibit higher node densities in urban or built-up areas than in rural regions. These clustering phenomena can be well approximated by various types of \textit{Poisson cluster processes}, among which the \textit{Mat\'{e}rn cluster process (MCP)} and the \textit{Thomas cluster process (TCP)} are the most widely studied \cite{Poisson_cluster,Wireless_Network_TCP,Wireless_Network_MCP_TCP}; see \cite[Chapter~2]{Haenggi2012} for background. Both are special cases of the broader class of \textit{Neyman--Scott processes} \cite[Chapter~3.4.3]{Haenggi2012}.

In these models, the construction begins with an HPPP $\Phi^p = \sum_{i \in \mathcal{I}^{\Phi^p}}\delta_{X_i}$ of parent nodes in $\IR^d$ with intensity $\lambda_0$. For each parent node, the number of daughter nodes associated with it follows a Poisson distribution with mean $\bar{c}$, and each daughter node is then positioned around its parent according to a specified spatial scattering distribution. The final PP of interest is defined by the union of all daughter points, excluding the parent points themselves. Therefore, both MCP and TCP share the same overall density,
\begin{equation}\label{equ:density_MCP_TCP}
    \lambda = \lambda_0 \bar{c},
\end{equation}
and they differ only in the spatial arrangement of the daughter points relative to their parent nodes. In the MCP, the daughter nodes are independently and uniformly scattered within a ball $B_{X_i}(\rho_c)$ of radius $\rho_c$ centered at each parent node $X_i,i\in \mathcal{I}^{\Phi^p}$, where $\rho_c>0$ is the cluster radius. In contrast, the TCP independently scatters daughter nodes according to an isotropic multivariate normal distribution centered at the parent location, with covariance matrix $\sigma_c^2 I_d$, where $I_d$ is the $d$-dimensional identity matrix and $\sigma_c>0$. By employing these clustered PPs, we can better model spatial correlations and local aggregation effects, leading to more realistic and analytically tractable frameworks for analyzing network performance in heterogeneous environments.

We use $\Xi_1$ to denote an MCP with intensity given by \eqref{equ:density_MCP_TCP}, and use $\Xi_2$ to denote a TCP with the same intensity. We will use Theorem~\ref{convergence_thm} to show that the observable processes $\mathcal{N}_{\Xi_1}^\sig$ and $\mathcal{N}_{\Xi_2}^\sig$, induced by $\Xi_1$ and $\Xi_2$ respectively, converge in distribution to the same limiting process as the observable processes induced by HPPPs with the same intensity.

We first observe that both $\Xi_1$ and $\Xi_2$ are stationary and isotropic since the parent PPP $\Phi^p$ is stationary and isotropic, and the scattering of daughter nodes around each parent is uniform in direction. It suffices to verify the SRD condition \eqref{equ: SRD}.

The pair correlation function of the MCP $\Xi_1$ is given by \cite[Example~6.9]{Haenggi2012}
\[
h_{\mathrm{SI},1}(\rho)=1+\frac{1}{\lambda_0}\,\frac{A_d(\rho_c,\rho)}{V_d (\rho_c)^2},\qquad \rho>0 .
\]
Notice that when $\rho>2\rho_c$, we have $h_{\mathrm{SI},1}(\rho)=1$ and hence, the SRD condition \eqref{equ: SRD} is satisfied for MCP.

The pair correlation function of the TCP $\Xi_2$ is given by \cite[Example~5.3]{Statistical_inference_and_simulation_for_spatial}
\[
h_{\mathrm{SI},2}(\rho)=1+\frac{1}{\lambda_0 (4\pi\sigma_c^2)^{d/2}}\,e^{-\frac{\rho^2}{4\sigma_c^2}},\qquad \rho>0.
\]
Therefore, the SRD condition \eqref{equ: SRD} is also satisfied for TCP.

Theorem~\ref{convergence_thm} yields
\[
\lim_\sigmatoinf \NN^\sig_{\Xi_1}\dequal \lim_\sigmatoinf \NN^\sig_{\Xi_2}\dequal \lim_\sigmatoinf \NN^\sig_{\Phi},
\]
where $\NN^\sig_\Phi$ is the observable process induced by an HPPP $\Phi$ with intensity $\lambda$ given by \eqref{equ:density_MCP_TCP}.

\section{Finite Homogeneous Poisson Model}\label{Section: Poisson model}
Theorem~\ref{convergence_thm} justifies the use of an HPPP for modelling the sensor network when the underlying sensor configuration is not ``too far'' from HPPP; see Assumption~\ref{assumption}. From the perspective of the fusion center—which only observes the RSS and AOA measurements collected from the sensor nodes—the specific arrangement of sensors becomes immaterial: whether the nodes follow an HPPP or another stationary isotropic PP satisfying the SRD condition~\eqref{equ: SRD}, their observable behaviour is statistically indistinguishable in the noisy regime. Consequently, we focus our analysis on target localization using an HPPP sensor network. We consider both the simple average estimator defined in \eqref{equ: targetEstimator_PPP} and the weighted average estimator defined in \eqref{equ: weighted_estimator}.

\subsection{Main Results}\label{Section: Poisson model main result}

We state the assumptions of this section before presenting our results.

\begin{assumption}\label{assumption: finite_Poisson}
    The sensor locations are modeled by an HPPP $\Phi_\R$ on the ball $\BallD$ of radius $\R$ centered at the origin. We denote by $|\Phi_\R|$ the number of sensors in this observation region, and write $\Phi_\R = \sum_{i=1}^{|\Phi_\R|}\delta_{X_i}$.

    For each point $X_i$, $1\leq i \leq |\Phi_\R|$, let $\lb  R_i,  \lcb  \Psi_{i}^{(j)}\rcb_{j=1}^{d-1} \rb $ denote its hyperspherical coordinate representation. Assume that the AOAs $\lcb \vtj_i\rcb_{1\leq i \leq |\Phi_\R|,\,1\le j\le d-1}$ defined in \eqref{equ: wrapped normal AOA} are conditionally independent given $\Phi_\R$. 
    
    Assume that the distance estimator $\lcb \hatr_i\rcb_{1\leq i \leq |\Phi_\R|}$ are conditionally i.i.d. given $|\Phi_\R|$, and are conditionally independent of the AOAs $\lcb \vtj_i\rcb_{1\leq i \leq |\Phi_\R|,\,1\le j\le d-1}$ given $\Phi_\R$. Moreover, assume that
    \begin{equation}\label{assumption: cov_radial_positive}
        \Cov\!\lb \hatr_1,R_1 \rb \geq0.
    \end{equation}
\end{assumption}

\begin{remark}
Under Assumption~\ref{assumption: finite_Poisson}, the distance estimator $\hatr_i$ need not be a function of the RSS $P_i$. In fact, the distance to the target may be estimated using any other signal-derived statistic, provided that \eqref{assumption: cov_radial_positive} is satisfied. For instance, one may use the TOA $T_i$ measured at sensor $X_i$ and estimate the distance via $\hatr_i = T_i \mathscr{C}$, where $\mathscr{C}$ denotes the known signal propagation speed.
\end{remark}

When $\hatr_i$ is a non-increasing function of the RSS $P_i$, where $P_i$ is given by \eqref{equ: relation between P and R}, Assumption~\eqref{assumption: cov_radial_positive} is immediately satisfied. Specifically, let
\[
\hatr_i = l(P_i) = l\big( \ell(R_i) S_i(\sigma) \big),
\]
where $l:\mathbb{R}_+ \to \mathbb{R}_+$ and $\ell$ are non-increasing, and $S_i(\sigma)$ are i.i.d. positive RVs. Given $|\Phi_\R| = n$, the sensor locations $X_i$ are i.i.d. uniformly distributed over $\BallD$ (see \cite[Chapter~2.4.2]{Haenggi2012}). Hence, the radial distances $R_i$ are also i.i.d. with density given in \eqref{equ: marginal_r}. Since $S_i(\sigma)$ are i.i.d. and independent of the sensor locations, it follows that $\hatr_i$ are i.i.d. as well. For any fixed value of $S_i(\sigma)$, the quantity $\hatr_i$ is a non-decreasing function of $R_i$, since both $l$ and $\ell$ are non-increasing. Consequently, $\hatr_i$ and $R_i$ are non-negatively correlated. More precisely, let $F_\sigma$ denote the distribution function of $S_1(\sigma)$. Then it follows that
\begin{align*}
    \Cov\!\lb R_1,\hatr_1\rb
    &=\IE\left[ R_1 \;l\left( \ell(R_1)S_1{(\sigma)}\right)  \right]-\IE\left[ R_1 \right] \IE\left[ l\left( \ell(R_1)S_1{(\sigma)}\right)\right]\\
    &= \int_{\mathbb{R}_+^0} \IE\left[ R_1 \;l\left( \ell(R_1)s\right)\right] \dx F_\sigma(s) -\int_{\mathbb{R}_+^0}\IE\left[ R_1 \right] \IE\left[ l\left( \ell(R_1)s\right)\right] \dx F_\sigma(s)\\
    &= \int_{\mathbb{R}_+^0} \Cov\left( R_1 ,l\left( \ell(R_1)s\right)\right) \dx F_\sigma(s)\\
    &\geq 0.
\end{align*}

Recall that the target estimator $\X$ is defined as the (simple) average of the individual estimates \eqref{equ: individual_estimate_AOA_vec}:
\begin{equation}\label{equ: targetEstimator_PPP}
    \X :=\frac{1}{|\Phi_\R|} \sum_{i=1}^{|\Phi_\R|} \tilde{X}_{i}.
\end{equation}

\begin{lemma}\label{lemma: unbiasedEstimator}
    Under Assumption~\ref{assumption: finite_Poisson}, the estimator $\X$ is unbiased.
\end{lemma}

Lemma~\ref{lemma: unbiasedEstimator} shows that $\X$ is unbiased \emph{regardless} of the choice of the distance estimator $\hatr_i$. Intuitively, each sensor at $X_i$ forms an individual location estimate $\tilde{X}_i$ using its (possibly biased) range estimate $\hatr_i$ together with the measured AOAs. The global estimate $\X$ is the average of these individual estimates. Because the sensor bearings $\lcb \boldsymbol{\Psi}_{i}\rcb_{i=1}^{|\Phi_\R|}$ are uniformly distributed in angle for the HPPP model, any offsets induced by the bias in $\hatr_i$ symmetrically cancel across sensors, yielding an unbiased overall estimator of the target position.

We are interested in both the CMSE, given the number of sensors, and the unconditional MSE of the overall estimate $\X$. These are defined respectively as:
\[
\mathrm{CMSE}_\R(n) := \IE\!\lsb  |\X|^2 \given |\Phi_\R| = n \rsb = \IE\!\lsb \sum_{j=1}^d\X_j^2 \given |\Phi_\R| = n\rsb,
\]
\[
\mathrm{MSE}_\R := \IE \mathrm{CMSE}(|\Phi_\R|).
\]

\begin{theorem}\label{thm: CMSD and MSD}
Under Assumption~\ref{assumption: finite_Poisson} and the target estimator defined in \eqref{equ: targetEstimator_PPP}, we have
\[
\mathrm{CMSE}_\R(n) \leq \frac{1}{n} \lb  \frac{d}{d+2} \R^2 + \IE\!\lsb  \hatr_1^2 \rsb - \frac{2d \R}{d+1} e^{-\frac{d-1}{2}\E(\R)} \IE \hatr_1  \rb ,
\]

\begin{align}
\mathrm{MSE}_\R
&\le e^{-\lambda V_d(\R)} \lb \sum_{n=1}^\infty \frac{(\lambda V_d(\R))^n}{n\cdot n!} \rb \lb \frac{d}{d+2}\R^2 + \IE\!\lsb  \hatr_1^2 \rsb - \frac{2d \R}{d+1} e^{-\frac{d-1}{2}\E(\R)} \IE \hatr_1  \rb  \label{eq:new_bound}\\
&\leq \frac{2}{\lambda V_d(\R)}\lb  \frac{d}{d+2} \R^2 + \IE\!\lsb  \hatr_1^2 \rsb - \frac{2d \R}{d+1} e^{-\frac{d-1}{2}\E(\R)} \IE \hatr_1  \rb ,\label{eq:old_bound}
\end{align}
where $V_d(\R)$ is the volume of a $d$ dimensional ball $\BallD$ of radius $\R$.
\end{theorem}

\begin{remark}
The bound in \eqref{eq:new_bound} is tighter than that in \eqref{eq:old_bound}, as it provides a more accurate characterization of $\mathrm{MSE}_\R$ across the full range of system parameters. However, the bound in \eqref{eq:old_bound} admits a simpler closed-form expression, which makes the decay rate in $\lambda$ explicit. For this reason, we retain both bounds: the tighter bound is used for numerical evaluation, while the simpler bound provides clearer insight into the asymptotic scaling behaviour. The difference between the two bounds will be illustrated in the simulation results in Section~\ref{Section: simulate_HPPP}.
\end{remark}

Theorem~\ref{thm: CMSD and MSD} provides an explicit upper bound for the CMSE and MSE between the estimated and true location of the target, expressed in terms of the number of sensors within the ball $\BallD$, the density of the sensor placements, and the estimation range $\R$. From this result, we observe that for a given estimation range $\R$, the CMSE decreases as the number of sensors increases. Similarly, the MSE decreases as the sensor density increases. However, for the CMSE, increasing the estimation range $\R$ while keeping the number of sensors fixed leads to a larger localization error. These monotonicity properties are consistent with intuition: increasing information—via additional sensors or higher spatial density—improves accuracy, whereas enlarging the estimation region with a fixed sensor count degrades it.

We note that the quantities $\IE\hatr_1$ and $\IE[\hatr_1^2]$ appearing in the bounds are not explicitly expressed in terms of fundamental system parameters (e.g., sensor density, path-loss exponent, noise variance, or observation radius). This is a consequence of the minimal constraints on the distance estimator, $\hatr_1$, which is only assumed to be positively correlated with $R_1$, without imposing a specific parametric form. This level of abstraction allows the result to remain applicable to a broad class of practical distance estimators, rather than being restricted to a particular signal model or inversion scheme. As a result, the derived bounds provide a unified performance characterization that is robust across different implementations of distance estimation.

For specific choices of the distance estimator, such as those induced by standard path-loss models (see \eqref{equ: regression_estimator}), closed-form expressions for $\IE\hatr_1$ and $\IE[\hatr_1^2]$ can be obtained in terms of the underlying system parameters; see \eqref{equ: mean_hatr_system_parameter} and \eqref{equ: mean_hatrsq_system_parameter}. Such explicit characterizations will be presented in Section~\ref{section: 2d_system_model}. In addition, from a practical perspective, the quantities $\IE\hatr_1$ and $\IE[\hatr_1^2]$ can be consistently estimated via sample moments from the observed sensor measurements, provided that a sufficiently large number of sensors are available within the observation region. This ensures that the proposed bounds remain directly computable in real-world implementations.

To the best of our knowledge, the closest result to Theorem~\ref{thm: CMSD and MSD} is due to \cite{BergelNoam2018}, who consider an \emph{infinite} sensor field modeled as an HPPP and derive CRB-type \emph{lower bounds} on the variance (or equivalently MSE) of any unbiased estimator. In their asymptotic regimes, the bound scales with density as $\lambda^{-\beta/2}$ for wideband localization and $\lambda^{-\beta/2-1}$ for narrowband localization, with $\beta>2$ the path-loss exponent. In contrast, Theorem~\ref{thm: CMSD and MSD} provides an \emph{upper bound} on the MSE (for narrowband localization) that decays as $\lambda^{-1}$ for a fixed estimation range $\mathcal{R}$.

While both works adopt an HPPP model for sensor locations, the bounds are not directly comparable due to fundamental differences in modeling assumptions and analytical frameworks. Specifically, \cite{BergelNoam2018} considers an infinite network with an infinitely countable number of sensors over $\mathbb{R}^d$, whereas our analysis is restricted to a finite observation window containing a random but finite number of sensors, leading to different scaling regimes and boundary effects. Moreover, their CRB is derived from a continuous-time waveform model and captures the full signal structure, providing limits for \emph{any unbiased estimator}. In contrast, our approach is based on summary statistics of the received signals (e.g., RSS and AOA) and analyzes a specific, implementable localization procedure (Algorithm~\ref{Algo1}).

Consequently, the two results should be viewed as complementary rather than forming a tight upper--lower performance bracket. The gap between the bounds primarily reflects the differences in network scale, signal modeling, and estimator assumptions, rather than a direct performance discrepancy under a common framework. Understanding the precise relationship between these regimes, and identifying conditions under which the bounds may become comparable, remains an interesting direction for future work.

Theorem~\ref{convergence_thm} considers stationary isotropic sensor networks in the entire space \(\mathbb{R}^d\) and states that, as the noise level \(\sigma \to \infty\), the fusion center cannot distinguish—from the observed measurements—whether the underlying sensor deployment follows an HPPP or a more general stationary isotropic PP. The result can be interpreted as a joint asymptotic statement for a large observation radius \(\mathcal{R}\) and a large noise parameter \(\sigma\). In particular, when the fusion center collects measurements from a sufficiently large region and those measurements are strongly affected by random propagation effects, the influence of short-range spatial dependence among sensor locations becomes negligible. Consequently, one may employ a finite HPPP model with the same intensity even when the actual sensor configuration is not exactly an HPPP. The performance bounds in Theorem~\ref{thm: CMSD and MSD}, derived under the finite HPPP assumption, therefore serve as reasonable approximation bounds for a wide class of sensor deployment models satisfying \eqref{equ: SRD}. This demonstrates the robustness of the HPPP model as a practical and analytically tractable benchmark for evaluating localization performance in realistic wireless networks.

\subsection{Weighted Fusion}\label{Section: weighted function result}
In the previous section, we considered the simple averaging estimator \eqref{equ: targetEstimator_PPP} for the target position. While this estimator is easy to implement and analytically tractable, its performance is not necessarily optimal. In practice, the measurement errors across sensors can vary significantly; for instance, the RSS and AOA errors for distant sensors are typically larger than those for sensors located closer to the target. Consequently, assigning equal weight to all sensors may not be appropriate.

A more refined approach is to perform weighted fusion based on the estimated distances \cite{Data_Fusion_and_Filtering_Techniques}. Intuitively, sensors with smaller estimated distances are more likely to be closer to the target and hence provide more reliable information. Therefore, assigning larger weights to such sensors may, on average, improve the localization accuracy. Motivated by this observation, we extend our analysis from the simple averaging estimator to a general class of weighted estimators that exploit the estimated distance information.

A weighted estimator of the target location is defined as
\begin{equation}\label{equ: weighted_estimator}
    \X^W := \sum_{i=1}^{|\Phi_\R|} W_i \, \tilde X_i,
\end{equation}
where the weights $W_i := W_i\!\lb \hatr_1,\dots,\hatr_{|\Phi_\R|}\rb $ satisfy
\[
W_i \ge 0, \qquad \sum_{i=1}^{|\Phi_\R|} W_i = 1.
\]
For example, one may consider weights of the form
\begin{equation}\label{equ: general_weighting_function}
    W_i = \frac{w\!\lb  \hatr_i \rb }{\sum_{j=1}^{|\Phi_\R|} w\!\lb  \hatr_j \rb },
\end{equation}
for some non-increasing function $w:\posRealLine\to\posRealLine$. 

The inverse-variance weighting scheme considered in \cite{Data_Fusion_and_Filtering_Techniques} can be viewed as a special case of \eqref{equ: general_weighting_function}. In this approach, weights are assigned based on the variance of the individual estimates $\tilx_i$, where estimates with higher variance (and thus greater uncertainty) receive lower weights. The conditional variance of $\tilx_i$ given the sensor location $X_i$ is typically an increasing function of the true distance $R_i$. Since $R_i$ is not directly observable, the variance is commonly approximated using the estimated distance $\hatr_i$. Consequently, the weighting function $\omega$ can be taken as the inverse of the estimated variance, which is equivalent to choosing $\omega$ as a non-increasing function of $\hatr_i$.

\begin{lemma}\label{lemma: unbiasedEstimator_weighted}
    Under Assumption~\ref{assumption: finite_Poisson}, the weighted estimator $\X^W$ is unbiased.
\end{lemma}

Lemma~\ref{lemma: unbiasedEstimator_weighted} shows that $\X^W$ is unbiased \emph{regardless} of the specific choice of the distance estimator $\hatr_i$ and the weighting function $W_i$. This follows from the same symmetry argument used for the simple averaging estimator (see Lemma~\ref{lemma: unbiasedEstimator} and the subsequent discussion), together with the assumption that the weights depend only on the estimated distances and not on the measured AOAs.

Direct characterization of the MSE and CMSE for weighted estimators is generally intractable, due to the loss of independence structure between the individual estimates provided by different sensors, except for a few special choices of weighting functions. Even in these special cases, obtaining informative and closed-form upper bounds on the CMSE and MSE in terms of system parameters remains challenging.

From the above consideration, we consider a class of weighted estimators that assign weights according to the order statistics of the estimated distances. Given $|\Phi_\R| = n$, let $\hat R_{(1)} \le \cdots \le \hat R_{(n)}$ denote the order statistics of $\hat R_1,\dots,\hat R_n$, and let $\kappa$ be the corresponding random permutation such that
\[
\hat R_{(i)} = \hat R_{\kappa(i)}, \qquad 1\leq i\leq n.
\]
For a fixed $a\in (0,1]$, define the weights
\[
W_{\kappa(i)}
=
\frac{a^i}{\sum_{j=1}^n {a^j}},
\qquad 1\leq i\leq n.
\]
The resulting weighted estimator is given by
\begin{equation}\label{equ: weighted_estimator_order}
    \X^W
= \sum_{i=1}^n W_{\kappa(i)}\tilde X_{\kappa(i)}
= \sum_{i=1}^n
\frac{a^i}{\sum_{j=1}^n a^j}
\,  \lb X_{\kappa(i)}+\hatr_{(i)}\eta(\boldsymbol{\Theta}_{\kappa(i)})\rb .
\end{equation}
When $a=1$, the weighted estimator \eqref{equ: weighted_estimator_order} reduces to the simple average estimator \eqref{equ: targetEstimator_PPP}.

The CMSE of the weighted estimator $\X^W$ is defined by
\[
\mathrm{CMSE}^W\!(n) := \IE\!\lsb  |\X^W|^2 \given |\Phi_\R| = n \rsb.
\]

\begin{lemma}\label{lemma: weighted_CMSE_rate}
    Under Assumption~\ref{assumption: finite_Poisson} and the weighted estimator defined in \eqref{equ: weighted_estimator_order}, for any $a\in(0,1]$, we have
    \[
        \mathscr{X}_{n,a} \leq \mathscr{U}_{n,a} + \mathscr{V}_{n,a}- 2 e^{-\frac{1}{2}\lim_{r\to0^+}\E(r)}\mathscr{W}_{n,a}\leq\mathrm{CMSE}^W\!(n)  \leq \mathscr{U}_{n,a} + \mathscr{V}_{n,a}- 2 e^{-\frac{d-1}{2}\E(\R)} \mathscr{W}_{n,a},
    \]
where
\begin{equation*}
    \mathscr{U}_{n,a}:= \frac{1}{A_n^2} \sum_{i=1}^n a^{2i}\IE\!\lsb  R_{\kappa(i)}^2\rsb,
\end{equation*}

\begin{equation*}
    \mathscr{V}_{n,a}:=\frac{1}{A_n^2} \sum_{i=1}^n a^{2i}\IE\!\lsb  \hatr_{(i)}^2\rsb,
\end{equation*}

\begin{equation*}
    \mathscr{W}_{n,a}:= \frac{1}{A_n^2} \sum_{i=1}^n a^{2i}\IE\!\lsb R_{\kappa(i)} \hatr_{(i)} \rsb,
\end{equation*}
\begin{equation*}
    \mathscr{X}_{n,a}:= \frac{1}{A_n^2} \sum_{i=1}^n a^{2i}\IE\!\lsb (R_{\kappa(i)}- \hatr_{(i)})^2 \rsb,
\end{equation*}
and $A_n:=\sum_{j=1}^n a^j$.
\end{lemma}
Lemma~\ref{lemma: weighted_CMSE_rate} implies that when $a=1$, the CMSE of the simple average estimator in \eqref{equ: targetEstimator_PPP} satisfies
\[
\mathrm{CMSE}(n)  \leq \bigo\!\lb n^{-1}\rb.
\]

We further characterize the CMSE of this weighted estimator under the following assumptions.

\begin{assumption}\label{assumption weighted estimator}
Suppose that the RSS follows a path loss model with log-normal shadowing, such that
\[
P_i = \ell(R_i) S_i(\sigma),
\]
where $\ell(r) = (Cr)^{-\beta}$, and
\[
S_i(\sigma) = \exp\{\sigma B_i - \sigma^2/\beta\},
\]
for some $C, \beta, \sigma > 0$, and where $\{B_i\}_{i=1}^n$ are i.i.d. standard normal RVs.

We assume that the system model and parameters are known exactly to the sensors, so that they can invert the path loss model to obtain the distance estimate
\[
\hatr_i = \ell^{-1}(P_i) = R_i\, S_i(\sigma)^{-\frac{1}{\beta}}.
\]
\end{assumption}

\begin{theorem}\label{thm: weighted_CMSE_rate_example}
    Under Assumptions~\ref{assumption: finite_Poisson} and \ref{assumption weighted estimator}, for any $a\in (0,1)$, the CMSE of the weighted estimator defined in \eqref{equ: weighted_estimator_order} satisfies
    \[
    \mathrm{CMSE}^W\!(n)\asymp n^{-\frac{2}{d}}.
    \]
\end{theorem}

Theorem~\ref{thm: weighted_CMSE_rate_example} shows that, under the RSS and distance estimation model specified in Assumption~\ref{assumption weighted estimator}, the decay rate of the CMSE for the weighted estimator defined in \eqref{equ: weighted_estimator_order} does not improve for $d \geq 2$. In particular, when $d > 2$, the decay rate is slower than that of the simple average estimator \eqref{equ: targetEstimator_PPP}. This indicates that, asymptotically, the simple average estimator \textit{outperforms} the weighted estimator as $n$ becomes large.

Combined with the observations in Section~\ref{Section: weighted fusion}, this suggests that weighted estimators may achieve improved CMSE when $n$ is small, provided that the weighting function is appropriately chosen. However, for general weighting schemes, such improvement is not guaranteed, especially in the large-$n$ regime.

The underlying reason is that the weighting function typically assigns the largest weights to a small subset of sensors that report the smallest estimated distances. There is no guarantee that these sensors provide accurate individual estimates; nevertheless, they dominate the fusion process. As more sensors are added, they receive negligible weight unless they report an even smaller estimated distance---an event whose probability becomes vanishingly small as $n$ increases. Consequently, newly added sensors are effectively ignored by the weighted estimator and do not contribute meaningfully to improving localization accuracy.

In contrast, the simple average estimator assigns equal weights to all sensors. While it may not outperform the weighted estimator for small $n$, it eventually does so as $n$ grows. This is because each additional sensor contributes equally, allowing the estimation error to average out, leading to a law-of-large-numbers-type behavior. As a result, the simple average estimator effectively exploits the information from all sensors.

Figure~\ref{fig:bad_sensor_large_weight} illustrates this phenomenon. Consider a scenario with three sensors, where $X_1$ is a poor-quality sensor that happens to report the smallest estimated distance. The weighted estimator assigns a disproportionately large weight to $\tilx_1$, resulting in a poor overall estimate $\X^W$. In contrast, the simple average estimator $\X$ yields a more accurate estimate by treating all sensors equally. Viewing this as a dynamic process in which sensors are added sequentially, the CMSE of the simple average estimator decays faster than that of the weighted estimator, due to its effective use of additional information. This conclusion holds in general, even when $X_1$ is not an outlier.

In summary, while weighted estimators may offer improved performance for small $n$ with carefully designed weights, such gains are not guaranteed and are difficult to characterize analytically. On the other hand, the simple average estimator is asymptotically superior, with performance guarantees provided by the bounds in Theorem~\ref{thm: CMSD and MSD}. A detailed comparison between the simple estimator \eqref{equ: targetEstimator_PPP} and the weighted estimator \eqref{equ: weighted_estimator_order} is presented in Section~\ref{Section: weighted fusion}.

\begin{figure}[!t]
\centering
\begin{tikzpicture}[scale=1]

  \draw[->] (-4.2,0) -- (4.4,0) node[right] {$x$};
  \draw[->] (0,-3.4) -- (0,4.0) node[above] {$y$};

  \coordinate (O) at (0,0);
  \node at (O) [star, star points=5, star point ratio=2.25,
                fill=black, minimum size=10pt, inner sep=0pt] {};
  \node[below right] at (O) {};

  \coordinate (Sone) at (3.2,2.4);
  \coordinate (Stwo) at (-3.0,2.2);
  \coordinate (Sthree) at (2.8,-2.6);

  \coordinate (Eone) at (0.25, 0.5);    
  \coordinate (Etwo) at (-0.50,-0.25);   
  \coordinate (Ethree) at (1.95,-1.75); 

  \coordinate (W) at (1.45,-1.20);
  \coordinate (S) at (0.5,-0.5);

  \draw[dotted, thick] (Sone) -- (Eone)
    node[midway, above left] {$\hat R_3$};
  \draw[dotted, thick] (Stwo) -- (Etwo)
    node[midway, above right] {$\hat R_2$};
  \draw[dotted, very thick] (Sthree) -- (Ethree)
    node[midway, below left] {$\hat R_1$};

  \fill (Sone) circle (2.2pt) node[above right] {$X_3$};
  \fill (Stwo) circle (2.2pt) node[above left] {$X_2$};
  \fill (Sthree) circle (2.2pt) node[right] {$X_1$};

  \node at (Eone) {$\times$};
  \node[below right] at (Eone) {$\tilde X_3$};

  \node at (Etwo) {$\times$};
  \node[left] at (Etwo) {$\tilde X_2$};

  \node at (Ethree) {$\times$};
  \node[right] at (Ethree) {$\tilde X_1$};

  \node at (W) [circle, draw=black, fill=white, inner sep=1.5pt] {};
  \node[above right] at (W) {$\X^{W}$};
  \node at (S) [circle, draw=black, fill=white, inner sep=1.5pt] {};
  \node[above right] at (S) {$\X$};

  \node[align=center] at (-2.4,3.15) {\scriptsize good sensor\\[-1mm]\scriptsize accurate estimate};
  \node[align=center] at (3.1,3.35) {\scriptsize good sensor\\[-1mm]\scriptsize accurate estimate};
  \node[align=center] at (2.5,-3.15) {\scriptsize bad sensor\\[-1mm]\scriptsize receives large weight};

\end{tikzpicture}
\caption{Weighted estimator dominated by a bad sensor with a small estimated distance.}
\label{fig:bad_sensor_large_weight}
\end{figure}

\section{Simulation Studies}\label{Section: Simulation}
In this section, we conduct simulation studies to evaluate the quality of the bounds in Theorem~\ref{thm: CMSD and MSD} under realistic parameter settings. We also introduce a practical model that captures the effects of signal propagation on both the RSS and AOA measurements, and develop a linear regression–type estimator \(\hatr_i\) for inferring the distance \(R_i\) to the target from the observed RSS \(P_i\). In addition, we conduct further simulations to assess the impact of repulsive and clustered sensor placements on localization performance, and to investigate how they differ from uniform (HPPP) sensor placement. Finally, we numerically compare the performance of the simple average estimator with a class of weighted estimators defined in \eqref{equ: weighted_estimator_order}.

\subsection{System Parameters}\label{section: 2d_system_model}
We choose system parameters that represent a generic indoor wireless sensor network localization setting using RSS and AOA measurements \cite{Ding2025,Miranda2010}. We assume that the sensors are drawn from a stationary and isotropic PP $\Xi$ on $\IR^2$ with intensity $\lambda$ sensors per square meter, restricted to the disc $\BallD$ of radius $\R=30$ m centered at the target, which is taken to be the origin of the coordinate system.

To model the RSS, we adopt the standard path-loss function \cite{KeelerRossXia2018,7745970,BlaszczyszynKarrayKeeler2013,BlaszczyszynKarrayKeeler2015}
\[
\ell(r) = (C r)^{-\beta}.
\]
The parameter $\beta$ denotes the path-loss exponent, which depends on the propagation environment. Empirical studies in wireless sensor networks report that $\beta$ typically ranges from $2$ (free space) to $4$ or higher in cluttered environments, with values around $2.5$--$3.5$ commonly observed in indoor office scenarios \cite{Miranda2010}. Accordingly, we choose
\[
\beta = 2.7,
\]
which is consistent with the indoor RSS-based localization setting considered in \cite{Ding2025}, where a smartphone (target node) is localized using routers (sensor nodes).

The constant $C$ is determined from the received power $P_0$ measured at a prescribed reference distance $d_0$ under ideal (free-space) propagation conditions for a chosen transmission power, via
\[
P_0 = \ell(d_0) = (C d_0)^{-\beta}.
\]
Following \cite{Ding2025}, we choose the transmission power such that the received power satisfies $P_0 = 0.1$ mW at $d_0 = 1$ m, which yields
\[
C = \frac{(P_0)^{-1/\beta}}{d_0} \approx 2.345~\mathrm{m}^{-1}.
\]

The propagation effect is modeled using log-normal shadowing:
\begin{equation}\label{equ: noise_model}
    S_i(\sigma) = \exp\!\lcb -\frac{\sigma^2}{\beta} + \sigma B_i\rcb,
\end{equation}
where $B_i$ are i.i.d. standard normal RVs. The shadowing standard deviation $\sigma$ is typically specified in logarithmic scale as $\sigma_{\mathrm{dB}}$, with empirical values ranging from $4$ to $13$ dB depending on the environment \cite{sigma_values}. The conversion between the logarithmic and linear scales is given by
\[
\sigma = \frac{\sigma_{\mathrm{dB}}}{10}\log 10.
\]
We choose $\sigma_{\mathrm{dB}} = 12\,\mathrm{dB}$, corresponding to $\sigma \approx 2.76$, which represents a moderately strong indoor shadowing condition.

Therefore, the RSS at location $X_i \in \Xi_\R$ is given by
\begin{equation}\label{equ: simulation_RSS_model}
    P_i = \ell(R_i) S_i(\sigma) = (C R_i)^{-\beta} S_i(\sigma).
\end{equation}

To account for AOA measurement noise, we adopt a wrapped normal model to capture the periodic nature of angular measurements:
\[
\Theta^{(1)}_i = \til{\Theta}^{(1)}_i \bmod 2\pi,
\qquad
\lb  \til{\Theta}^{(1)}_i \given \Psi^{(1)}_i, R_i \rb  \overset{d}{=} \N\!\lb\Psi^{(1)}_i - \pi,\, \E(R_i)\rb,
\]
where the AOA standard deviation $\sqrt{\mathcal{E}(r)}$ is a logistic-type function defined in radians:
\[
\sqrt{\mathcal{E}(r)} = \tau_{\min} + 
\frac{\tau_{\max} - \tau_{\min}}{1 + \exp\{-a(r - r_0)\}},
\]
where 
\[
\tau_{\min} = \frac{\pi}{90},~
\tau_{\max} = \frac{\pi}{36},~
a = 0.2,~
r_0 = 20.
\]
The AOA standard deviation $\sqrt{\mathcal{E}(r)}$ is an increasing function bounded between $\tau_{\min}$ and $\tau_{\max}$, and $a$ and $r_0$ control the rate of increase.

The choice of the model and the parameters is motivated by the following reasons. Empirical studies of indoor AOA-based localization systems report angular estimation errors typically on the order of a few degrees. In particular, experimental results in \cite{Diagne2020} show that the average AOA error is around $5^\circ$, with most errors below $10^\circ$ and worst-case deviations reaching approximately $15^\circ$ in indoor environments. These errors arise primarily due to multipath propagation and non-line-of-sight conditions, which are inherent to indoor wireless environments. Motivated by these empirical findings, we choose
\[
\tau_{\min} = \frac{\pi}{90} \approx 2^\circ,
\qquad
\tau_{\max} = \frac{\pi}{36} \approx 5^\circ,
\]
which capture high-quality short-range measurements and degraded conditions at larger distances. In particular, with $\tau_{\max} = \pi/36$, three standard deviations correspond to $\pi/12$, so that approximately $99.7\%$ of angular deviations lie within $\pm 15^\circ$ of the true direction under the Gaussian approximation. This is consistent with the indoor experimental observations reported in \cite{Diagne2020}.

We further set
\[
a = 0.2, \qquad r_0 = 20,
\]
so that the angular uncertainty remains close to its minimum at short distances, increases most rapidly in the medium-range regime, and approaches the upper bound for distant sensors. This reflects the expected deterioration of AOA estimation accuracy with propagation distance in indoor environments.

The logistic form ensures that the AOA uncertainty increases smoothly with distance. The resulting shape of $\E(r)$ used in the experiments is shown in Figure~\ref{fig:variance_function}.

\begin{figure}[htbp]
\centering
\begin{tikzpicture}
\begin{axis}[
    width=0.5\linewidth,
    xlabel={$r$ (m)},
    ylabel={$\sqrt{\mathcal{E}(r)}$ (radians)},
    grid=major,
    domain=0:40,
    samples=200,
    thick
]
\addplot [blue, thick] {rad(2) + (rad(5)-rad(2)) / (1 + exp(-0.2*(x - 20)))};
\end{axis}
\end{tikzpicture}
\caption{Logistic-type AOA standard deviation function $\sqrt{\mathcal{E}(r)}$ (in radians).}
\label{fig:variance_function}
\end{figure}

Given the propagation model, we construct an estimator $\hatr_i$ for the true distance to the target $R_i$ based on the received signal strength $P_i$. Taking logarithms of both sides of \eqref{equ: simulation_RSS_model} and use the model in \eqref{equ: noise_model} yields:
\begin{align}
    \ln P_i 
&= -\beta \ln C - \beta \ln R_i - \frac{\sigma^2}{\beta} + \sigma B_i \nonumber\\
&= \alpha + \gamma \ln R_i + \sigma B_i,\label{equ: relation between ln(P) and ln(R)}
\end{align}
where 
\[
\alpha := -\beta \ln C - \frac{\sigma^2}{\beta}, 
\quad 
\gamma := -\beta.
\]
This reveals a linear relationship between the logarithm of the RSS and the logarithm of the distance, which allows the system parameters to be estimated using a least squares approach. We assume that prior data are available before the main experiment and can be leveraged to estimate the parameters $\alpha$ and $\gamma$. These prior data sets can be interpreted as calibration measurements collected under controlled conditions, thus providing reliable estimates of the underlying system parameters. Let $\hat{\alpha}$ and $\hat{\gamma}$ denote the least squares estimates derived from this calibration data.

In all subsequent simulations, we use
\[
\hat{\alpha} = -5.3073, \quad \hat{\gamma} = -2.6277,
\]
which are the least squares estimates based on $10{,}000$ simulated pairs of $(P_i, R_i)$.

We can then define the distance estimator according to the relation in \eqref{equ: relation between ln(P) and ln(R)}

\begin{equation}\label{equ: regression_estimator}
    \hatr_i = \exp\!\lcb  \frac{\ln P_i - \hat{\alpha}}{\hat{\gamma}} \rcb.
\end{equation}

Under this regression-based estimator, the bounds in Theorem~\ref{thm: CMSD and MSD} can be expressed explicitly in terms of fundamental system parameters. Based on the RSS model \eqref{equ: relation between ln(P) and ln(R)}, we obtain
\[
\hatr_1
=
\exp\!\lcb \frac{\alpha - \hat{\alpha}}{\hat{\gamma}}\rcb
R_1^{\gamma/\hat{\gamma}}
\exp\!\lcb \frac{\sigma}{\hat{\gamma}} B_1\rcb.
\]
The first and second moments can be computed using the moment generating function of the standard normal RV, together with the density \eqref{equ: marginal_r} of the radial component in the finite HPPP model. This yields
\begin{equation}\label{equ: mean_hatr_system_parameter}
\IE \hatr_1 
=
\exp\!\lcb 
\frac{\alpha - \hat{\alpha}}{\hat{\gamma}}
+
\frac{\sigma^2}{2\hat{\gamma}^2}
\rcb
\frac{2\R^{\gamma/\hat{\gamma}}}{\gamma/\hat{\gamma}+2},
\end{equation}
and
\begin{equation}\label{equ: mean_hatrsq_system_parameter}
\IE\!\lsb  \hatr_1^2 \rsb
=
\exp\!\lcb 
\frac{2(\alpha - \hat{\alpha})}{\hat{\gamma}}
+
\frac{2\sigma^2}{\hat{\gamma}^2}
\rcb
\frac{\R^{2\gamma/\hat{\gamma}}}{\gamma/\hat{\gamma}+1}.
\end{equation}
Substituting these expressions into Theorem~\ref{thm: CMSD and MSD} with $d=2$ yields explicit bounds for the CMSE and MSE.

In practice, the underlying system parameters $\alpha$ and $\gamma$ (or equivalently $\beta,K,$ and $\sigma$) may not be directly observable. In such cases, the quantities $\IE\hatr_1$ and $\IE[\hatr_1^2]$ can be approximated using their empirical counterparts, obtained via sample moments of the distance estimates.

\subsection{Simulation results for Finite HPPP Model}\label{Section: simulate_HPPP}

We now present simulation results for the finite HPPP model and illustrate the impact of key system parameters on the CMSE and MSE. All reported results are obtained by averaging over 5000 independent simulation runs.

\begin{figure*}[!t]
\centering
\includegraphics[width=5in]{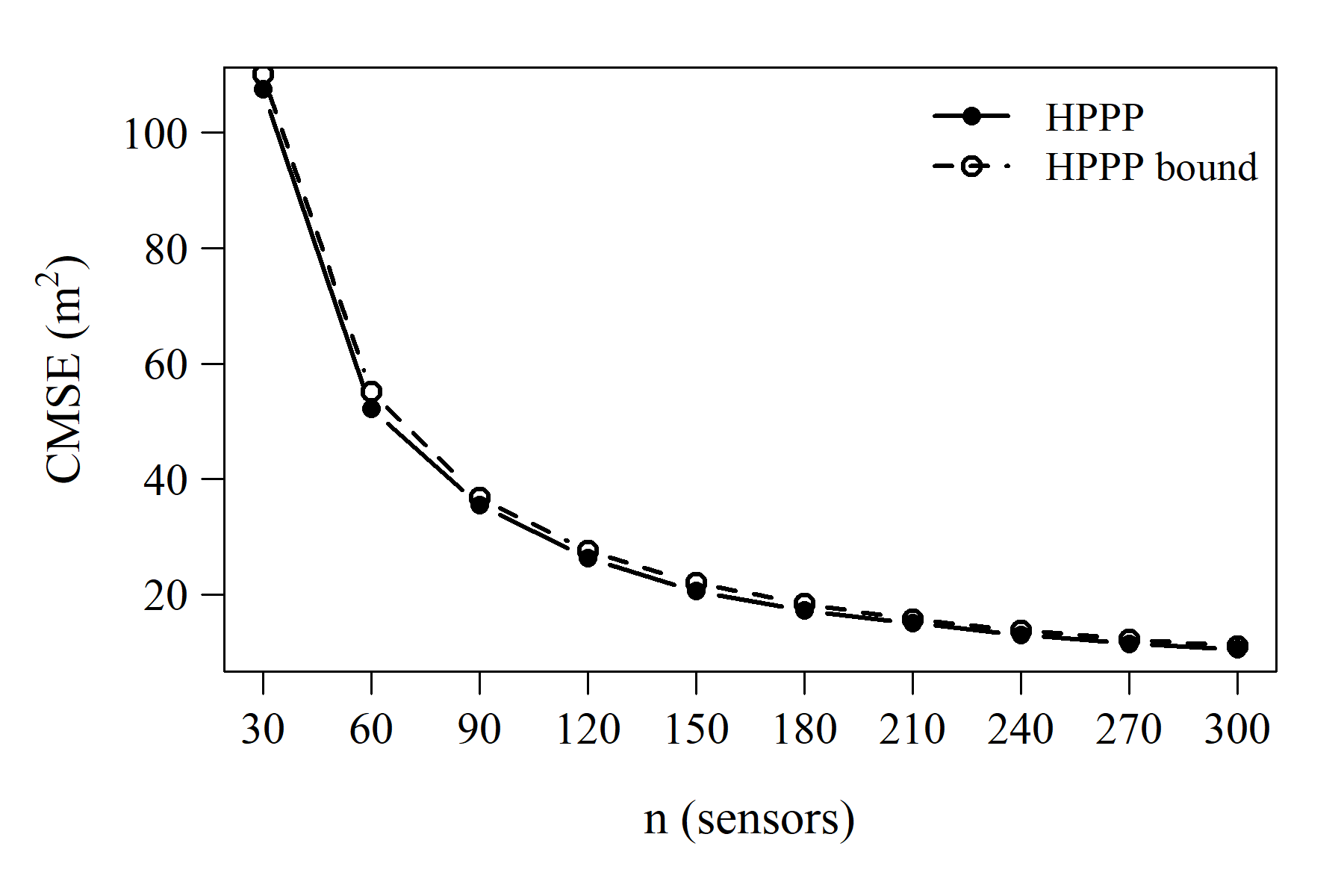}
\caption{CMSE and its theoretical bound versus the number of sensors $n$ for $\R = 30$ km under the HPPP network model.}
\label{fig:CMSD_vs_n_R30}
\end{figure*}

Figure~\ref{fig:CMSD_vs_n_R30} and Table~\ref{tab:cmse_ppp} shows the simulated CMSE and its theoretical bound as functions of the number of sensors $n$ for a fixed estimation range of $\R = 30$ m. The number of sensors is varied from 30 to 300. The results confirm that the simulated CMSE decreases as $n$ increases, consistent with the $n^{-1}$ scaling predicted by Theorem~\ref{thm: CMSD and MSD}. Moreover, the simulated CMSE remains consistently below the theoretical bound. This is further quantified in Table~\ref{tab:cmse_ppp}, where the ratio between the simulated CMSE and the theoretical bound is close to one (typically between $0.93$ and $0.98$), indicating that the bound is tight and accurately captures the conditional performance.

\begin{table}
\centering
\caption{Simulated CMSE and CMSE bound versus $n$ for the HPPP network.} 
\label{tab:cmse_ppp}
\begin{tabular}{rrrr}
  \toprule
$n$ & Simulated CMSE & CMSE Bound & Simulated CMSE / CMSE Bound \\ 
  \midrule
30 & 107.50 & 110.22 & 0.9754 \\ 
  60 & 52.20 & 55.11 & 0.9472 \\ 
  90 & 35.49 & 36.74 & 0.9661 \\ 
  120 & 26.34 & 27.55 & 0.9561 \\ 
  150 & 20.59 & 22.04 & 0.9339 \\ 
  180 & 17.25 & 18.37 & 0.9392 \\ 
  210 & 15.04 & 15.75 & 0.9555 \\ 
  240 & 12.99 & 13.78 & 0.9426 \\ 
  270 & 11.50 & 12.25 & 0.9389 \\ 
  300 & 10.48 & 11.02 & 0.9504 \\ 
   \bottomrule
\end{tabular}
\end{table}

\begin{figure*}[!t]
\centering
\includegraphics[width=5in]{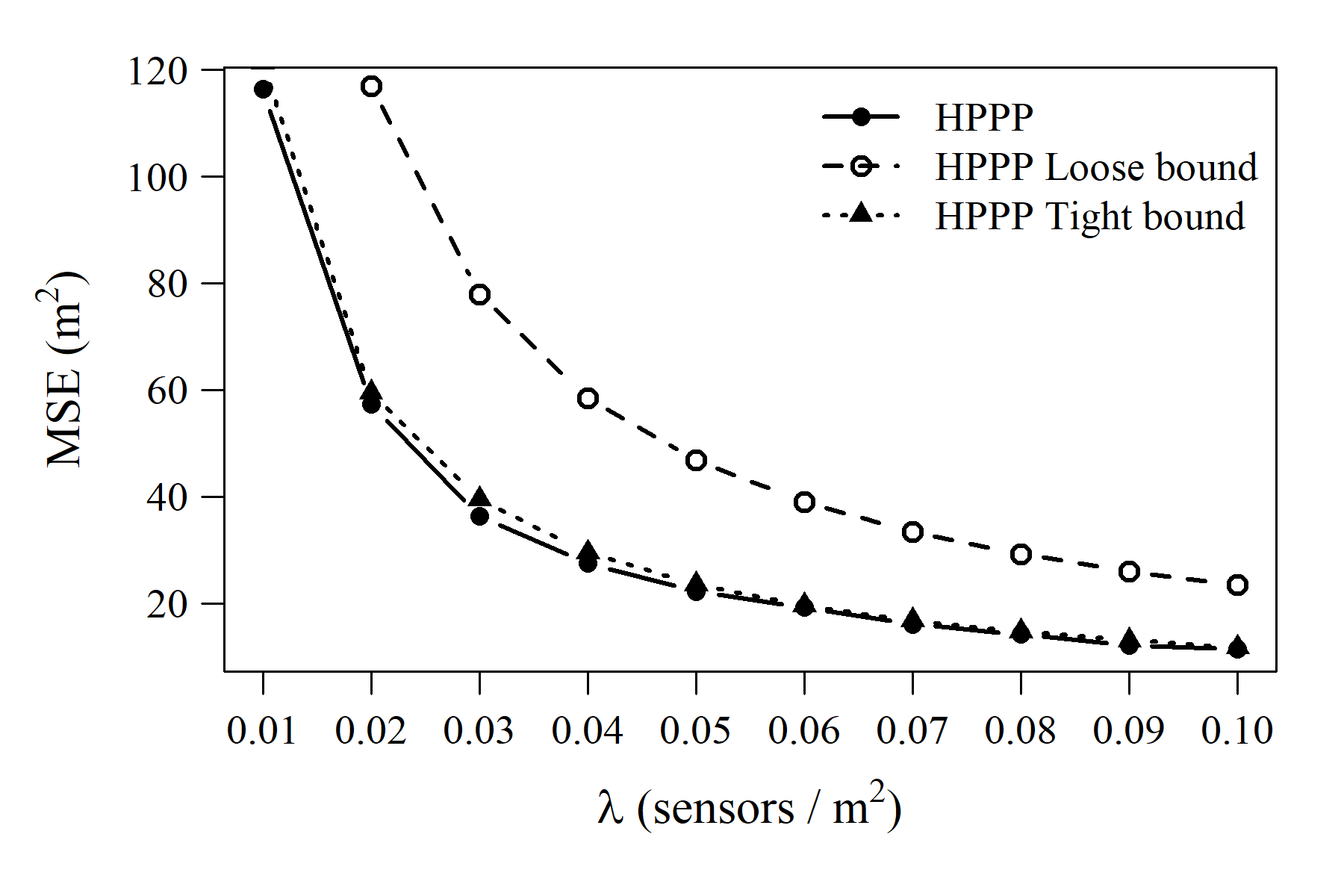}
\caption{MSE together with the tighter and looser bounds versus the sensor density $\lambda$ for $\R = 30$ km under the HPPP network model.}
\label{fig:MSD_vs_density_R30}
\end{figure*}

Figure~\ref{fig:MSD_vs_density_R30} and Table~\ref{tab:mse_ppp_improved} shows the simulated MSE together with both the tighter and looser bounds as functions of the sensor density $\lambda$, which ranges from $0.01$ to $0.1$ sensors per square meter. This corresponds to an average of approximately $30$ to $300$ sensors within the observation disk. As $\lambda$ increases, the simulated MSE decreases, which is consistent with the theoretically predicted scaling rate of $\lambda^{-1}$. The tighter bound \eqref{eq:new_bound} closely tracks the simulated MSE across the entire range of $\lambda$, while the looser bound \eqref{eq:old_bound} is approximately twice as large. This behavior is further quantified in Table~\ref{tab:mse_ppp_improved}, where the ratio $\text{Simulated MSE} / \text{Tight Bound}$ is consistently close to one (around $0.92$--$0.98$), whereas the ratio with respect to the loose bound is around $0.46$--$0.50$.

\begin{table}
\centering
\caption{Simulated MSE and MSE bounds versus $\lambda$ for the HPPP network.} 
\label{tab:mse_ppp_improved}
\begin{tabular}{rrrrrrr}
  \toprule
$\lambda$ & $\IE\lvert \Phi_{\mathcal{R}} \rvert$ & Simulated MSE & Tight Bound & Loose Bound & Simulated MSE / Tight Bound & Simulated MSE / Loose Bound \\ 
  \midrule
0.01 & 28.27 & 116.31 & 121.41 & 233.89 & 0.9580 & 0.4973 \\ 
  0.02 & 56.55 & 57.27 & 59.54 & 116.94 & 0.9617 & 0.4897 \\ 
  0.03 & 84.82 & 36.31 & 39.45 & 77.96 & 0.9204 & 0.4657 \\ 
  0.04 & 113.10 & 27.47 & 29.50 & 58.47 & 0.9314 & 0.4699 \\ 
  0.05 & 141.37 & 22.22 & 23.56 & 46.78 & 0.9434 & 0.4751 \\ 
  0.06 & 169.65 & 19.21 & 19.61 & 38.98 & 0.9796 & 0.4927 \\ 
  0.07 & 197.92 & 16.09 & 16.79 & 33.41 & 0.9581 & 0.4815 \\ 
  0.08 & 226.19 & 14.18 & 14.68 & 29.24 & 0.9654 & 0.4849 \\ 
  0.09 & 254.47 & 12.06 & 13.05 & 25.99 & 0.9247 & 0.4642 \\ 
  0.10 & 282.74 & 11.35 & 11.74 & 23.39 & 0.9674 & 0.4854 \\ 
   \bottomrule
\end{tabular}
\end{table}
Overall, these simulation results demonstrate that the proposed bounds are both theoretically sound and practically relevant. The CMSE bound accurately captures the conditional performance as a function of the number of sensors, while the MSE bounds quantify the impact of sensor density on overall localization accuracy. The comparison between the tighter and looser bounds highlights the trade-off between numerical precision and analytical simplicity: the tighter bound provides an accurate approximation to the true MSE, whereas the looser bound offers a clear and interpretable expression of the $\lambda^{-1}$ decay rate. These results provide useful guidance for the design and deployment of wireless localization systems under realistic conditions.

\subsection{Simulation results for Type II MHP Model}\label{section: simulate_MHP}

To the best of the authors’ knowledge, there is currently no computationally efficient and exact method for simulating repulsive or dependent PPs—such as the MHP, DPP, or MCP—\emph{conditioned on having exactly \(n\) sensors} within a bounded observation window. 
Since the CMSE is defined under conditioning on a fixed number of sensors, it cannot be reliably simulated for these non-HPPP models. 
We therefore restrict our numerical investigations for these models to the MSE performance, where the number of sensors is random and governed by the point intensity. 

A Type~II MHP sensor network \(\Xi_\R\) with density ranging from 0.03 to 0.07 sensors per square meter is simulated within a disk of radius \(\R = 30\) m in \(\IR^2\). 
For comparison, we also simulate a reference HPPP network \(\Phi_\R\) with the same intensity, and include the corresponding theoretical MSE bound derived under the HPPP assumption.

\begin{figure*}[!t]
\centering
\includegraphics[width=5in]{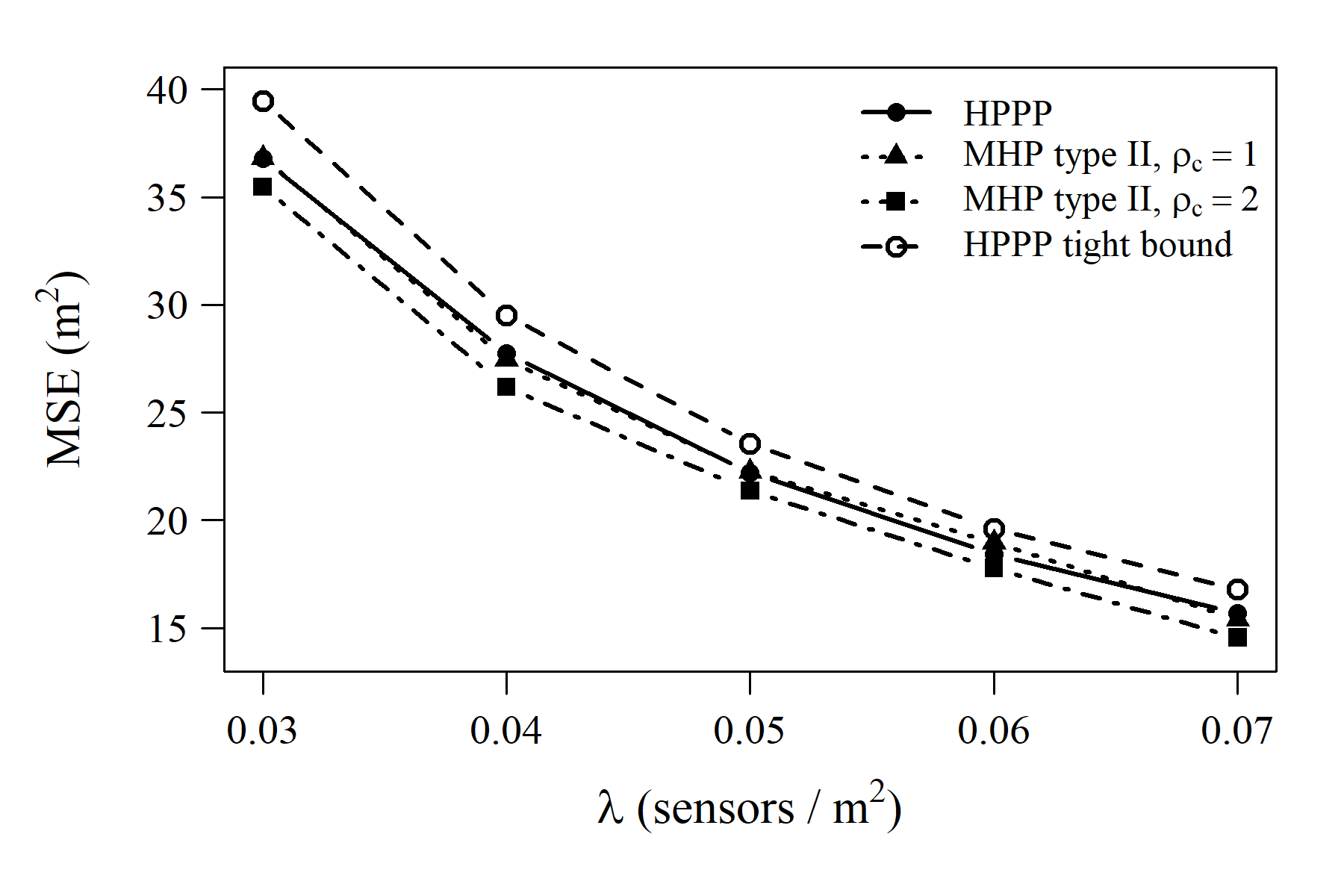}
\caption{MSE versus $\lambda$ for the HPPP and Type~II MHP networks with different hardcore radius \(\rho_c\).}
\label{fig:MSD_Poisson_vs_Matern_R30}
\end{figure*}

\begin{table}
\centering
\caption{Simulated MSE for the HPPP and Type II MHP networks with different hardcore radii.} 
\label{tab:ppp_vs_matern2_rho_sensitivity}
\begin{tabular}{rrrrrr}
  \toprule
$\lambda$ & $\lambda \pi \R^2$ & HPPP & MHP Type II $(\rho_c=1)$ & MHP Type II $(\rho_c=2)$ & HPPP Tight Bound \\ 
  \midrule
0.03 & 84.82 & 36.78 & 36.82 & 35.48 & 39.45 \\ 
  0.04 & 113.10 & 27.74 & 27.47 & 26.19 & 29.50 \\ 
  0.05 & 141.37 & 22.20 & 22.26 & 21.36 & 23.56 \\ 
  0.06 & 169.65 & 18.43 & 18.98 & 17.77 & 19.61 \\ 
  0.07 & 197.92 & 15.69 & 15.40 & 14.54 & 16.79 \\ 
   \bottomrule
\end{tabular}
\end{table}

Figure~\ref{fig:MSD_Poisson_vs_Matern_R30} and Table~\ref{tab:ppp_vs_matern2_rho_sensitivity} present the simulated MSE for the reference HPPP and Type~II MHP networks under different hardcore radii \(\rho_c\). 
The intensity of the underlying parent HPPP in the MHP construction is selected according to~\eqref{equ:density_MHP_II} so that the resulting MHP \(\Xi_\R\) has the same point density as the reference HPPP \(\Phi_\R\).

We observe that when the hardcore radius is small (e.g., \(\rho_c = 1\) m), the MSE of the MHP network is nearly identical to that of the HPPP. 
In this regime, the repulsion between points is weak, and the spatial configuration of the MHP closely resembles that of an HPPP, resulting in similar estimation performance.

As the hardcore radius increases (e.g., \(\rho_c = 2\) m), the repulsion becomes stronger, leading to a more evenly spaced sensor deployment. 
This increased spatial regularity improves coverage and reduces clustering effects, which in turn yields a consistently lower MSE compared to the HPPP across all considered density levels.

While the repulsive deployments lead to a more regular spatial configuration and thus improve the MSE performance compared to the uniform (HPPP) deployment, the MSE of both MHP networks remains close to that of the HPPP across all density levels. These results are consistent with the theoretical convergence result in Theorem~\ref{convergence_thm}, which implies that when both the observation range $\R$ and the system noise parameter $\sigma$ are sufficiently large, the fusion center cannot distinguish between the observable processes induced by the HPPP and the MHP. Consequently, the corresponding MSE curves appear almost identical, which confirms that the bounds established in Theorem~\ref{thm: CMSD and MSD} apply to a broader class of random sensor networks beyond the HPPP assumption, provided that both $\R$ and $\sigma$ are large enough.

\subsection{Simulation results for MCP Model}
We simulate several MCP networks $\Xi_\R$ with varying cluster radius $\rho_c$ and mean cluster size $\bar{c}$ within a fixed observation region $\R = 30$ m in $\IR^2$, together with a reference HPPP network $\Phi_\R$. The parent intensity $\lambda_0$ of each MCP is selected such that the overall density $\lambda$ matches that of the reference HPPP, according to \eqref{equ:density_MCP_TCP}.

As discussed in Section~\ref{section: simulate_MHP}, there is currently no computationally efficient method for simulating an MCP network conditioned on having exactly $n$ sensors within a bounded observation window. Consequently, we restrict our numerical study of the MCP model to the MSE performance.

\begin{figure*}[!t]
\centering
\includegraphics[width=5in]{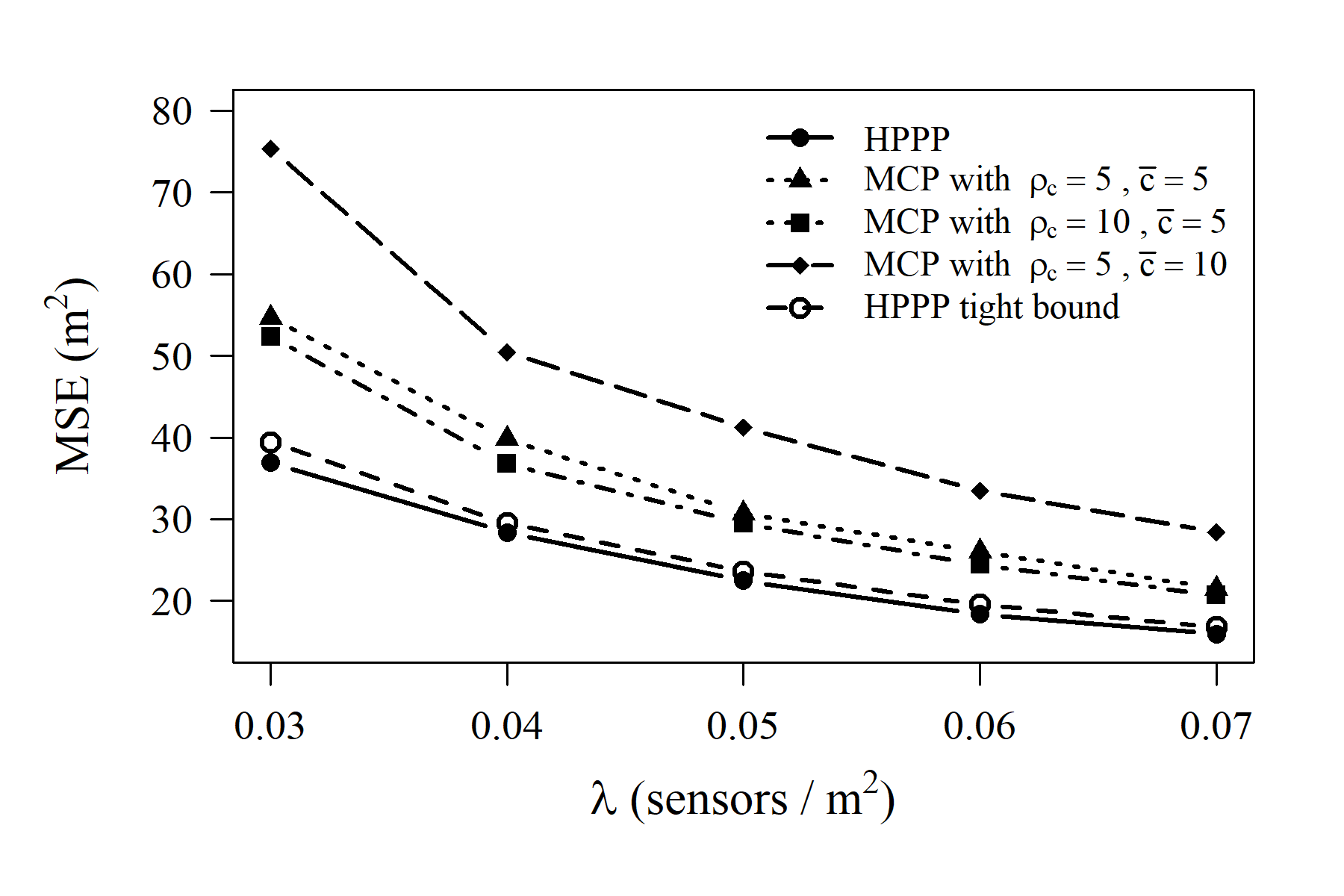}
\caption{MSE versus $\lambda$ for the HPPP and MCP networks with different cluster radius \(\rho_c\) and mean cluster size $\bar{c}$.}
\label{fig:MSD_Poisson_vs_MaternCluster_R30}
\end{figure*}

\begin{table}
\centering
\caption{Simulated MSE for the HPPP and MCP networks under different clustering settings.} 
\label{tab:ppp_vs_mcp_sensitivity}
\begin{tabular}{rrrrrrr}
  \toprule
$\lambda$ & $\lambda \pi \R^2$ & HPPP & MCP $(\bar{c}=5,\rho_c=5)$ & MCP $(\bar{c}=5,\rho_c=10)$ & MCP $(\bar{c}=10,\rho_c=5)$ & HPPP Tight Bound \\ 
  \midrule
0.03 & 84.82 & 36.92 & 54.62 & 52.32 & 75.38 & 39.45 \\ 
  0.04 & 113.10 & 28.39 & 39.81 & 36.82 & 50.44 & 29.50 \\ 
  0.05 & 141.37 & 22.46 & 30.72 & 29.47 & 41.21 & 23.56 \\ 
  0.06 & 169.65 & 18.36 & 26.02 & 24.44 & 33.46 & 19.61 \\ 
  0.07 & 197.92 & 15.94 & 21.50 & 20.72 & 28.39 & 16.79 \\ 
   \bottomrule
\end{tabular}
\end{table}

Figure~\ref{fig:MSD_Poisson_vs_MaternCluster_R30} and Table~\ref{tab:ppp_vs_mcp_sensitivity} presents the simulated MSE for both the HPPP $\Phi_\R$ and MCP $\Xi_\R$ networks, together with the theoretical upper bound derived for the HPPP, as functions of the sensor density. We observe that the MSE of all MCP configurations is consistently higher than that of the HPPP, indicating that clustered deployments generally yield inferior global localization performance compared to HPPP deployments.

The impact of clustering parameters is also evident. For a fixed cluster radius $\rho_c$, increasing the mean cluster size $\bar{c}$ (and thus reducing the parent intensity $\lambda_0$) leads to worse performance, as sensors become more tightly concentrated within clusters. Conversely, for a fixed $\bar{c}$ (and hence fixed $\lambda_0$), increasing the cluster radius $\rho_c$ improves performance, since the sensors are more spatially dispersed and the overall deployment becomes closer to uniform.

Despite this degradation in global performance, clustered deployments may offer advantages in localized regions. In particular, when the target lies within or near a cluster, the higher local sensor density can lead to improved localization accuracy. This highlights an inherent trade-off: while uniform deployments provide more balanced performance across the entire region, clustered deployments can offer superior accuracy in selected areas at the expense of performance elsewhere.

Finally, we note that the MSE of the MCP remains relatively close to that of the HPPP despite the differing spatial structures. This suggests that when the observation range $\R$ and the system noise parameter $\sigma$ are sufficiently large, the observable processes induced by MCP and PPP become increasingly similar. We expect this agreement to strengthen further as $\sigma$ increases. These findings provide additional numerical evidence supporting the applicability of the bounds in Theorem~\ref{thm: CMSD and MSD} to a broader class of random sensor networks, beyond the HPPP setting, under sufficiently large observation ranges and noise levels.

\subsection{Weighted fusion}\label{Section: weighted fusion}

In this section, we compare the performance of the simple average estimator \eqref{equ: targetEstimator_PPP} with a class of weighted average estimators defined in \eqref{equ: weighted_estimator_order} under the HPPP network model. The weighted estimator assigns higher weights to sensors that report smaller estimated distances to the target, where the parameter $0<a\leq 1$ controls the sharpness of the weighting function. Smaller values of $a$ place more emphasis on ``nearby'' sensors, while larger values distribute more weight toward distant sensors. The case $a=1$ reduces to the simple average estimator, which assigns equal weight to all sensors.

\begin{figure*}[!t]
\centering
\includegraphics[width=5in]{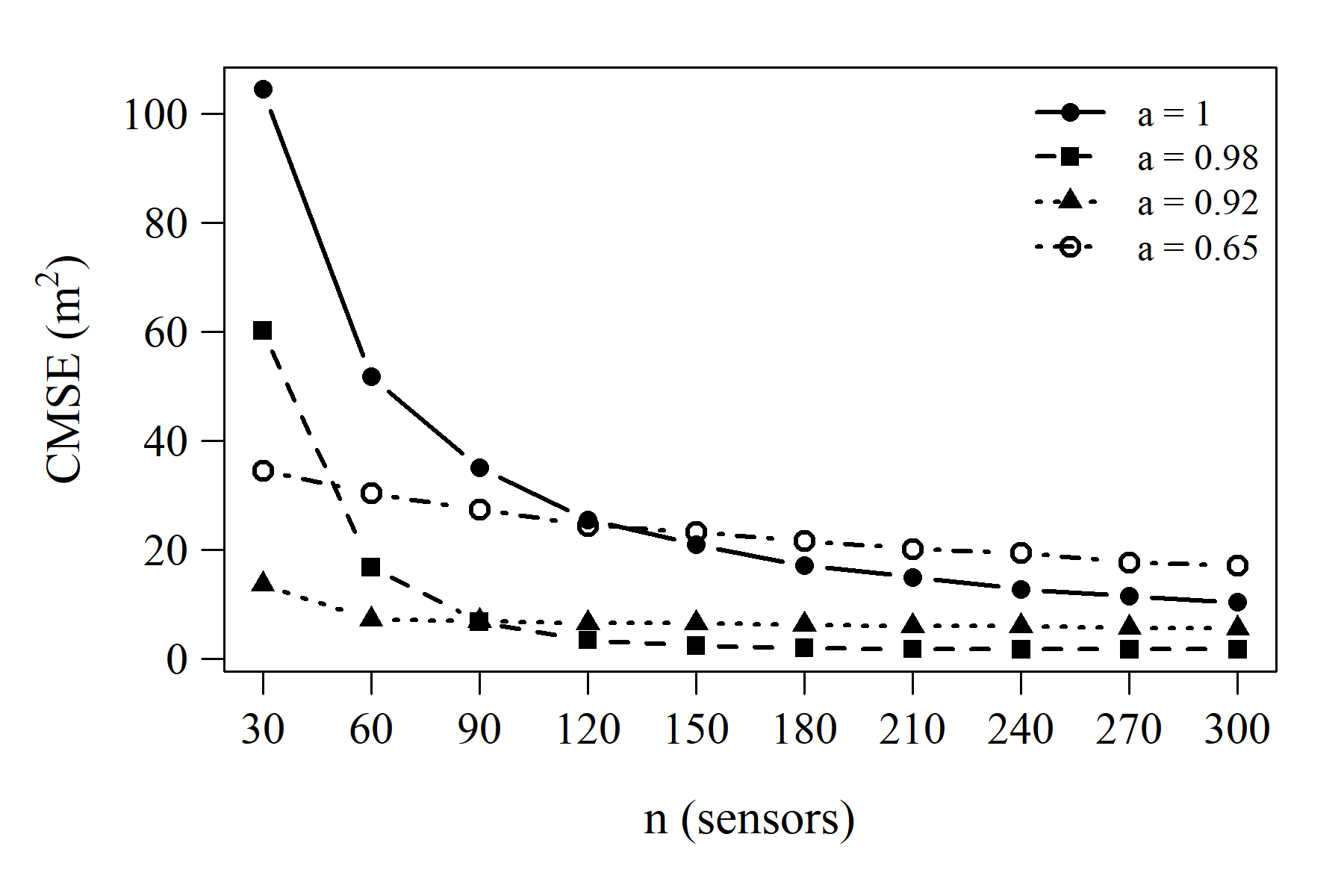}
\caption{CMSE versus $n$ of the weighted estimator with different weighting parameters under the HPPP model.}
\label{fig:CMSE_weighted_different_a}
\end{figure*}

\begin{table}
\centering
\caption{Simulated CMSE of weighted estimators with different weighting parameters under the HPPP model.} 
\label{tab:cmse_different_a_values}
\begin{tabular}{rrrrr}
  \toprule
$n$ & $a=1$ & $a=0.98$ & $a=0.92$ & $a=0.65$ \\ 
  \midrule
30 & 104.40 & 60.12 & 13.62 & 34.48 \\ 
  60 & 51.70 & 16.79 & 7.23 & 30.32 \\ 
  90 & 35.05 & 6.76 & 6.88 & 27.38 \\ 
  120 & 25.51 & 3.40 & 6.49 & 24.40 \\ 
  150 & 20.90 & 2.33 & 6.53 & 23.21 \\ 
  180 & 17.06 & 1.92 & 6.22 & 21.56 \\ 
  210 & 14.97 & 1.76 & 5.97 & 20.17 \\ 
  240 & 12.76 & 1.74 & 5.93 & 19.37 \\ 
  270 & 11.42 & 1.70 & 5.66 & 17.64 \\ 
  300 & 10.36 & 1.69 & 5.57 & 17.15 \\ 
   \bottomrule
\end{tabular}
\end{table}

Figure~\ref{fig:CMSE_weighted_different_a} and Table~\ref{tab:cmse_different_a_values} present the simulated CMSE of the simple and weighted estimators as functions of the number of sensors $n$. For small values of $n$, all weighted estimators with different choices of $a$ outperform the simple average estimator. However, as $n$ increases, the performance of some weighted estimators (e.g., $a=0.65$) becomes worse than that of the simple estimator. 

It is also evident that the CMSE of all weighted estimators decays more slowly than that of the simple average estimator. This suggests that the simple estimator will eventually outperform all weighted estimators as $n \to \infty$. For instance, when $n=5000$, the simple estimator outperforms the weighted estimator with $a=0.98$
\[
\mathrm{CMSE}(5000)\approx 0.6242\ \mathrm{m}^2,\qquad 
\mathrm{CMSE}^W\!(5000)\approx 0.7707\ \mathrm{m}^2,
\]
demonstrating its superior asymptotic performance.

On the other hand, for moderate values of $n$ and suitably chosen weighting parameters, weighted estimators can provide improved localization accuracy. For example, when $n=3000$, the weighted estimator with $a=0.98$ still outperforms the simple estimator:
\[
\mathrm{CMSE}(3000)\approx 1.1110\ \mathrm{m}^2,\qquad 
\mathrm{CMSE}^W\!(3000)\approx 0.9866\ \mathrm{m}^2,
\]
which is more than sufficient for practical purposes.

Overall, while some weighted estimators can offer gains in finite-sample regimes, their performance is not guaranteed in general—particularly for large $n$—due to the inefficient use of information from distant sensors.

\section{Discussion and Future Work}\label{Section: Discussion}
This paper has presented a comprehensive analysis of wireless target localization in \(d\)-dimensional sensor networks, with particular emphasis on the role of spatial PP models. By combining analytical derivations with simulation studies, we have characterized the MSE and CMSE under a broad range of network configurations. In particular, we established an approximation theorem showing that when the observation region is sufficiently large and the propagation noise is sufficiently significant, the fusion center---which aggregates measurements from all sensors within this finite observation region---cannot statistically distinguish whether the underlying sensor placements follow an HPPP or a more general stationary isotropic PP satisfying the SRD condition~\eqref{equ: SRD}. Consequently, the MSE and CMSE of this broader class of networks are expected to be close to those of the HPPP case.

Focusing on the finite HPPP model, we analyzed the performance of an unbiased estimator \eqref{equ: targetEstimator} of the target position and then derived explicit upper bounds for the MSE and CMSE in terms of fundamental system parameters. These bounds can serve as approximate performance bounds for a wide range of non-HPPP deployment models satisfying the SRD condition~\eqref{equ: SRD}. From a practical perspective, if the network designers believe that the underlying sensor configuration falls within this class, then the derived bounds can offer quantitative guidance for deployment decisions: for example, determining how many sensors are required to achieve a desired level of accuracy, or evaluating how different spatial deployment strategies (uniform, repulsive, or clustered) affect localization performance.

As emphasized in Section~\ref{Section: Intro}, the objective of this work is not to analyze the performance of a particular realization of a sensor network at a specific geographic location. Such an approach would require repeating the analysis for each distinct configuration. Instead, our goal has been to characterize the \emph{average performance across all possible realizations}, thereby providing general guidelines and insights for the design and deployment of wireless sensor networks. By interpreting different geographical networks as realizations of the same spatial PP, we leveraged tools from stochastic geometry and PP theory to analyze the performance of the localization algorithm (Algorithm~\ref{Algo1}) that utilizes RSS and AOA as inputs.

The findings of this paper open several promising directions for future research:

\begin{itemize}
\item \textbf{CMSE and MSE bounds for stationary isotropic sensor networks:}\\
More refined CMSE and MSE bounds for general stationary isotropic sensor networks should ideally be expressed in terms of intrinsic characteristics of the underlying network. In this direction, there are two main aspects to investigate. First, one may study the limiting distribution of the properly scaled second-order Wasserstein distance between the estimator and the target location, since such a limit law would form the basis for statistical inference. A natural candidate is the chi-squared distribution, as it arises from the sum of squares of independent standard normal RVs. Second, it is of interest to establish quantitative bounds on the second-order Wasserstein distance between the estimator and its limiting distribution. Developing suitable tools for controlling these approximation errors presents an intriguing research challenge, with potential implications for improving the accuracy and reliability of localization methods in sensor networks.

    \item \textbf{Spatially correlated propagation effects:} \\
    Our convergence results assume i.i.d. signal strength distortions and conditional independence of angular deviations across sensors. While analytically convenient, these assumptions may be unrealistic in practice \cite{correlated_shadowing1,correlated_shadowing2}. For example, signals arriving from similar directions are expected to exhibit spatially correlated fading and angular errors. In this context, \cite{7745970} extended the convergence results of \cite{KeelerRossXia2018} for the inverse signal strength process by modeling the fading variables as a log-Gaussian random field with a spatial correlation structure, rather than as i.i.d. positive RVs. Extending our results to incorporate spatially correlated propagation effects would therefore be of both theoretical and practical interest.
    
    \item \textbf{Alternative localization strategies:} \\
    This work focused exclusively on localization strategies that use RSS and AOA as inputs. 
    An interesting future direction is to study other strategies that rely on different 
    measurement modalities, such as TOA, time difference of arrival, 
    or combinations of multiple signal features. Analyzing their performance within the 
    stochastic geometry framework could provide deeper insights into the trade-offs between 
    different sensing modalities.
\end{itemize}

\section{Proofs}\label{Section: proof}

\begin{proof}[Proof of Theorem~\ref{thm: dtv_thm}]
    The proof follows similar lines of reasoning as in \cite[Theorem~2.2]{KeelerRossXia2018}. Let $\ZZ_i^\sig$ be independent PPPs on $\posRealLine \times [0,2\pi)^{d-1}$ with mean measure
\[
\M_i^\sig(\dx t, \dx \boldsymbol{\nu})
= p_i^\sig(\dx t)\IP\!\lb \boldsymbol{\Theta}_1 \in \dx \boldsymbol{\nu} \given X_1=x_i\rb ,
\]
and let $\NN_i^\sig:=\delta_{\lb  N_i,\boldsymbol{\Theta}_i\rb }$ be the process that places a single point at $\lb  N_i,\boldsymbol{\Theta}_i\rb $. Then
\[
\sum_{i\in\mathcal{I}^\xi} \ZZ_i^\sig \dequal \ZZ_\xi^\sig,
\quad
\sum_{i\in\mathcal{I}^\xi} \NN_i^\sig = \NN^\sig_\xi.
\]

Restricting to $(0,\tau]\times[0,2\pi)^{d-1}$, we denote the corresponding processes by $\ZZ_i^\sig|_\tau$ and $\NN_i^\sig|_\tau$. By independence and a standard tensorization argument for total variation distance \cite[(1.4)]{Kontorovich2025}, we obtain
\begin{equation}\label{equ: dtv_tensor}
\dtv\!\lb \law\!\lb \NN^\sig_\xi |_\tau\rb ,\law\!\lb \ZZ^\sig_\xi |_\tau\rb \rb 
\le \sum_{i\in\mathcal{I}^\xi}
\dtv\!\lb \law\lb \NN_i^\sig |_\tau\rb ,\law\lb \ZZ_i^\sig |_\tau\rb \rb .
\end{equation}

Using the coupling characterization for the total variation distance \cite[Chapter~A.1]{BarbourHolstJanson1992}, we have
\[
\dtv\!\lb \law\lb \NN_i^\sig |_\tau\rb ,\law\lb \ZZ_i^\sig |_\tau\rb \rb 
= \inf_{\lb  \til\NN_i^\sig |_\tau ,\til \ZZ_i^\sig |_\tau\rb } \IP\!\lb  \til\NN_i^\sig |_\tau \neq \til \ZZ_i^\sig |_\tau \rb ,
\]
where $\lb  \til\NN_i^\sig |_\tau ,\til \ZZ_i^\sig |_\tau\rb $ ranges over all couplings of $\lb  \NN_i^\sig |_\tau , \ZZ_i^\sig |_\tau\rb $.

For the PPP $\ZZ_i^\sig$, conditional on $\ZZ_i^\sig(\tau)=1$ (i.e., there is exactly one point in $(0,\tau]\times[0,2\pi)^{d-1}$), this point is distributed according to the density
\[
\frac{\M_i^\sig(\dx t, \dx \boldsymbol{\nu})}{\M_i^\sig(\tau)}
= \frac{p_i^\sig(\dx t)\IP\!\lb \boldsymbol{\Theta}_1 \in \dx \boldsymbol{\nu} \given X_1=x_i\rb}{p_i^\sig(\tau)}.
\]

Similarly, conditional on $\NN_i^\sig(\tau)=1$, the point $\lb  N_i,\boldsymbol{\Theta}_i\rb $ has the same distribution.

Since $\NN_i^\sig(\tau)\in\{0,1\}$, whenever $\NN_i^\sig(\tau)=\ZZ_i^\sig(\tau)$, the processes $\ZZ_i^\sig|_\tau$ and $\NN_i^\sig|_\tau$ can be coupled exactly, as they share the same conditional distribution (the case $\NN_i^\sig(\tau)=\ZZ_i^\sig(\tau)=0$ is trivial since both processes are empty).

Notice that
\begin{align*}
\IP\!\lb  \til\NN_i^\sig |_\tau \neq \til \ZZ_i^\sig |_\tau \rb  
= \IP&\lb  \til\NN_i^\sig |_\tau \neq \til \ZZ_i^\sig |_\tau \given \til\NN_i^\sig (\tau) = \til \ZZ_i^\sig (\tau) \rb  \IP\!\lb  \til\NN_i^\sig (\tau) = \til \ZZ_i^\sig (\tau) \rb  +\IP\!\lb  \til\NN_i^\sig (\tau) \neq \til \ZZ_i^\sig (\tau) \rb 
\end{align*}

Since an optimal coupling matches $\til\ZZ_i^\sig|_\tau$ and $\til\NN_i^\sig|_\tau$ whenever $\til\NN_i^\sig(\tau)=\til\ZZ_i^\sig(\tau)$, it follows that
\begin{align}
\dtv\!\lb \law\lb \NN_i^\sig |_t\rb ,\law\lb \ZZ_i^\sig |_t\rb \rb 
&= \inf_{\lb  \til\NN_i^\sig(\tau),\til\ZZ_i^\sig(\tau)\rb } \IP\!\lb  \til\NN_i^\sig(\tau) \neq \til\ZZ_i^\sig(\tau) \rb \nonumber\\
&= \dtv\!\lb \law\lb \NN_i^\sig(\tau)\rb ,\law\lb \ZZ_i^\sig(\tau)\rb \rb ,
\label{equ: dtv_reduce_to_PoissonApproximation}
\end{align}
where $\lb  \til\NN_i^\sig(\tau),\til\ZZ_i^\sig(\tau)\rb $ ranges over all couplings of $\lb  \NN_i^\sig(\tau),\ZZ_i^\sig(\tau)\rb $.

Finally, since $\ZZ_i^\sig(\tau)$ is a Poisson RV with mean $p_i^\sig(\tau)$ and $\NN_i^\sig(\tau)$ is a Bernoulli RV with success probability $p_i^\sig(\tau)$, it follows from \cite[(1.8)]{BarbourHolstJanson1992} that
\[
\dtv\!\lb \law\lb \NN_i^\sig(\tau)\rb ,\law\lb \ZZ_i^\sig(\tau)\rb \rb 
\leq p_i^\sig(\tau)^2.
\]

Combining \eqref{equ: dtv_tensor} and \eqref{equ: dtv_reduce_to_PoissonApproximation} yields the first upper bound. The second upper bound follows directly from
\[
p_i^\sig(\tau)\leq \sup_{j\in\mathcal{I}^\xi} p_j^\sig(t).
\]

For the lower bound, note that the number of points of $\ZZ_\xi^\sig$ and $\NN_\xi^\sig$ falling inside $(0,\tau]\times[0,2\pi)^{d-1}$ is a measurable function of $ \ZZ_\xi^\sig|_\tau$ and $ \NN_\xi^\sig|_\tau$. Since total variation distance cannot increase under measurable mappings, it follows that
\[
\dtv\!\lb \law\!\lb \NN^\sig_\xi(\tau)\rb ,\law\!\lb \ZZ^\sig_\xi(\tau)\rb \rb  \leq \dtv\!\lb \law\!\lb \NN^\sig_\xi|_\tau\rb ,\law\!\lb \ZZ^\sig_\xi|_\tau\rb \rb .
\]

Observe that $\NN^\sig_\xi(\tau) = \sum_{i\in\mathcal{I}^\xi} \I\!\lb  N_i\in (0,\tau]\rb $ is a sum of independent indicator RVs, while $\ZZ^\sig_\xi(\tau)$ is a Poisson RV, both having the same mean $\M_\xi^\sig (\tau)$. We can therefore apply \cite[Corollary~3.D.1]{BarbourHolstJanson1992}, which yields the desired lower bound.
\end{proof}

\begin{proof}[Proof of Corollary~\ref{corollary: cox_approximation}]
    The result follows directly from Theorem~\ref{thm: dtv_thm} and Campbell's formula (see \cite[Theorem 4.1]{Haenggi2012}) once we establish that
    \begin{align*}
        \dtv\!\lb \law\lb \NN^\sig_\Xi|_\tau\rb ,\law\lb \ZZ^\sig_\Xi|_\tau\rb \rb
        &=\sup_{A \in \BB(\HH)}\left| \IE\!\lsb \I\!\lb \NN^\sig_\Xi|_\tau \in A\rb - \I\! \lb\ZZ^\sig_\Xi|_\tau \in A\rb \rsb\right|\\
        &\leq\IE\!\sup_{A \in \BB(\HH)}\left| \IE\! \lsb \I\!\lb \NN^\sig_\Xi|_\tau \in A\rb - \I\! \lb\ZZ^\sig_\Xi|_\tau \in A\rb \given\Xi\rsb\right|\\
        &=\IE \dtv\!\lb \law\lb \NN^\sig_\Xi|_\tau \given\Xi\rb ,\law\lb \ZZ^\sig_\Xi|_\tau \given\Xi\rb \rb,
    \end{align*}
    where the inequality follows from conditioning and Jensen’s inequality.
\end{proof}

\begin{proof}[Proof of Theorem~\ref{theorem: mean measure converge}]
    We apply \cite[Theorem~4.2]{Kallenberg1983}, which states that for random measures defined on a Polish space (i.e., a complete and separable metric space), convergence of random measures is characterized by the convergence of integrals against a suitable class of continuous test functions.

    The state space $\posRealLine\times[0,2\pi)^{d-1}$ can be equipped with a metric under which it is complete and separable. For example, for $x=(x_1,\dots,x_d)$ and $y=(y_1,\dots,y_d)$, define
\[
d(x,y)
=
\sqrt{(x_1-y_1)^2+\left(\frac 1{x_1}-\frac1{y_1}\right)^2}
+
\sum_{i=2}^d
\frac{|x_i-y_i|}{(2\pi-x_i)(2\pi-y_i)}.
\]
Under this metric, distances diverge as points approach the open boundary, preventing Cauchy sequences from converging to the boundary and ensuring that all limits remain within the space. 
Therefore, $\posRealLine\times[0,2\pi)^{d-1}$ is a Polish space, and Theorem~4.2 applies.

To show the convergence of the $\M^\sig_\Xi$, it is enough to check that for any continuous function 
$f:\posRealLine\times[0,2\pi)^{d-1}\to\mathbb{R}_+$ with compact support and bounded continuous partial derivatives of order $d$,
\begin{equation*}
    \lim_\sigmatoinf\int_{\posRealLine\times[0,2\pi)^{d-1}} f\lb  t,\boldsymbol{\nu} \rb  \M_\Xi^\sig\!\lb  \dx t,\dx \boldsymbol{\nu} \rb  
    \dequal \lim_\sigmatoinf \int_{\posRealLine\times[0,2\pi)^{d-1}} f\lb  t,\boldsymbol{\nu} \rb  \M^\sig\!\lb  \dx t,\dx \boldsymbol{\nu} \rb ,
\end{equation*}
since this class of functions is dense in the space of continuous functions. The main idea of the proof is to represent the left-hand side in terms of the right-hand side plus some error terms, and then show that these error terms converge to zero as $\sigma\to\infty$.

For any $\rho>0$ and $\boldsymbol{\nu}:=\lcb  \nu^{(j)}\rcb_{j=1}^{d-1}$ satisfying $0<\nu^{(j)}<2\pi$, we define 
\[
|\Xi|(\rho):=\sum_{i \in \mathcal{I}^\Xi} \I\!\lb 0< R_i \leq \rho\rb 
\]
as the number of points of $\Xi$ that lie within distance $\rho$ from the origin, and
\begin{align}
    |\Xi|_{\boldsymbol{\Theta}}\!\lb  \rho, \boldsymbol{\nu} \rb 
    &:= \sum_{i \in \mathcal{I}^\Xi} \IP\!\lb  R_i \leq \rho, \lcb \vtj_i \leq \nu^{(j)}\rcb_{j=1}^{d-1} \given X_i \rb  \label{equ: representation|Xi|}\\
    &=\sum_{i \in \mathcal{I}^\Xi} \I\!\lb  R_i \leq \rho\rb  \PP(X_i,\boldsymbol{\nu}),\nonumber
\end{align}
where
\[
\PP(X_i,\boldsymbol{\nu}):=\prod_{j=1}^{d-1} \IP\!\lb \vtj_i \le \nu^{(j)} \given X_i\rb .
\]

Accordingly, we define 
\[
|\Lambda|(\rho):=\IE|\Xi|(\rho) = \lambda V_d(\rho),
\]
where we recall $V_d(\rho)$ denotes the volume of a $d$-dimensional ball with radius $\rho$, and
\begin{align}
    |\Lambda|_{\boldsymbol{\Theta}}\!\lb  \rho, \boldsymbol{\nu} \rb 
    &:= \IE|\Xi|_{\boldsymbol{\Theta}}\lb  \rho, \boldsymbol{\nu} \rb \nonumber \label{equ: new_rep_E|Xi|}
\end{align}
as their expectations. 

We observe that
\[
|\Xi|_{\boldsymbol{\Theta}}\!\lb  \rho, \boldsymbol{\nu} \rb  \leq |\Xi|(\rho)
\]
almost surely, and hence also
\begin{equation}\label{equ: LambdaInequality}
    |\Lambda|_{\boldsymbol{\Theta}}\!\lb  \rho, \boldsymbol{\nu} \rb  \leq |\Lambda|(\rho).
\end{equation}

We let $F_\sigma$ be the distribution function of $S\sig$, then it follows from the representation \eqref{equ: representation|Xi|} that
\begin{align*}
    \M_\Xi^\sig\!\lb  t,\boldsymbol{\nu} \rb  
    &:=\M_\Xi^\sig\!\lb  (0,t],
    \prod_{j=1}^{d-1} \lb 0,\nu^{(j)} \rsb\rb  \\
    &=  \sum_{i \in \mathcal{I}^\Xi} \IP\!\lb \frac{g(R_i)}{S(\sigma)} \leq t, \lcb \vtj_i \leq \nu^{(j)}\rcb_{j=1}^{d-1} \given X_i \rb \\
    &=  \sum_{i \in \mathcal{I}^\Xi} \int_{\posRealLine}\IP\!\lb \frac{g(R_i)}{s} \leq t, \lcb \vtj_i \leq \nu^{(j)}\rcb_{j=1}^{d-1} \given X_i \rb  \dx F_\sigma(s)\\
    &= \int_{\posRealLine} \sum_{i \in \mathcal{I}^\Xi} \IP\!\lb  R_i \leq g^{-1}\!\lb  st \rb , \lcb \vtj_i \leq \nu^{(j)}\rcb_{j=1}^{d-1} \given X_i \rb  \dx F_\sigma(s)\\
    &= \int_{\posRealLine} |\Xi|_{\boldsymbol{\Theta}}\! \lb  g^{-1}\!\lb  st \rb   , \boldsymbol{\nu} \rb  \dx F_\sigma(s),
\end{align*}
and, by taking expectation,
\begin{equation}\label{equ: same_mean_measure_CDF}
    \M^\sig\!\lb  t,\boldsymbol{\nu} \rb  :=\IE \M_\Xi^\sig\!\lb  t,\boldsymbol{\nu} \rb = \int_{\posRealLine} |\Lambda|_{\boldsymbol{\Theta}}\! \lb  g^{-1}\!\lb  st \rb   , \boldsymbol{\nu} \rb  \dx F_\sigma(s).
\end{equation}

Recall that $f$ is continuous with compact support and bounded continuous $d^{\text{th}}$ partial derivative. Hence there exist constants $0 < a < b < \infty$ and $0< c_j<e_j<2\pi,1\leq j\leq d-1$ such that the support of $f$ is contained in 
\begin{equation}\label{equ: support_f}
    \mathcal{S}:=[a, b]\times\prod_{j=1}^{d-1}[c_j,e_j]
\end{equation}
and $f\lb  t,\boldsymbol{\nu} \rb  = 0$ for $\lb  t,\boldsymbol{\nu} \rb \notin (a, b)\times\prod_{j=1}^{d-1}(c_j,e_j)$. Using Fubini's theorem, we have
\begin{align}
    &\int_{\posRealLine\times[0,2\pi)^{d-1}} f\lb  t,\boldsymbol{\nu} \rb  \M_\Xi^\sig\!\lb  \dx t,\dx \boldsymbol{\nu} \rb \nonumber\\ 
    &= \int_{t=a}^{t=b} \lb  \prod_{j=1}^{d-1} \int_{\nu^{(j)}=c_j}^{\nu^{(j)}=e_j} \rb  \lsb \int_{x=a}^{x=t} \lb  \prod_{j=1}^{d-1} \int_{u^{(j)}=c_j}^{u^{(j)}=\nu^{(j)}} \rb  \Dmixd{f}\lb  x, \boldsymbol{u}\rb  \;\dx \boldsymbol{u}\;\dx x\;\rsb \M_\Xi^\sig\!\lb  \dx t,\dx \boldsymbol{\nu} \rb  \nonumber \\
    &= \int_{x=a}^{x=b} \lb  \prod_{j=1}^{d-1} \int_{u^{(j)}=c_j}^{u^{(j)}=e_j} \rb  \lsb  \int_{t=x}^{t=b} \lb  \prod_{j=1}^{d-1} \int_{\nu^{(j)}=u^{(j)}}^{\nu^{(j)}=e_j} \rb  \M_\Xi^\sig\!\lb  \dx t,\dx \boldsymbol{\nu} \rb \rsb \Dmixd{f}\lb  x, \boldsymbol{u}\rb  \;\dx \boldsymbol{u}\;\dx x \label{equ: fubini_before} \\
    &= (-1)^d
    \int_{x=a}^{x=b} \lb  \prod_{j=1}^{d-1} \int_{u^{(j)}=c_j}^{u^{(j)}=e_j} \rb  \M_\Xi^\sig\!\lb  x, \boldsymbol{u}\rb \; \Dmixd{f}\lb  x, \boldsymbol{u}\rb  \;\dx \boldsymbol{u}\;\dx x  \label{equ: (-1)^d}\\
    &= (-1)^d  \int_{t=a}^{t=b} \lb  \prod_{j=1}^{d-1} \int_{\nu^{(j)}=c_j}^{\nu^{(j)}=e_j} \rb  \int_0^\infty |\Xi|_{\boldsymbol{\Theta}} \!\lb  g^{-1}\!\lb  st \rb   , \boldsymbol{\nu} \rb  \; \dx F_\sigma(s)\; \Dmixd{f}\lb  t, \boldsymbol{\nu}\rb   \;\dx \boldsymbol{\nu}\;\dx t \nonumber\\
    &= (-1)^d \int_0^\infty \int_\mathcal{S}  |\Xi|_{\boldsymbol{\Theta}}\! \lb  g^{-1}\!\lb  st \rb   , \boldsymbol{\nu} \rb  \Dmixd{f}\lb  t, \boldsymbol{\nu}\rb  \;\dx \boldsymbol{\nu}\;\dx t \; \dx F_\sigma(s),\nonumber
\end{align}
where we define $\Dmixd{f}\lb  x, \boldsymbol{u}\rb  = \frac{\partial^d f}{\partial x  \prod_{j=1}^{d-1}\partial u^{(j)}}\lb  x, \boldsymbol{u}\rb $ as the mixed partial derivative of $f$. To see why \eqref{equ: fubini_before} reduces to \eqref{equ: (-1)^d}, we first consider the case $d=2$. The integral in \eqref{equ: fubini_before} becomes
\begin{align}
    & \int_{x=a}^{x=b} \int_{u^{(1)}=c_1}^{u^{(1)}=e_1} \lsb \int_{t=x}^{t=b} \int_{\nu^{(1)}=u^{(1)}}^{\nu^{(1)}=e_1} \M_\Xi^\sig\!\lb  \dx t,\dx \nu^{(1)} \rb \rsb \frac{\partial^2 f}{\partial x  \partial u^{(1)}}\!\lb  x, u^{(1)} \rb  \;\dx u^{(1)}\;\dx x \nonumber \\
    &= \M_\Xi^\sig(b,e_1)\int_{x=a}^{x=b} \int_{u^{(1)}=c_1}^{u^{(1)}=e_1}  \frac{\partial^2 f}{\partial x  \partial u^{(1)}}\!\lb  x, u^{(1)} \rb  \;\dx u^{(1)}\;\dx x \label{equ: fubini 2d int1}\\
    &\quad - \int_{u^{(1)}=c_1}^{u^{(1)}=e_1}  \M_\Xi^\sig\!\lb  b,u^{(1)}\rb  \int_{x=a}^{x=b} \frac{\partial^2 f}{\partial x  \partial u^{(1)}}\!\lb  x, u^{(1)} \rb  \;\dx x\;\dx u^{(1)} \label{equ: fubini 2d int2}\\
    &\quad- \int_{x=a}^{x=b} \M_\Xi^\sig(x,e_1) \int_{u^{(1)}=c_1}^{u^{(1)}=e_1}  \frac{\partial^2 f}{\partial x  \partial u^{(1)}}\!\lb  x, u^{(1)} \rb  \;\dx u^{(1)}\;\dx x \label{equ: fubini 2d int3}\\
    &\quad + \int_{x=a}^{x=b} \int_{u^{(1)}=c_1}^{u^{(1)}=e_1} \M_\Xi^\sig\!\lb  x,u^{(1)} \rb \; \frac{\partial^2 f}{\partial x  \partial u^{(1)}}\!\lb  x, u^{(1)} \rb  \;\dx u^{(1)}\;\dx x  \nonumber\\
    &= (-1)^2 \int_{x=a}^{x=b} \int_{u^{(1)}=c_1}^{u^{(1)}=e_1} \M_\Xi^\sig\lb x,u^{(1)}\rb\; \frac{\partial^2 f}{\partial x  \partial u^{(1)}}\!\lb  x, u^{(1)} \rb  \;\dx u^{(1)}\;\dx x,  \nonumber
\end{align}
where the first three integrals in \eqref{equ: fubini 2d int1}, \eqref{equ: fubini 2d int2}, and \eqref{equ: fubini 2d int3} vanish due to the support assumption on $f$; see \eqref{equ: support_f}. The same argument extends directly to arbitrary dimension $d$.

Using the SRD condition \eqref{equ: SRD}, we will show that
\begin{equation}\label{equ: sup_var_to0}
    \sup_{\boldsymbol{\nu}\in \prod_{j=1}^{d-1}[c_j,e_j]}\frac{\mathrm{Var}|\Xi|_{\boldsymbol{\Theta}}\!\lb  \rho, \boldsymbol{\nu} \rb }{|\Lambda|_{\boldsymbol{\Theta}}\lb  \rho, \boldsymbol{\nu} \rb ^2} \to 0,
\end{equation}
as $\rho\toinf$. Then there exists a $\rho_0>0$ such that 
\[
\sup_{\boldsymbol{\nu}\in \prod_{j=1}^{d-1}[c_j,e_j]}\frac{\mathrm{Var}|\Xi|_{\boldsymbol{\Theta}}\!\lb  \rho, \boldsymbol{\nu} \rb }{|\Lambda|_{\boldsymbol{\Theta}}\lb  \rho, \boldsymbol{\nu} \rb ^2} \leq 1, \quad \forall \rho \geq \rho_0.
\]
Since $g$ is left continuous and non-decreasing, we can choose $s_0>0$ such that
\[
\sup_{\boldsymbol{\nu}\in \prod_{j=1}^{d-1}[c_j,e_j]}\frac{\mathrm{Var}|\Xi|_{\boldsymbol{\Theta}}\!\lb  g^{-1}(sa), \boldsymbol{\nu} \rb }{|\Lambda|_{\boldsymbol{\Theta}}\lb  g^{-1}(sa), \boldsymbol{\nu} \rb ^2} \leq 1, \quad \forall s \geq s_0.
\]
Therefore, we may decompose
\begin{align*}
    &\int_{\posRealLine\times[0,2\pi)^{d-1}} f\lb  t,\boldsymbol{\nu} \rb  \M_\Xi^\sig\!\lb  \dx t,\dx \boldsymbol{\nu} \rb  \\
    &=  (-1)^d\int_{s_0}^\infty \int_\mathcal{S}  \lsb |\Xi|_{\boldsymbol{\Theta}} \!\lb  g^{-1}\!\lb  st \rb   , \boldsymbol{\nu} \rb  -\IE|\Xi|_{\boldsymbol{\Theta}} \!\lb  g^{-1}\!\lb  st \rb   , \boldsymbol{\nu} \rb  \rsb\Dmixd{f}\lb  t, \boldsymbol{\nu}\rb  \;\dx \boldsymbol{\nu}\;\dx t \; \dx F_\sigma(s)\\
    &\quad + (-1)^d\int_0^{\infty} \int_\mathcal{S}  \IE|\Xi|_{\boldsymbol{\Theta}}\! \lb  g^{-1}\!\lb  st \rb   , \boldsymbol{\nu} \rb  \Dmixd{f}\lb  t, \boldsymbol{\nu}\rb  \;\dx \boldsymbol{\nu}\;\dx t \; \dx F_\sigma(s) \\
    &\quad - (-1)^d\int_0^{s_0} \int_\mathcal{S}  \IE|\Xi|_{\boldsymbol{\Theta}} \!\lb  g^{-1}\!\lb  st \rb   , \boldsymbol{\nu} \rb  \Dmixd{f}\lb  t, \boldsymbol{\nu}\rb  \;\dx \boldsymbol{\nu}\;\dx t \; \dx F_\sigma(s) \\
    &\quad + (-1)^d\int_0^{s_0} \int_\mathcal{S}  |\Xi|_{\boldsymbol{\Theta}} \!\lb  g^{-1}\!\lb  st \rb   , \boldsymbol{\nu} \rb  \Dmixd{f}\lb  t, \boldsymbol{\nu}\rb  \;\dx \boldsymbol{\nu}\;\dx t \; \dx F_\sigma(s) \\
    &=: (\mathrm{I})_\Xi + (\mathrm{II})_\Xi + (\mathrm{III})_\Xi + (\mathrm{IV})_\Xi.
\end{align*}

We now complete the proof by showing that  
(a) $(\mathrm{II})_\Xi = \int_{\posRealLine\times[0,2\pi)^{d-1}} f\lb  t,\boldsymbol{\nu} \rb  \M^\sig\!\lb  \dx t,\dx \boldsymbol{\nu} \rb $ for any $\sigma\geq0$;  
(b) $(\mathrm{I})_\Xi \xrightarrow{\IP} 0$;  
(c) $(\mathrm{III})_\Xi\to 0$;  
(d) $(\mathrm{IV})_\Xi\xrightarrow{\IP} 0$ as $\sigma \to \infty$;  
(e) the claim in \eqref{equ: sup_var_to0} holds.

\smallskip
\noindent(a) Applying Fubini's theorem and using the representation \eqref{equ: same_mean_measure_CDF}, similar to \eqref{equ: fubini_before} and \eqref{equ: (-1)^d}, we have

\begin{align*}
    &\int_{\posRealLine\times[0,2\pi)^{d-1}} f\lb  t,\boldsymbol{\nu} \rb  \M^\sig\!\lb  \dx t,\dx \boldsymbol{\nu} \rb\\
    &= (-1)^d \int_0^\infty \int_\mathcal{S}  |\Lambda|_{\boldsymbol{\Theta}} \!\lb  g^{-1}\!\lb  st \rb   , \boldsymbol{\nu} \rb  \Dmixd{f}\lb  t, \boldsymbol{\nu}\rb  \;\dx \boldsymbol{\nu}\;\dx t \; \dx F_\sigma(s)\\
    &=(\mathrm{II})_\Xi
\end{align*}
for any $\sigma\geq0$.

\smallskip
\noindent(b) This is the only part of the proof that relies crucially on the SRD condition \eqref{equ: SRD}. To show $(\mathrm{I})_\Xi \xrightarrow{\IP} 0$, we show that its variance goes to $0$ as $\sigmatoinf$, and then apply Markov's inequality. Let $v_\sigma$ denote the variance of $(\mathrm{I})_\Xi$. Then
\begin{align*}
    v_\sigma 
    &= \IE\!\lsb \lcb    \int_{s_0}^\infty \int_\mathcal{S}    \lsb |\Xi|_{\boldsymbol{\Theta}}\! \lb  g^{-1}\!\lb  st \rb   , \boldsymbol{\nu} \rb  -|\Lambda|_{\boldsymbol{\Theta}}\! \lb  g^{-1}\!\lb  st \rb   , \boldsymbol{\nu} \rb  \rsb\Dmixd{f}\lb  t, \boldsymbol{\nu}\rb  \;\dx \boldsymbol{\nu}\;\dx t \; \dx F_\sigma(s) \rcb^2 \rsb\\
    &= \IE   \int_{s_0}^\infty \int_\mathcal{S}   \int_{s_0}^\infty \int_\mathcal{S}    \lb  \frac{|\Xi|_{\boldsymbol{\Theta}}\! \lb  g^{-1}\!\lb  s_1t_1 \rb   , \boldsymbol{\nu}_1 \rb }{|\Lambda|_{\boldsymbol{\Theta}} \!\lb  g^{-1}\!\lb  s_1t_1 \rb   , \boldsymbol{\nu}_1 \rb } -1 \rb \Dmixd{f}\lb  t_1, \boldsymbol{\nu}_1\rb \nonumber\\
    &\quad\times\lb  \frac{|\Xi|_{\boldsymbol{\Theta}} \!\lb  g^{-1}\!\lb  s_2t_2 \rb   , \boldsymbol{\nu}_2 \rb }{|\Lambda|_{\boldsymbol{\Theta}}\! \lb  g^{-1}\!\lb  s_2t_2 \rb   , \boldsymbol{\nu}_2 \rb } -1 \rb  \Dmixd{f}\lb  t_2, \boldsymbol{\nu}_2\rb \;\nonumber\\
    &\quad\times|\Lambda|_{\boldsymbol{\Theta}} \!\lb  g^{-1}\!\lb  s_1t_1 \rb   , \boldsymbol{\nu}_1 \rb   \;\dx \boldsymbol{\nu}_1\;\dx t_1 \; \dx F_\sigma(s_1)\; |\Lambda|_{\boldsymbol{\Theta}}\! \lb  g^{-1}\!\lb  s_2t_2 \rb   , \boldsymbol{\nu}_2 \rb \;\dx \boldsymbol{\nu}_2\;\dx t_2 \; \dx F_\sigma(s_2).\nonumber
\end{align*}

Using the geometric-arithmetic mean inequality $AB \leq (A^2 + B^2)/2$ and symmetry, we obtain
\begin{align*}
    v_\sigma 
    &\leq \IE   \int_{s_0}^\infty \int_\mathcal{S}   \int_{s_0}^\infty \int_\mathcal{S}   \Bigg[ \frac{1}{2}  \lb \frac{|\Xi|_{\boldsymbol{\Theta}}\! \lb  g^{-1}\!\lb  s_1t_1 \rb   , \boldsymbol{\nu}_1 \rb }{|\Lambda|_{\boldsymbol{\Theta}}\! \lb  g^{-1}\!\lb  s_1t_1 \rb   , \boldsymbol{\nu}_1 \rb } -1 \rb ^2 \Dmixd{f}\lb  t_1, \boldsymbol{\nu}_1\rb ^2   \\
    &\quad+ \frac{1}{2}\lb  \frac{|\Xi|_{\boldsymbol{\Theta}}\! \lb  g^{-1}\!\lb  s_2t_2 \rb   , \boldsymbol{\nu}_2 \rb }{|\Lambda|_{\boldsymbol{\Theta}}\! \lb  g^{-1}\!\lb  s_2t_2 \rb   , \boldsymbol{\nu}_2 \rb } -1 \rb  \Dmixd{f}\lb  t_2, \boldsymbol{\nu}_2\rb ^2\Bigg] \\
    &\quad\times \; |\Lambda|_{\boldsymbol{\Theta}}\! \lb  g^{-1}\!\lb  s_1t_1 \rb   , \boldsymbol{\nu}_1 \rb   \;\dx \boldsymbol{\nu}_1\;\dx t_1 \; \dx F_\sigma(s_1)\; |\Lambda|_{\boldsymbol{\Theta}}\! \lb  g^{-1}\!\lb  s_2t_2 \rb   , \boldsymbol{\nu}_2 \rb \;\dx \boldsymbol{\nu}_2\;\dx t_2 \; \dx F_\sigma(s_2)\\
    &= \IE   \int_{s_0}^\infty \int_\mathcal{S}   \int_{s_0}^\infty \int_\mathcal{S}  \lb \frac{|\Xi|_{\boldsymbol{\Theta}}\! \lb  g^{-1}\!\lb  s_1t_1 \rb   , \boldsymbol{\nu}_1 \rb }{|\Lambda|_{\boldsymbol{\Theta}}\! \lb  g^{-1}\!\lb  s_1t_1 \rb   , \boldsymbol{\nu}_1 \rb } -1 \rb ^2 \Dmixd{f}\lb  t_1, \boldsymbol{\nu}_1\rb ^2   \\
    &\quad\times \; |\Lambda|_{\boldsymbol{\Theta}}\! \lb  g^{-1}\!\lb  s_1t_1 \rb   , \boldsymbol{\nu}_1 \rb   \;\dx \boldsymbol{\nu}_1\;\dx t_1 \; \dx F_\sigma(s_1)\; |\Lambda|_{\boldsymbol{\Theta}}\! \lb  g^{-1}\!\lb  s_2t_2 \rb   , \boldsymbol{\nu}_2 \rb \;\dx \boldsymbol{\nu}_2\;\dx t_2 \; \dx F_\sigma(s_2)\\
    &= \int_{s_0}^\infty \int_\mathcal{S}      \frac{\mathrm{Var}|\Xi|_{\boldsymbol{\Theta}}\! \lb  g^{-1}\!\lb  s_1t_1 \rb   , \boldsymbol{\nu}_1 \rb }{|\Lambda|_{\boldsymbol{\Theta}}\! \lb  g^{-1}\!\lb  s_1t_1 \rb   , \boldsymbol{\nu}_1 \rb ^2} \Dmixd{f}\lb  t_1, \boldsymbol{\nu}_1\rb ^2 \; |\Lambda|_{\boldsymbol{\Theta}}\! \lb  g^{-1}\!\lb  s_1t_1 \rb   , \boldsymbol{\nu}_1 \rb   \;\dx \boldsymbol{\nu}_1\;\dx t_1 \; \dx F_\sigma(s_1)\\
    &\quad\times \int_{s_0}^\infty \int_\mathcal{S}   |\Lambda|_{\boldsymbol{\Theta}}\! \lb  g^{-1}\!\lb  s_2t_2 \rb   , \boldsymbol{\nu}_2 \rb \;\dx \boldsymbol{\nu}_2\;\dx t_2 \; \dx F_\sigma(s_2)\\   
    &\leq \left\|\Dmixd{f} \right \|^2 \int_{s_0}^\infty |\Lambda|_{\boldsymbol{\Theta}}\! \lb  g^{-1}\!\lb  s_1b \rb , \lcb  e_j \rcb_{j=1}^{d-1}  \rb  \int_\mathcal{S}      \frac{\mathrm{Var}|\Xi|_{\boldsymbol{\Theta}}\! \lb  g^{-1}\!\lb  s_1t_1 \rb   , \boldsymbol{\nu}_1 \rb }{|\Lambda|_{\boldsymbol{\Theta}}\! \lb  g^{-1}\!\lb  s_1t_1 \rb   , \boldsymbol{\nu}_1 \rb ^2}   \;\dx \boldsymbol{\nu}_1\;\dx t_1 \; \dx F_\sigma(s_1)\\
    &\quad\times V(\mathcal{S})\int_{s_0}^\infty |\Lambda|_{\boldsymbol{\Theta}}\! \lb  g^{-1}\!\lb  s_2b \rb , \lcb  e_j \rcb_{j=1}^{d-1}  \rb  \; \dx F_\sigma(s_2)\\  
    &\leq \left\|\Dmixd{f} \right \|^2 V(\mathcal{S}) \IE|\Lambda|_{\boldsymbol{\Theta}}\! \lb  g^{-1}\!\lb  S\sig b \rb , \lcb  e_j \rcb_{j=1}^{d-1}  \rb \\
    &\quad\times\int_{s_0}^\infty |\Lambda|_{\boldsymbol{\Theta}}\! \lb  g^{-1}\!\lb  sb \rb , \lcb  e_j \rcb_{j=1}^{d-1}  \rb  \int_\mathcal{S}      \frac{\mathrm{Var}|\Xi|_{\boldsymbol{\Theta}}\! \lb  g^{-1}\!\lb  st \rb   , \boldsymbol{\nu} \rb }{|\Lambda|_{\boldsymbol{\Theta}}\! \lb  g^{-1}\!\lb  st \rb   , \boldsymbol{\nu} \rb ^2}   \;\dx \boldsymbol{\nu}\;\dx t \; \dx F_\sigma(s),
\end{align*}
where $\left\|\Dmixd{f} \right \| := \sup_{\lb  t,\boldsymbol{\nu} \rb \in\posRealLine\times[0,2\pi)^{d-1}} \left|\Dmixd{f}\lb  t, \boldsymbol{\nu}\rb  \right |$ and $V(\mathcal{S})=(b-a)\prod_{j=1}^{d-1}(e_j-c_j)$.
 
For each $\epsilon > 0$, let $T_\epsilon > s_0$ be such that
\[
\sup_{\boldsymbol{\nu}\in \prod_{j=1}^{d-1}[c_j,e_j]}\frac{\mathrm{Var}|\Xi|_{\boldsymbol{\Theta}}\! \lb  g^{-1}\!\lb  sa \rb   , \boldsymbol{\nu} \rb }{|\Lambda|_{\boldsymbol{\Theta}}\! \lb  g^{-1}\!\lb  sa \rb   , \boldsymbol{\nu} \rb ^2} \leq \epsilon, \quad \forall s \geq T_\epsilon .
\]
Then
\[
\frac{\mathrm{Var}|\Xi|_{\boldsymbol{\Theta}}\! \lb  g^{-1}\!\lb  st \rb   , \boldsymbol{\nu} \rb }{|\Lambda|_{\boldsymbol{\Theta}}\! \lb  g^{-1}\!\lb  st \rb   , \boldsymbol{\nu} \rb ^2} \leq 1, \quad \forall s_0<s < T_\epsilon, t\in[a,b], \boldsymbol{\nu}\in \prod_{j=1}^{d-1}[c_j,e_j],
\]
and
\[
\frac{\mathrm{Var}|\Xi|_{\boldsymbol{\Theta}}\! \lb  g^{-1}\!\lb  st \rb   , \boldsymbol{\nu} \rb }{|\Lambda|_{\boldsymbol{\Theta}}\! \lb  g^{-1}\!\lb  st \rb   , \boldsymbol{\nu} \rb ^2} \leq \epsilon, \quad \forall s \geq T_\epsilon, t\in[a,b], \boldsymbol{\nu}\in \prod_{j=1}^{d-1}[c_j,e_j].
\]
It follows that
\begin{align*}
    v_\sigma 
    &\leq \left\|\Dmixd{f} \right \|^2 V(\mathcal{S})\IE|\Lambda|_{\boldsymbol{\Theta}}\! \lb  g^{-1}\!\lb  S\sig b \rb , \lcb  e_j \rcb_{j=1}^{d-1}  \rb  \\
    &\quad\times \lb \int_{s_0}^{T_\epsilon} + \int_{T_\epsilon}^\infty \rb  |\Lambda|_{\boldsymbol{\Theta}}\! \lb  g^{-1}\!\lb  sb \rb , \lcb  e_j \rcb_{j=1}^{d-1}  \rb  \int_\mathcal{S}      \frac{\mathrm{Var}|\Xi|_{\boldsymbol{\Theta}}\! \lb  g^{-1}\!\lb  st \rb   , \boldsymbol{\nu} \rb }{|\Lambda|_{\boldsymbol{\Theta}}\! \lb  g^{-1}\!\lb  st \rb   , \boldsymbol{\nu} \rb ^2}   \;\dx \boldsymbol{\nu}\;\dx t \; \dx F_\sigma(s)\\
    &\leq \left\|\Dmixd{f} \right \|^2 V(\mathcal{S})\IE|\Lambda|_{\boldsymbol{\Theta}}\! \lb  g^{-1}\!\lb  S\sig b \rb , \lcb  e_j \rcb_{j=1}^{d-1}  \rb  \\
    &\quad\times \Biggl(\int_{s_0}^{T_\epsilon}  |\Lambda|_{\boldsymbol{\Theta}}\! \lb  g^{-1}\!\lb  sb \rb , \lcb  e_j \rcb_{j=1}^{d-1}  \rb  \int_\mathcal{S}         \;\dx \boldsymbol{\nu}\;\dx t \; \dx F_\sigma(s) \\
    &\qquad+ \int_{T_\epsilon}^\infty |\Lambda|_{\boldsymbol{\Theta}}\! \lb  g^{-1}\!\lb  sb \rb , \lcb  e_j \rcb_{j=1}^{d-1}  \rb  \int_\mathcal{S}      \epsilon   \;\dx \boldsymbol{\nu}\;\dx t \; \dx F_\sigma(s)\Biggr)\\
    &\leq \left\|\Dmixd{f} \right \|^2 V(\mathcal{S})^2\IE|\Lambda|_{\boldsymbol{\Theta}}\! \lb  g^{-1}\!\lb  S\sig b \rb , \lcb  e_j \rcb_{j=1}^{d-1}  \rb  \\
    &\quad\times \lcb  \IE\!\lsb |\Lambda|_{\boldsymbol{\Theta}}\! \lb  g^{-1}\!\lb  S\sig b \rb , \lcb  e_j \rcb_{j=1}^{d-1}  \rb  \I(S\sig\leq T_\epsilon)\rsb  + \epsilon \IE|\Lambda|_{\boldsymbol{\Theta}}\! \lb  g^{-1}\!\lb  S\sig b \rb , \lcb  e_j \rcb_{j=1}^{d-1}  \rb  \rcb.
\end{align*}

From the observation in \eqref{equ: LambdaInequality}, we have
\begin{align*}
    &\IE\!\lsb |\Lambda|_{\boldsymbol{\Theta}}\! \lb  g^{-1}\!\lb  S\sig b \rb , \lcb  e_j \rcb_{j=1}^{d-1}  \rb  \I(S\sig\leq T_\epsilon)\rsb 
    \leq \IE\!\lsb |\Lambda| \lb  g^{-1}\!\lb  S\sig b \rb  \rb  \I(S\sig\leq T_\epsilon)\rsb
\end{align*}
and
\[
\IE|\Lambda|_{\boldsymbol{\Theta}}\! \lb  g^{-1}\!\lb  S\sig b \rb , \lcb  e_j \rcb_{j=1}^{d-1}  \rb  
\leq 
\IE|\Lambda| \lb  g^{-1}\!\lb  S\sig b \rb  \rb  
=
\lambda\IE\!\lsb  V_d\lb  g^{-1}\!\lb  S\sig b \rb \rb  \rsb.
\]

Using that $S(\sigma) \xrightarrow{\IP} 0$, $\lim_{t \to 0} g^{-1}(t) = 0$, $\lim_{\rho \to 0} |\Lambda|(\rho) = 0$, and noting that $|\Lambda| \lb  g^{-1}\!\lb  S\sig b \rb  \rb  \I(S\sig\leq T_\epsilon)\leq |\Lambda| (g^{-1}\!\lb  T_\epsilon b\rb ) = \lambda V_d( g^{-1}\!\lb  T_\epsilon b\rb ) <\infty$ is bounded, we may apply the bounded convergence theorem and obtain
\begin{align}
    &\lim_\sigmatoinf\IE\!\lsb |\Lambda|_{\boldsymbol{\Theta}}\! \lb  g^{-1}\!\lb  S\sig b \rb , \lcb  e_j \rcb_{j=1}^{d-1}  \rb  \I(S\sig\leq T_\epsilon)\rsb \label{equ: BCT_(a)}\\
    &\leq\lim_\sigmatoinf\IE\!\lsb |\Lambda| \lb  g^{-1}\!\lb  S\sig b \rb  \rb  \I(S\sig\leq T_\epsilon)\rsb \nonumber\\
    &= 0.\nonumber
\end{align}
Therefore,
\[
\limsup_{\sigma \to \infty} v_\sigma \leq \left\|\Dmixd{f} \right \|^2 V(\mathcal{S})^2\; \epsilon\; \underbrace{\limsup_\sigmatoinf\lambda^2 \IE\!\lsb  V_d\lb  g^{-1}\!\lb  S\sig b \rb \rb  \rsb^2}_{<\infty \text{ by \eqref{equ: finite assumption}}}.
\]
Since $\epsilon >0$ is arbitrary, we conclude that
\[
\lim_\sigmatoinf v_\sigma = 0.
\]
Now $(\mathrm{I})_\Xi\convergeP0$ follows directly from Markov's inequality.

\smallskip
\noindent(c) Using the same argument as in \eqref{equ: BCT_(a)}, we have
\begin{align*}
    |(\mathrm{III})_\Xi|&\leq \int_0^{s_0} \int_\mathcal{S}    |\Lambda|_{\boldsymbol{\Theta}}\! \lb  g^{-1}\!\lb  st \rb   , \boldsymbol{\nu} \rb  \left|\Dmixd{f}\lb  t, \boldsymbol{\nu}\rb \right| \;\dx \boldsymbol{\nu}\;\dx t \; \dx F_\sigma(s)\\
    &\leq \left\|\Dmixd{f} \right \| V(\mathcal{S})\IE\!\lsb |\Lambda|_{\boldsymbol{\Theta}}\! \lb  g^{-1}\!\lb  S\sig b \rb , \lcb  e_j \rcb_{j=1}^{d-1}  \rb  \I(S\sig\leq s_0)\rsb\nonumber\\
    &\to 0\nonumber
\end{align*}
as $\sigmatoinf$.

\smallskip
\noindent(d) Following the same reasoning as in (c),
\begin{align*}
    \IE|(\mathrm{IV})_\Xi|&\leq \IE\int_0^{s_0} \int_\mathcal{S}    |\Xi|_{\boldsymbol{\Theta}}\! \lb  g^{-1}\!\lb  st \rb   , \boldsymbol{\nu} \rb  \left|\Dmixd{f}\lb  t, \boldsymbol{\nu}\rb \right| \;\dx \boldsymbol{\nu}\;\dx t \; \dx F_\sigma(s)\\
    &= \int_0^{s_0} \int_\mathcal{S}    |\Lambda|_{\boldsymbol{\Theta}}\! \lb  g^{-1}\!\lb  st \rb   , \boldsymbol{\nu} \rb  \left|\Dmixd{f}\lb  t, \boldsymbol{\nu}\rb \right| \;\dx \boldsymbol{\nu}\;\dx t \; \dx F_\sigma(s)\nonumber\\
    &\to0\nonumber
\end{align*}
as $\sigmatoinf$.

\smallskip
\noindent(e) Let $\boldsymbol{\nu}\in \prod_{j=1}^{d-1}[c_j,e_j]$. We prove that under the SRD condition \eqref{equ: SRD},
\[
\frac{\Var\!|\Xi|_{\boldsymbol{\Theta}}\!\lb  \rho,\boldsymbol{\nu}\rb }
{\;|\Lambda|_{\boldsymbol{\Theta}}\!\lb  \rho,\boldsymbol{\nu}\rb ^2}
\;\to\;0
\]
uniformly in $\boldsymbol{\nu}$ as $\rho\to\infty$. Recall that
\[
|\Xi|_{\boldsymbol{\Theta}}\!\lb  \rho,\boldsymbol{\nu}\rb 
=\sum_{i\in \mathcal{I}^\Xi} \I\!\lb  |X_i| \le \rho\rb \, \PP(X_i,\boldsymbol{\nu})=\int_{\mathbb{R}^d}\I\!\lb |x|\le\rho\rb \,\PP(x,\boldsymbol{\nu})\,\Xi(\dx x).
\]
By Campbell’s formula (see \cite[Theorem 4.1]{Haenggi2012}),
\[
|\Lambda|_{\boldsymbol{\Theta}}\!\lb  \rho,\boldsymbol{\nu}\rb 
=\lambda\int_{\mathbb{R}^d}\I\!\lb |x|\le\rho\rb \,\PP(x,\boldsymbol{\nu})\,\dx x .
\]
We notice that $\PP(X_i,\boldsymbol{\nu})$ is bounded away from zero and infinity. Indeed,
\[
0<\prod_{j=1}^{d-1}\IP\!\lb \vtj_i\le c_j \given X_i\rb 
\;\le\;
\PP(X_i,\boldsymbol{\nu})
\;\le\;1 ,
\]
as the angular components follow a wrapped Normal distribution in $[0,2\pi)$ and $c_j\in(0,2\pi),\forall1\leq j\leq d-1$. 

Therefore, we have
\[
|\Lambda|_{\boldsymbol{\Theta}}\!\lb  \rho,\boldsymbol{\nu}\rb 
\ \asymp\
\lambda\!\int_{\mathbb{R}^d}\I\!\lb 0<|x|\le\rho\rb  \dx x
= \lambda\,V_d(\rho)\asymp\rho^d,
\]
where we write $f_1(\rho) \asymp f_2(\rho)$ if $\frac{f_1(\rho)}{f_2(\rho)}$ tends to a positive constant as $\rho \to \infty$.

Similarly, again using (second-order) Campbell’s formula,
\begin{align*}
&\Var|\Xi|_{\boldsymbol{\Theta}}\!\lb  \rho,\boldsymbol{\nu}\rb \\
&= \Var  \sum_{i\in\mathcal{I}^\Xi} \I(|X_i|\le \rho)\,\PP(X_i,\boldsymbol{\nu}) \\
&= \IE\!\lsb \lb \sum_{i\in\mathcal{I}^\Xi} \I(|X_i|\le \rho)\,\PP(X_i,\boldsymbol{\nu})\rb ^{\!2}\rsb
   - \IE\!\lsb \sum_{i\in\mathcal{I}^\Xi} \I(|X_i|\le \rho)\,\PP(X_i,\boldsymbol{\nu})\rsb^{\!2} \\
&= \IE\!\lsb \sum_{i\in\mathcal{I}^\Xi} \I(|X_i|\le \rho)\,\PP(X_i,\boldsymbol{\nu})^{2}\rsb
   + \IE\!\lsb \sum_{i,j\in\mathcal{I}^\Xi\ :\ i\neq j}
      \I(|X_i|\le \rho)\I(|X_j|\le \rho)\,\PP(X_i,\boldsymbol{\nu})\PP(X_j,\boldsymbol{\nu})\rsb \\
&\qquad - \lambda^{2}\!\lb \int_{\mathbb{R}^{d}} \I(|x|\le \rho)\,\PP(x,\boldsymbol{\nu})\,\dx x\rb ^{\!2} \\
&= \lambda \int_{\mathbb{R}^{d}} \I(|x|\le \rho)\,\PP(x,\boldsymbol{\nu})^{2}\,\dx x \\
&\qquad + \iint_{\mathbb{R}^{d}\times\mathbb{R}^{d}}
      \I(|x|\le \rho)\I(|y|\le \rho)\,\PP(x,\boldsymbol{\nu})\PP(y,\boldsymbol{\nu})\,
      \varrho^{(2)}(x,y)\,\dx y\,\dx x \\
&\qquad - \lambda^{2}\!\iint_{\mathbb{R}^{d}\times\mathbb{R}^{d}}
      \I(|x|\le \rho)\I(|y|\le \rho)\,\PP(x,\boldsymbol{\nu})\PP(y,\boldsymbol{\nu})\,\dx y\,\dx x \\
&= \lambda \int_{\mathbb{R}^{d}} \I(|x|\le \rho)\,\PP(x,\boldsymbol{\nu})^{2}\,\dx x \\
&\qquad + \lambda^{2}\!\iint_{\mathbb{R}^{d}\times\mathbb{R}^{d}}
      \I(|x|\le \rho)\I(|y|\le \rho)\,\PP(x,\boldsymbol{\nu})\PP(y,\boldsymbol{\nu})\,
      \lsb \frac{\varrho^{(2)}(x,y)}{\lambda^{2}}-1\rsb\,\dx y\,\dx x \\
&= \lambda \int_{\mathbb{R}^{d}} \I(|x|\le \rho)\,\PP(x,\boldsymbol{\nu})^{2}\,\dx x\\
&\qquad   + \lambda^{2}\!\iint_{\mathbb{R}^{d}\times\mathbb{R}^{d}}
      \I(|x|\le \rho)\I(|y|\le \rho)\,\PP(x,\boldsymbol{\nu})\PP(y,\boldsymbol{\nu})\,
      \lb h_{\mathrm{SI}}(|x-y|)-1\rb \,\dx y\,\dx x .
\end{align*}

Since $0< \PP(x,\boldsymbol{\nu})\le 1$, we have
\[
\int_{\mathbb{R}^{d}} \I(|x|\le \rho)\,\PP(x,\boldsymbol{\nu})^{2}\,dx
\asymp \rho^d
\]
as $\rho\to\infty$.
Using the change of variables $u=y-x$ and the bound $\PP(x,\boldsymbol{\nu})\le 1$, we obtain
\begin{align*}
&\left|\iint_{\mathbb{R}^{d}\times\mathbb{R}^{d}}
\I(|x|\le \rho)\I(|y|\le \rho)\,\PP(x,\boldsymbol{\nu})\PP(y,\boldsymbol{\nu})\,
\lb h_{\mathrm{SI}}(|x-y|)-1\rb \,\dx y\,\dx x\right| \\
&\quad=\left|\iint_{\mathbb{R}^{d}\times\mathbb{R}^{d}}
\I(|x|\le \rho)\I(|x+u|\le \rho)\,\PP(x,\boldsymbol{\nu})\PP(x+u,\boldsymbol{\nu})\,
\lb h_{\mathrm{SI}}(|u|)-1\rb \,\dx u\,\dx x\right| \\
&\quad\le \iint_{\mathbb{R}^{d}\times\mathbb{R}^{d}}
\I(|x|\le \rho)\I(|x+u|\le \rho)\,\big|h_{\mathrm{SI}}(|u|)-1\big|\,\dx u\,\dx x \\
&\quad\le \int_{\mathbb{R}^{d}} \big|h_{\mathrm{SI}}(|u|)-1\big|\,
\lb \int_{\mathbb{R}^{d}}\I(|x|\le \rho)\,\dx x\rb \,\dx u \\
&\quad= V_d(\rho)\int_{\mathbb{R}^{d}} \big|h_{\mathrm{SI}}(|u|)-1\big|\,\dx u\\
&\quad= V_d(\rho)\,\frac{2\pi^{d/2}}{\Gamma(d/2)}\!\underbrace{\int_{0}^{\infty} \bigl|h_{\mathrm{SI}}(r)-1\bigr|\, r^{d-1}\,\dx r}_{<\infty\ \text{ by SRD \eqref{equ: SRD}}}
\\
&\quad\asymp \rho^d,
\end{align*}
where we used hyperspherical coordinates \cite{Blumenson1960} in the second last equality, and $\Gamma$ is the Gamma function. Therefore,
\[
\Var |\Xi|_{\boldsymbol{\Theta}}\!\lb   \rho,\boldsymbol{\nu}\rb 
\asymp\rho^d 
\]
as $\rho\to\infty$, and hence
\[
\frac{\Var |\Xi|_{\boldsymbol{\Theta}}\!\lb \rho,\boldsymbol{\nu}\rb }
{\,|\Lambda|_{\boldsymbol{\Theta}}\!\lb \rho,\boldsymbol{\nu}\rb ^{2}}
\asymp\rho^{-d}\;\longrightarrow\;0
\]
uniformly in $\boldsymbol{\nu}$ as $\rho\to\infty$, which proves the claim.

\end{proof}

\begin{proof}[Proof of Theorem~\ref{convergence_thm}]

The convergence of the observable processes follows directly from the proof of \cite[Theorem~2.10]{KeelerRossXia2018} once the convergence of their mean measures has been established in Theorem~\ref{theorem: mean measure converge}. For completeness, we provide the proof here.

The proof is based on the natural metric for the vague topology introduced in \cite[(3.4)]{BARBOUR19929}. For any $\tau>0$, let $\chi_\tau$ denote the set of all finite (and hence locally finite) measures on $(0,\tau]\times[0,2\pi)^{d-1}$, and let $\HH_\tau \subset \chi_\tau$ denote the set of all finite integer-valued measures on the same space. Let $\mathcal{K}_\tau$ be the set of all Lipschitz functions on $(0,\tau]\times[0,2\pi)^{d-1}$ with respect to the metric
\[
d_0(x,y) = \min\{1, |x-y|\}, \quad x,y \in (0,\tau]\times[0,2\pi)^{d-1},
\]
that is,
\[
\mathcal{K}_\tau = \lcb  k : |k(x)-k(y)| \le d_0(x,y), \; x,y \in (0,\tau]\times[0,2\pi)^{d-1} \rcb.
\]

\cite[(3.2)]{BARBOUR19929} defines a Wasserstein metric $d_{1\tau}$ for finite measures $\zeta_1, \zeta_2 \in \chi_\tau$ as
\[
d_{1\tau}(\zeta_1,\zeta_2) =
\begin{cases}
1, & \text{if } \zeta_1(\tau) \ne \zeta_2(\tau), \\[6pt]
0, & \text{if } \zeta_1(\tau) = \zeta_2(\tau) = 0, \\[10pt]
\displaystyle \frac{
\sup_{k \in \mathcal{K}_\tau}
\left| \int_{(0,\tau]} k(x)\,\zeta_1(\dx x)
- \int_{(0,\tau]} k(x)\,\zeta_2(\dx x) \right|
}{\zeta_1(\tau)},
& \text{if } \zeta_1(\tau) = \zeta_2(\tau) > 0,
\end{cases}
\]
where $\zeta_i(\tau):=\zeta_i\!\lb  (0,\tau]\times [0,2\pi)^{d-1}\rb ,i=1,2$. 

Let $\mathcal{V}_\tau$ denote the set of all Lipschitz functions on $\HH_\tau$ with respect to $d_{1\tau}$. Then \cite[(3.4)]{BARBOUR19929} defines a Wasserstein metric $d_{2\tau}$ for two probability measures $Q_1$ and $Q_2$ on $\HH_\tau$ as
\[
d_{2\tau}(Q_1, Q_2)
=
\sup_{f\in\mathcal{V}_\tau}
\left|
\int_{\HH_\tau} f \, \dx Q_1
-
\int_{\HH_\tau} f \, \dx Q_2
\right|.
\]

The metric $d_{1\tau}$ quantifies the weak topology on $\HH_\tau$ and the metric $d_{2\tau}$ quantifies the weak topology on the space of probability measures on $\HH_\tau$; see \cite[Chapter~4.1 and 4.2]{Xia_survey}.

By the triangle inequality,
\[
d_{2\tau}\!\lb \law\lb \NN^\sig_\Xi|_\tau\rb ,\law\lb \NN^\sig_\Phi|_\tau\rb \rb  
\leq d_{2\tau}\!\lb \law\lb \NN^\sig_\Xi|_\tau\rb ,\law\lb \ZZ^\sig_\Xi|_\tau\rb \rb  + d_{2\tau}\!\lb \law\lb \ZZ^\sig_\Xi|_\tau\rb ,\law\lb \NN^\sig_\Phi|_\tau\rb \rb .
\]

Using the fact that for two probability measures $Q_1$ and $Q_2$, 
$d_{2\tau}(Q_1,Q_2)\leq \dtv(Q_1,Q_2)$ (see \cite[Chapter 4.2]{Xia_survey}), together with Corollary~\ref{corollary: cox_approximation}, we obtain
\begin{align*}
    d_{2\tau}\!\lb \law\lb \NN^\sig_\Xi|_\tau\rb ,\law\lb \ZZ^\sig_\Xi|_\tau\rb \rb 
    &\leq \dtv\!\lb \law\lb \NN^\sig_\Xi|_\tau\rb ,\law\lb \ZZ^\sig_\Xi|_\tau\rb \rb \\
    &\leq\lambda\int_{\IR^d} p_x^\sig(\tau)^2 \dx x.
\end{align*}

For any $r>0$, we further bound
\begin{align*}
    \int_{\IR^d} p_x^\sig(\tau)^2 \dx x
    &= \int_{|x|\leq r} p_x^\sig(\tau)^2 \dx x + \int_{|x|>r} p_x^\sig(\tau)^2 \dx x\\
    &\leq \int_{|x|\leq r}  \dx x + \IP\!\lb S\sig\geq\frac{g(r)}{\tau}\rb \int_{\IR^d} \IP\!\lb \frac{g(|x|)}{S(\sigma)}\le \tau\rb  \dx x\\
    &= V_d(r) + \IP\!\lb S\sig\geq\frac{g(r)}{\tau}\rb  \IE\int_{\IR^d} \I\!\lb |x|\le g^{-1}\!\lb  S(\sigma)\tau \rb \rb  \dx x\\
    &= V_d(r) + \IP\!\lb S\sig\geq\frac{g(r)}{\tau}\rb  \IE V_d\lb g^{-1}\!\lb  S(\sigma)\tau \rb \rb ,
\end{align*}
where the second last equality follows from Fubini’s theorem.

By the assumption that $S\sig\convergeP 0$ as $\sigmatoinf$ and \eqref{equ: finite assumption}, we obtain
\[
\limsup_{\sigma \to \infty}\int_{\IR^d} p_x^\sig(\tau)^2 \dx x \leq V_d(r).
\]
Since $r>0$ is arbitrary, it follows that
\[
\limsup_{\sigma \to \infty}\int_{\IR^d} p_x^\sig(\tau)^2 \dx x =0,
\]
and hence
\[
\lim_{\sigma \to \infty}d_{2\tau}\!\lb \law\lb \NN^\sig_\Xi|_\tau\rb ,\law\lb \ZZ^\sig_\Xi|_\tau\rb \rb  = 0.
\]

For the second term, we have
\begin{align*}
    d_{2\tau}\!\lb \law\lb \ZZ^\sig_\Xi|_\tau\rb ,\law\lb \NN^\sig_\Phi|_\tau\rb \rb 
    &=\sup_{f\in\mathcal{V}_\tau} \left| \IE f\!\lb  \ZZ^\sig_\Xi|_\tau \rb  - \IE f\!\lb  \NN^\sig_\Phi|_\tau \rb  \right|\\
    &\leq\IE\!\sup_{f\in\mathcal{V}_\tau} \left|\IE\! \lsb  f\!\lb  \ZZ^\sig_\Xi|_\tau \rb  -  f\!\lb  \NN^\sig_\Phi|_\tau \rb  \given \Xi\rsb \right|\\
    &=\IE d_{2\tau}\!\lb \law\lb \ZZ^\sig_\Xi|_\tau\given\Xi\rb ,\law\lb \NN^\sig_\Phi|_\tau\rb \rb ,
\end{align*}
where the inequality follows from conditioning and Jensen’s inequality.

Conditioning on $\Xi=\xi$, the process $\ZZ^\sig_\xi$ is a PPP on $\posRealLine \times [0,2\pi)^{d-1}$ with mean measure $\M_\xi^\sig$, while $\NN^\sig_\Phi$ is also a PPP on the same space with mean measure $\M^\sig$. Applying \cite[Theorem~1.5]{Xia1995OnMI}, we obtain

\[
d_{2\tau}\!\lb \law\lb \ZZ^\sig_\xi|_\tau\rb ,\law\lb \NN^\sig_\Phi|_\tau\rb \rb 
\leq d_{1\tau}\!\lb \M^\sig_\xi|_\tau/\M^\sig_\xi(\tau), \M^\sig|_\tau/\M^\sig(\tau)\rb  + 1\wedge\left| \M^\sig_\xi(\tau) - \M^\sig(\tau)\right|.
\]
Here, the first term in the upper bound of \cite[Theorem~1.5]{Xia1995OnMI} vanishes because $\ZZ^\sig_\xi$ is a PPP, and the term $1\wedge|\M^\sig_\xi(\tau) - \M^\sig(\tau)|$ arises from the fact that $d_{2\tau}$ is bounded by $1$. Note that the normalized measure $\M^\sig_\xi|_\tau/\M^\sig_\xi(\tau)$ defines a probability measure on $(0,\tau]$.

Taking expectations, we obtain 
\[
d_{2\tau}\!\lb \law\lb \ZZ^\sig_\Xi|_\tau\rb ,\law\lb \NN^\sig_\Phi|_\tau\rb \rb 
\leq \IE d_{1\tau}\!\lb \M^\sig_\Xi|_\tau/\M^\sig_\Xi(\tau), \M^\sig|_\tau/\M^\sig(\tau)\rb  + \IE 1\wedge\left| \M^\sig_\Xi(\tau) - \M^\sig(\tau)\right|.
\]
Theorem~\ref{theorem: mean measure converge} ensures that both $\M^\sig_\Xi|_\tau/\M^\sig_\Xi(\tau)$ and $\M^\sig|_\tau/\M^\sig(\tau)$ converge to the same limit with respect to the weak topology and $d_{1\tau}$ is a continuous function on the space of all distribution functions $(0,\tau]$ in terms of the weak topology, we have

\[
\lim_\sigmatoinf d_{1\tau}\!\lb \M^\sig_\Xi|_\tau/\M^\sig_\Xi(\tau), \M^\sig|_\tau/\M^\sig(\tau)\rb \dequal 0.
\]

It follows that
\[
\lim_\sigmatoinf d_{2\tau}\!\lb \law\lb \ZZ^\sig_\Xi|_\tau\rb ,\law\lb \NN^\sig_\Phi|_\tau\rb \rb =0.
\]

Therefore, for any $\tau>0$, we have
\[
\lim_\sigmatoinf\NN^\sig_\Xi|_\tau \dequal \lim_\sigmatoinf\NN^\sig_\Phi|_\tau,
\]
and it follows from \cite[Theorem~4.2]{Kallenberg1983} that
\[
\lim_\sigmatoinf\NN^\sig_\Xi \dequal \lim_\sigmatoinf\NN^\sig_\Phi.
\]
\end{proof}

\begin{proof}[Proof of Lemma~\ref{lemma: unbiasedEstimator}]

To show that $\X$ is an unbiased estimator of the target position (the origin), it suffices to prove that
\[
\IE \X  = \mathbf{0}.
\]
It is enough to show that, for any $n\geq 1$,
\[
\IE\!\lsb  \X \given |\Phi_\R|=n\rsb
\]
is invariant under rotations, since the only vector in $\IR^d$ that is invariant under all rotation transformations is $\mathbf{0}$.

For any $x \in \IR^d$, a rotation is represented by multiplication with a rotation matrix $Q_d \in \mathrm{SO}(d)$, where $\mathrm{SO}(d)$ is the $d\times d$ special orthogonal group; see \cite[Chapter~1.5]{Baker2002}. In general, $Q_d$ can be parameterized by $\frac{d(d-1)}{2}$ angular parameters. For example, when $d=2$, any rotation matrix takes the form
\[
Q_2(\vartheta)=
\begin{bmatrix}
\cos \vartheta & -\sin \vartheta \\
\sin \vartheta & \cos \vartheta
\end{bmatrix},
\qquad \vartheta \in \IR.
\]

We aim to show that, for any rotation matrix $Q_d$,
\begin{equation}\label{equ: zero_vector_claim}
\IE\!\lsb  \X \given |\Phi_\R|=n\rsb
=
Q_d \, \IE\!\lsb  \X \given |\Phi_\R|=n\rsb.
\end{equation}

Condition on the event $\{|\Phi_\R|=n\}$. Using the representation in \eqref{equ: individual_estimate_AOA_vec}, the estimator can be written as
\[
\X
=
\frac{1}{n}\sum_{i=1}^n \tilx_i
=
\frac{1}{n}\sum_{i=1}^n \lb X_i + \hatr_i \, \eta(\boldsymbol{\Theta}_i)\rb ,
\]
which is a measurable function of the data
\[
\lcb (X_i,\hatr_i,\boldsymbol{\Theta}_i)\rcb_{i=1}^n.
\]

Applying a rotation $Q_d$ yields
\[
Q_d \X
=
\frac{1}{n}\sum_{i=1}^n \lb Q_d X_i + \hatr_i \, Q_d \eta(\boldsymbol{\Theta}_i)\rb .
\]

Define $X_i' := Q_d X_i$ as the rotated sensor locations, and let $\boldsymbol{\Theta}_i'$ denote the AOA measurements associated with $X_i'$. Under the AOA model \eqref{equ: unwrapped normal AOA 1}--\eqref{equ: wrapped normal AOA}, the angular observations transform equivariantly under rotations, so that
\[
Q_d \eta(\boldsymbol{\Theta}_i) = \eta(\boldsymbol{\Theta}_i').
\]
Hence,
\[
Q_d \X
=
\frac{1}{n}\sum_{i=1}^n \lb X_i' + \hatr_i \, \eta(\boldsymbol{\Theta}_i')\rb ,
\]
which is precisely the estimator constructed from the rotated data
\[
\lcb (X_i',\hatr_i,\boldsymbol{\Theta}_i')\rcb_{i=1}^n.
\]

It is well known that, under the HPPP model, conditioning on having $n$ points in $\BallD$, the locations $\{X_i\}_{i=1}^n$ are i.i.d.\ uniformly distributed over $\BallD$; see \cite[Chapter~2.4.2]{Haenggi2012}. Since $\BallD$ is invariant under rotations, the joint distribution of $\{X_i\}_{i=1}^n$ is rotation invariant. Furthermore, the estimated distance $\hatr_i$ depends on $X_i$ only through its radial component $R_i$, and the AOA observations are centered at the true directions with dispersion depending only on $R_i$. Therefore, the joint distribution of the full data
\[
\lcb (X_i,\hatr_i,\boldsymbol{\Theta}_i)\rcb_{i=1}^n
\]
is invariant under rotations.

Since $\X$ is a measurable function of the data, it follows that
\[
\X \overset{d}{=} Q_d \X
\qquad \text{given } |\Phi_\R|=n.
\]
Taking expectations yields \eqref{equ: zero_vector_claim}, and hence
\[
\IE\!\lsb  \X \given |\Phi_\R|=n\rsb = \mathbf{0}.
\]
This completes the proof.
\end{proof}

\begin{proof}[Proof of Theorem~\ref{thm: CMSD and MSD}]
Given $|\Phi_\R|  = n$, the sensor positions $X_i$ are i.i.d. uniformly distributed over $\BallD$ (see \cite[Chapter~2.4.2]{Haenggi2012}), which implies that their hyperspherical coordinates $\lb  R_i, \boldsymbol{\Psi}_{i}\rb $ are also i.i.d. Together with the conditional independence assumption, it follows that the AOAs $\boldsymbol{\Theta}_{i}$ are i.i.d. as well. The estimated locations $\tilx_i$ are therefore i.i.d., since each $\tilx_i$ depends only on the corresponding sensor location $X_i$, estimated distance $\hatr_i$, and AOA vector $\boldsymbol{\Theta}_{i}$, all of which are i.i.d. across sensors. Therefore,
\begin{align*}
    \mathrm{CMSE}_\R(n) &= \IE\!\lsb \sum_{j=1}^d \lb  \frac{1}{n} \sum_{i=1}^n \tilde{X}_{i,j} \rb ^2  \rsb\\
    &= \frac{1}{n^2}\sum_{i=1}^n \IE\!\lsb  \sum_{j=1}^d\tilx_{i,j}^2  \rsb + \frac{1}{n^2}\sum_{i\neq m}  \sum_{j=1}^d\IE\!\lsb  \tilx_{i,j}\rsb \IE\!\lsb \tilx_{m,j} \rsb  \\
    &= \frac{1}{n} \IE\!\lsb  \sum_{j=1}^d\tilx_{1,j}^2 \rsb + \frac{n-1}{n}  \sum_{j=1}^d \IE\!\lsb  \tilx_{1,j} \rsb ^2.
\end{align*}
Since we have shown in Lemma~\ref{lemma: unbiasedEstimator} that the estimator $\X$ is unbiased, and
\[
\IE\X_j = \IE\!\lsb \frac{1}{n} \sum_{i=1}^n \tilde{X}_{i,j}\rsb = \IE  \tilx_{1,j} 
\]
we have
\begin{equation*}
    \IE \tilx_{1,j} =0
\end{equation*}
for $1\leq j\leq d$, and thus
\begin{align}
    \mathrm{CMSE}_\R(n) &= \frac{1}{n} \IE\!\lsb  \sum_{j=1}^d\tilx_{1,j}^2 \rsb \label{equ: CMSE_derivation1} \\
    &= \frac{1}{n} \IE\Bigg[\sum_{j=1}^{d-1} \lb  X_{1,j} + \hatr_1 \cos \vtj_1  \prod_{k=1}^{j-1} \sin \vtk_1 \rb ^2\nonumber\\
    &\qquad\qquad+ \lb  X_{1,d} + \hatr_1 \sin \vtzero_1 \prod_{k=1}^{d-2} \sin \vtk_1  \rb ^2\Bigg]\nonumber\\
    &=\frac{1}{n} \Bigg( \IE\!\lsb  \sum_{j=1}^dX_{1,j}^2 \rsb + \IE\!\lsb  \hatr_1^2 \rsb + 2\sum_{j=1}^{d-1} \IE\!\lsb  X_{1,j}  \hatr_1 \cos \vtj_1  \prod_{k=1}^{j-1} \sin \vtk_1 \rsb\nonumber \\
    &\qquad\qquad + 2\IE\!\lsb  X_{1,d}  \hatr_1 \sin \vtzero_1 \prod_{k=1}^{d-2} \sin \vtk_1 \rsb\Bigg).\nonumber
\end{align}
We first notice that 
\[
\IE\!\lsb  \sum_{j=1}^dX_{1,j}^2 \rsb = \IE\!\lsb R_1^2\rsb=\frac{d}{d+2}\R^2 
\]
can be directly computed using the marginal density of $R_1$:
\begin{equation}\label{equ: marginal_r}
    f_R(r) = \frac{dr^{d-1}}{\R^d}\I(0\leq r\leq \R).
\end{equation}

Using the tower property, we obtain, for $1\le j\le d-1$,
\begin{align}
    &\IE\!\lsb  X_{1,j}  \hatr_1 \cos \vtj_1  \prod_{k=1}^{j-1} \sin \vtk_1 \rsb\label{equ: tower property argument}\\
    &=\IE\!\lsb  X_{1,j} \IE\!\lsb  \hatr_1 \given R_1\rsb \IE\!\lsb  \cos \vtj_1  \given R_1,\vpj_1\rsb \prod_{k=1}^{j-1}\IE\!\lsb   \sin \vtk_1 \given R_1,\vpk_1\rsb  \rsb\nonumber
\end{align}
as $\hatr_1$ and $\lcb  \vtk_1 \rcb_{k=1}^{j}$ are conditionally independent given $R_1 $ and $ \lcb  \vpk_1 \rcb_{k=1}^{j}$.

Using the assumption that $\vtj_1 = \tilvtj_1 \bmod 2\pi $, where 
\begin{equation*}
    \begin{cases}
        \lb  \tilvtj_1 \given \vpj_1, R_1 \rb  \overset{d}{=} \N\!\lb \pi-\vpj_1, \Efirst\rb ,1\leq j\leq d-2,\\
        \lb  \tilvtzero_1 \given \vpzero_1, R_1 \rb  \overset{d}{=} \N\!\lb \vpzero_1-\pi, \Efirst\rb ,
    \end{cases}
\end{equation*}
we can compute $\IE\!\lsb \cos\vtj_1 \given R_1,\vpj_1\rsb$ and $\IE\!\lsb \sin\vtj_1 \given R_1,\vpj_1\rsb$ via the characteristic function of normal RVs. For $1\leq j\leq d-2,$
\begin{align}\label{equ: charactistic}
    \IE\!\lsb \cos\vtj_1 \given R_1,\vpj_1\rsb 
    &=\IE\!\lsb \cos\tilvtj_1 \given R_1,\vpj_1\rsb\\
    &= \frac{1}{2} \IE\!\lsb  e^{\sqrt{-1}\tilvtj_1} \given R_1,\vpj_1\rsb + \frac{1}{2} \IE\!\lsb  e^{-\sqrt{-1}\tilvtj_1} \given R_1,\vpj_1\rsb\nonumber\\
    &= -e^{-\frac{1}{2}\Efirst} \cos\vpj_1,\nonumber
\end{align}
and
\begin{align*}
    \IE\!\lsb \sin\vtj_1 \given R_1,\vpj_1\rsb 
    &= e^{-\frac{1}{2}\Efirst} \sin\vpj_1.
\end{align*}
Similarly,
\[
\IE\!\lsb  \cos\vtzero_1  \given R_1,\vpzero_1\rsb = -e^{-\frac{1}{2}\Efirst} \cos\vpzero_1,
\]
\begin{equation}\label{equ: sin(theta d-1)}
    \IE\!\lsb  \sin\vtzero_1  \given R_1,\vpzero_1\rsb = -e^{-\frac{1}{2}\Efirst} \sin\vpzero_1.
\end{equation}
It follows that for $1\le j\le d-1$,
\begin{align}
    &\IE\!\lsb  X_{1,j}  \hatr_1 \cos \vtj_1 \prod_{k=1}^{j-1} \sin \vtk_1 \rsb\nonumber\\
    &= \IE\Bigg[  R_1 \cos \vpj_1 \lb  \prod_{k=1}^{j-1} \sin \vpk_1 \rb  \IE\!\lsb  \hatr_1 \given R_1\rsb\; \nonumber\\
    &\qquad\qquad\times\lb  -e^{-\frac{1}{2}\Efirst} \cos \vpj_1 \rb  \lb \prod_{k=1}^{j-1} e^{-\frac{1}{2}\Efirst} \sin \vpk_1 \rb   \Bigg]\nonumber\\
    &=\IE\!\lsb  - R_1 \IE\!\lsb  \hatr_1 \given R_1\rsb e^{-\frac{j}{2}\Efirst}  \rsb \IE\!\lsb  \cos^2 \vpj_1 \rsb \prod_{k=1}^{j-1}\IE\!\lsb \sin^2 \vpk_1\rsb\nonumber\\
    &\leq - \IE\!\lsb  R_1 \;\hatr_1 \; e^{-\frac{d-1}{2}\Efirst} \rsb \IE\!\lsb  \cos^2 \vpj_1 \rsb \prod_{k=1}^{j-1}\IE\!\lsb \sin^2 \vpk_1\rsb,\label{equ: iterative1}
\end{align}
and similarly,
\begin{align}
    &\IE\!\lsb  X_{1,d}  \hatr_1 \sin \vtzero_1 \prod_{k=1}^{d-2} \sin \vtk_1 \rsb\nonumber\\
    &= - \IE\!\lsb  R_1 \;\hatr_1 \; e^{-\frac{d-1}{2}\Efirst} \rsb \IE\!\lsb  \sin^2 \vpzero_1 \rsb \prod_{k=1}^{d-2}\IE\!\lsb  \sin^2 \vpk_1 \rsb,\label{equ: iterative2}
\end{align}
where the negative sign is due to \eqref{equ: sin(theta d-1)}.

By applying the identity 
\[
\IE\!\lsb \sin^2\vpj_1\rsb + \IE\!\lsb \cos^2\vpj_1\rsb =1
\]
iteratively to the sum of \eqref{equ: iterative1} for $1\leq j \leq d-1$ and \eqref{equ: iterative2}, it follows that
\begin{align}
    \mathrm{CMSE}_\R(n)
    &\leq \frac{1}{n} \lb  \frac{d}{d+2} \R^2 + \IE\!\lsb  \hatr_1^2 \rsb - 2 \IE\!\lsb  R_1 \hatr_1  e^{-\frac{d-1}{2}\Efirst} \rsb  \rb \nonumber\\
    &\leq\frac{1}{n} \lb  \frac{d}{d+2} \R^2 + \IE\!\lsb  \hatr_1^2 \rsb - 2 e^{-\frac{d-1}{2}\E(\R)} \IE  R_1  \IE \hatr_1  \rb \nonumber\\
    &=\frac{1}{n} \lb  \frac{d}{d+2} \R^2 + \IE\!\lsb  \hatr_1^2 \rsb - \frac{2d \R}{d+1} e^{-\frac{d-1}{2}\E(\R)} \IE \hatr_1  \rb ,\label{CMSDP01}
\end{align}
where $\IE  R_1  = \frac{d}{d+1} \R$ is computed using the marginal density \eqref{equ: marginal_r}, and the last inequality is because 
\begin{equation}\label{equ: cov>0}
    \Cov(R_1,\hatr_1)=\IE\!\lsb  R_1 \hatr_1  \rsb-\IE R_1  \IE\hatr_1\geq0
\end{equation}
by Assumption~\eqref{assumption: cov_radial_positive}.

We now derive bounds for $\mathrm{MSE}_\R$. Recall that
\[
\mathrm{MSE}_\R
=
\IE  \mathrm{CMSE}_\R(|\Phi_\R| )
=
\sum_{n=0}^\infty \IP(|\Phi_\R|  = n)\,\mathrm{CMSE}_\R(n).
\]
By convention, we define $\mathrm{CMSE}_\R(0) := 0$ to reflect the case where no sensors are present in the region. Hence, the summation begins effectively at $n = 1$:
\begin{align}
    \mathrm{MSE}_\R 
    &= \sum_{n=1}^\infty \IP(|\Phi_\R|  = n)\,\mathrm{CMSE}_\R(n) \nonumber\\
    &\le
\lb 
\frac{d}{d+2}\R^2
+
\IE\!\lsb  \hatr_1^2 \rsb
-
\frac{2d \R}{d+1} e^{-\frac{d-1}{2}\E(\R)} \IE \hatr_1 
\rb 
\sum_{n=1}^\infty \frac{1}{n}\IP(|\Phi_\R|  = n) \label{equ: general MSE formula}\\
&= e^{-\lambda V_d(\R)}
\lb 
\sum_{n=1}^\infty \frac{(\lambda V_d(\R))^n}{n\cdot n!}
\rb 
\lb 
\frac{d}{d+2}\R^2
+
\IE\!\lsb  \hatr_1^2 \rsb
-
\frac{2d \R}{d+1} e^{-\frac{d-1}{2}\E(\R)} \IE \hatr_1 
\rb ,\nonumber
\end{align}
where the last equality is because $|\Phi_\R|  \dequal \Po(\lambda V_d(\R))$. This yields the sharper upper bound \eqref{eq:new_bound}.

To obtain a simpler bound with an explicit decay rate in $\lambda$, note that for all $n \ge 1$,
\[
\frac{1}{n} \le \frac{2}{n+1}.
\]
Therefore,
\begin{align*}
\sum_{n=1}^\infty \frac{1}{n}\IP(|\Phi_\R|  = n)
&\le
\sum_{n=1}^\infty \frac{2}{n+1}
\frac{(\lambda V_d(\R))^n}{n!} e^{-\lambda V_d(\R)} \\
&=
\frac{2 e^{-\lambda V_d(\R)}}{\lambda V_d(\R)}
\sum_{n=1}^\infty \frac{(\lambda V_d(\R))^{n+1}}{(n+1)!} \\
&\le
\frac{2}{\lambda V_d(\R)}.
\end{align*}
Substituting this into \eqref{equ: general MSE formula} yields the other bound \eqref{eq:old_bound}.

\end{proof}

\begin{proof}[Proof of Lemma~\ref{lemma: unbiasedEstimator_weighted}]
Condition on the event $\{|\Phi_\R|=n\}$. Using the representation in \eqref{equ: individual_estimate_AOA_vec}, the estimator can be written as
\[
\X^W
=
\sum_{i=1}^n W_i\tilx_i
=
\sum_{i=1}^n W_i\lb \hatr_1,\dots,\hatr_n\rb  \lb X_i + \hatr_i \, \eta(\boldsymbol{\Theta}_i)\rb ,
\]
which is a measurable function of the data
\[
\lcb (X_i,\hatr_i,\boldsymbol{\Theta}_i)\rcb_{i=1}^n.
\]
By the same symmetry argument used in the proof of Lemma~\ref{lemma: unbiasedEstimator}, and noting that the weights depend only on the estimated distances $\hatr_1,...,\hatr_n$ and not on the angular measurements, it follows that the estimator $\X^W$ is unbiased.
\end{proof}

\begin{proof}[Proof of Lemma~\ref{lemma: weighted_CMSE_rate}]
    From direct computation, we have

    \begin{align*}
        \mathrm{CMSE}^W\!(n)
        &= \IE\!\lsb  \sum_{j=1}^d \lb  \sum_{i=1}^n W_{\kappa(i)} \tilx_{\kappa(i),j}  \rb ^2 \rsb\\
        &=\frac{1}{A_n^2} \lcb  \sum_{i=1}^n a^{2i} \IE\!\lsb  \sum_{j=1}^d \tilx_{\kappa(i),j}^2 \rsb + \sum_{i\neq m} a^{i+m} \sum_{j=1}^d \IE\!\lsb  \tilx_{\kappa(i),j}\tilx_{\kappa(m),j}  \rsb  \rcb.
    \end{align*}

    From \eqref{equ: individual_estimate_AOA_vec}, for $1\leq j\leq d-1$, we have
    \begin{align*}
        \IE\!\lsb  \tilx_{\kappa(i),j}\tilx_{\kappa(m),j} \rsb
        &= \IE\!\lsb  X_{\kappa(i),j}X_{\kappa(m),j} \rsb\\
        &\qquad+\IE\!\lsb  X_{\kappa(i),j} \hatr_{(m)} \cos \vtj_{\kappa(m)} \prod_{k=1}^{j-1} \sin \vtk_{\kappa(m)} \rsb\\
        &\qquad+\IE\!\lsb X_{\kappa(m),j} \hatr_{(i)} \cos \vtj_{\kappa(i)} \prod_{k=1}^{j-1} \sin \vtk_{\kappa(i)} \rsb\\
        &\qquad+\IE\!\lsb   \hatr_{(i)} \cos \vtj_{\kappa(i)} \prod_{k=1}^{j-1} \sin \vtk_{\kappa(i)} \hatr_{(m)} \cos \vtj_{\kappa(m)} \prod_{k=1}^{j-1} \sin \vtk_{\kappa(m)} \rsb.
\end{align*}
    where each term can be shown to be 0, following \eqref{equ: tower property argument}, the conditional independence of $\lcb  \vtk_{\kappa(i)} \rcb_{k=1}^{j}$ given $R_{(i)} $ and $ \lcb  \vpk_{\kappa(i)} \rcb_{k=1}^{j}$, \eqref{equ: charactistic}, and the property of HPPP that $\vpj_{\kappa(i)}$ is independent of $R_{(i)}$ satisfying 
    \begin{equation}\label{equ: PPP_cos_zero}
        \IE \cos\vpj_{\kappa(i)}=0.
    \end{equation}

    To justify \eqref{equ: PPP_cos_zero}, we notice that the hyperspherical coordinates $\lb  R_i, \boldsymbol{\Psi}_{i}\rb $ of the points of an HPPP are independent of each other, as their joint density (computed from the density of a uniformly chosen point in $\BallD$ via change of variables) factorizes:
\begin{align*}
&f_{\lb  R_i, \boldsymbol{\Psi}_{i}\rb }\!\lb  r,\lcb  \psi^{(j)}\rcb_{j=1}^{d-1}\rb \\
&\propto r^{d-1}\I(0\leq r\leq \R)\; \I(0\leq \psi^{(d-1)}< 2\pi) \lb \prod_{j=1}^{d-2} \sin^{d-1-j}\!\lb  \psi^{(j)}\rb  \I\!\lb  0\leq \psi^{(j)}\leq\pi\rb  \rb ,\nonumber
\end{align*}
where $\propto$ means that they differ by a constant scaling factor. The marginal densities of the angular coordinates are given by
\begin{equation*}
    f_{\vpzero_1}\!\lb  \psi^{(d-1)} \rb  = \frac{1}{2\pi}\I\!\lb  0\leq \psi^{(d-1)}< 2\pi \rb ,
\end{equation*}
and for $1\leq j\leq d-2$,
\begin{equation*}
    f_{\vpj_1}\!\lb  \psi^{(j)} \rb  \propto  \sin^{d-1-j}\!\lb  \psi^{(j)}\rb  \I\!\lb  0\leq \psi^{(j)}\leq\pi\rb .
\end{equation*}
Since the distributions of the angular coordinates are not affected by the ordering according to the estimated distance, we have 
\[
\IE \cos\vpj_{\kappa(i)}=\IE \cos\vpj_1=0.
\]

    Using similar arguments with $\IE \sin \vpzero_{\kappa(i)} =0$, we also have
    \[
    \IE\!\lsb  \tilx_{\kappa(i),d}\tilx_{\kappa(m),d} \rsb=0.
    \]

    Therefore,
    \begin{equation*}
        \mathrm{CMSE}^W\!(n)
        =\frac{1}{A_n^2} \sum_{i=1}^n a^{2i} \IE\!\lsb  \sum_{j=1}^d \tilx_{\kappa(i),j}^2 \rsb.
    \end{equation*}

    Following the same argument as in \eqref{equ: CMSE_derivation1}, \eqref{equ: tower property argument}, \eqref{equ: iterative1}, \eqref{equ: iterative2}, and \eqref{CMSDP01}, we obtain
    \[
    \IE\!\lsb  \sum_{j=1}^d \tilx_{\kappa(i),j}^2 \rsb
    \leq \IE\!\lsb R_{\kappa(i)}^2\rsb + \IE\!\lsb \hatr_{(i)}^2\rsb- 2 e^{-\frac{d-1}{2}\E(\R)} \IE\!\lsb  R_{\kappa(i)} \hatr_{(i)}\rsb
    \]
    and similarly,
    \[
    \IE\!\lsb  \sum_{j=1}^d \tilx_{\kappa(i),j}^2 \rsb
    \geq \IE\!\lsb R_{\kappa(i)}^2\rsb + \IE\!\lsb \hatr_{(i)}^2\rsb- 2 e^{-\frac{1}{2}\lim_{r\to0^+}\E(r)} \IE\!\lsb  R_{\kappa(i)} \hatr_{(i)}\rsb.
    \]

\end{proof}

\begin{proof}[Proof of Theorem~\ref{thm: weighted_CMSE_rate_example}]
From Lemma~\ref{lemma: weighted_CMSE_rate}, it suffices to show that
\begin{equation}\label{equ: asymptotic_u}
    \mathscr{U}_{n,a}= \frac{1}{A_n^2} \sum_{i=1}^n a^{2i}\IE\!\lsb  R_{\kappa(i)}^2\rsb \asymp n^{-\frac{2}{d}},
\end{equation}

\begin{equation}\label{equ: asymptotic_v}
    \mathscr{V}_{n,a}=\frac{1}{A_n^2} \sum_{i=1}^n a^{2i}\IE\!\lsb  \hatr_{(i)}^2\rsb \asymp n^{-\frac{2}{d}},
\end{equation}
for the upper bound and

\begin{equation}\label{equ: asymptotic_w}
    \mathscr{X}_{n,a}= \frac{1}{A_n^2} \sum_{i=1}^n a^{2i}\IE\!\lsb (R_{\kappa(i)}- \hatr_{(i)})^2 \rsb\asymp n^{-\frac{2}{d}}
\end{equation}
for the lower bound.

We first establish that \eqref{equ: asymptotic_v} holds. Recall that, under the finite HPPP model, the distances $R_i$ are i.i.d. with density given in \eqref{equ: marginal_r} (see Section~\ref{Section: Poisson model main result}). Therefore, the estimated distances 
\[
\hatr_i = R_i\, S_i(\sigma)^{-\frac{1}{\beta}}=R_i \exp\!\lcb \frac{\sigma^2}{\beta^2} - \frac{\sigma}{\beta} B_i\rcb
\]
are also i.i.d. RVs.

We write $\mu:= - \frac{\sigma}{\beta}$ and $\nu :=\frac{\sigma^2}{\beta^2}$, and define $Y_i := S_i(\sigma)^{-\frac{1}{\beta}}=\exp\!\lcb \mu B_i + \nu \rcb$. The density of $Y_i$ is given by
\begin{equation}\label{equ: density_Y}
    f_Y(y) = \frac{1}{|\mu|y}f_B\!\lb  \frac{\ln y -\nu}{\mu}\rb \I(y>0),
\end{equation}
where $f_B$ is the density of a standard normal RV.

The density of $\hatr_i$ can be computed using a change of variable $s = \ln r$
\begin{align}
    f_{\hatr} (t)
    &= \int_0^\R f_Y\!\lb  \frac{t}{r} \rb  f_R(r) \dx r \nonumber\\
    &= C_1dt^{d-1} F_B\!\lb  \frac{\ln\R-\ln t +\nu - d\mu^2}{|\mu|}\rb \I(t>0), \label{equ: density_hatr}
\end{align}
where $C_1:= \R^{-d}\exp\{ -\nu d+d^2\mu^2/2 \} $ and $F_B$ is the cumulative distribution function (CDF) of the standard normal RV.

Since $\hatr_{(1)},...,\hatr_{(n)}$ are the order statistics of i.i.d. RVs, their densities are given by
\begin{equation*}
    f_{\hatr_{(i)}}(t) = n\binom{n-1}{i-1} F_{\hatr} (t)^{i-1}\lb  1- F_{\hatr} (t) \rb ^{n-i} f_{\hatr} (t) ,
\end{equation*}
where $F_{\hatr}$ is the CDF of $\hatr_i$.

From direct computation, we have
\begin{align*}
    \mathscr{V}_{n,a}
    &= \frac{1}{A_n^2} \sum_{i=1}^n a^{2i} \int t^2  n\binom{n-1}{i-1} F_{\hatr} (t)^{i-1}\lb  1- F_{\hatr} (t) \rb ^{n-i} f_{\hatr} (t) \dx t \\
    &= \frac{na^2}{A_n^2} \sum_{i=1}^n a^{2(i-1)} \binom{n-1}{i-1} \IE\!\lsb  \hatr_1^2   F_{\hatr} \!\lb \hatr_1\rb ^{i-1}\lb  1- F_{\hatr}\! \lb \hatr_1\rb  \rb ^{n-i} \rsb \\
    &= \frac{na^2}{A_n^2}  \IE\!\lsb  \hatr_1^2  \sum_{i=1}^n \binom{n-1}{i-1} \lb a^2 F_{\hatr}\! \lb \hatr_1\rb  \rb ^{i-1}\lb  1- F_{\hatr}\! \lb \hatr_1\rb  \rb ^{n-i} \rsb \\
    &= \frac{na^2}{A_n^2}  \IE\!\lsb  \hatr_1^2\lb  1- \lb 1-a^2 \rb F_{\hatr}\! \lb \hatr_1\rb  \rb ^{n-1} \rsb ,
\end{align*}
where the last equality follows from the binomial theorem.

Using the density \eqref{equ: density_hatr} of $\hatr_1$, we have
\[
\mathscr{V}_{n,a} = dC_1\frac{na^2}{A_n^2} \int_0^\infty t^{d+1}\lb  1- \lb 1-a^2 \rb F_{\hatr}\! \lb t\rb  \rb ^{n-1} F_B\!\lb  \frac{\ln\R-\ln t +\nu - d\mu^2}{|\mu|}\rb  \dx t.
\]

We notice that when $t\to 0$, 
\[
F_B\!\lb  \frac{\ln\R-\ln t +\nu - d\mu^2}{|\mu|}\rb \to 1.
\]
Therefore, there exist $\epsilon>0$ such that for any $t<\epsilon$, we have
\[
F_{\hatr} (t) = \int_0^t f_{\hatr} (s)\dx s = C_1t^d(1+r(t)),
\]
with $|r(t)|\leq\frac{1}{2}$.

As $n$ becomes large, the term
\[
\lb  1-\lb 1-a^2 \rb F_{\hatr}\! \lb t\rb  \rb ^{n-1}
\]
is dominated by
\[
\lb 1-a^2 \rb F_{\hatr}\! \lb t\rb \asymp n^{-1},
\]
or equivalently,
\[
t\asymp n^{-\frac{1}{d}}.
\]

We define
\[
\eps_n:=\lb  \frac{\ln{n}}{(n-1)\lb 1-a^2 \rb C_1}  \rb  ^{\frac{1}{d}}.
\]

There exists $n_0\geq1$ such that $\eps_n<\epsilon$ for any $n>n_0$. For $n>n_0$, we split
\[
\mathscr{V}_{n,a} = dC_1\frac{na^2}{A_n^2} \lb \int_0^{\eps_n}+\int_{\eps_n}^\epsilon+\int_\epsilon^\infty\rb  t^{d+1}\lb  1- \lb 1-a^2 \rb F_{\hatr}\! \lb t\rb  \rb ^{n-1} F_B\!\lb  \frac{\ln\R-\ln t +\nu - d\mu^2}{|\mu|}\rb  \dx t.
\]

Using the inequality $1-x\leq e^{-x}$, $F_{\hatr} (t)\geq \frac{1}{2}C_1t^d$, and a change of variable $s = \frac{1}{2}(n-1)\lb 1-a^2 \rb C_1 t^d$, we have

\begin{align*}
    &\int_0^{\eps_n} t^{d+1}\lb  1- \lb 1-a^2 \rb F_{\hatr}\! \lb t\rb  \rb ^{n-1} F_B\!\lb  \frac{\ln\R-\ln t +\nu - d\mu^2}{|\mu|}\rb  \dx t\\
    &\leq \int_0^{\eps_n} t^{d+1}\exp\!\lcb  - (n-1)\lb 1-a^2 \rb \frac{1}{2}C_1t^d \rcb \dx t\\
    &= \frac{1}{d}\lb  (n-1)\lb 1-a^2 \rb \frac{1}{2}C_1 \rb ^{-\frac{d+2}{d}} \int_0^{0.5\ln{n}}s^{\frac{2}{d}}e^{-s}\dx s\\
    &\leq \frac{\Gamma\lb \frac{2}{d}+1\rb }{d}\lb  (n-1)\lb 1-a^2 \rb \frac{1}{2}C_1 \rb ^{-\frac{d+2}{d}}\\
    &\asymp n^{-\frac{d+2}{d}}.
\end{align*}

Using similar arguments, we have 
\begin{align*}
    &\int_{\eps_n}^\epsilon t^{d+1}\lb  1- \lb 1-a^2 \rb F_{\hatr}\! \lb t\rb  \rb ^{n-1} F_B\!\lb  \frac{\ln\R-\ln t +\nu - d\mu^2}{|\mu|}\rb  \dx t\\
    &\leq  \frac{1}{d}\lb  (n-1)\lb 1-a^2 \rb \frac{1}{2}C_1 \rb ^{-\frac{d+2}{d}} \underbrace{\int_{0.5\ln{n}}^\infty s^{\frac{2}{d}}e^{-s}\dx s}_{\to0 \text{ as } n\to\infty}\\
    & = \lito\!\lb  n^{-\frac{d+2}{d}}\rb ,
\end{align*}
where we write $f_1(n)=\lito(f_2(n))$ if $f_1(n)/f_2(n)\to 0$ as $n\to\infty$.

Finally,
\begin{align}
    &\int_\epsilon^\infty t^{d+1}\lb  1- \lb 1-a^2 \rb F_{\hatr}\! \lb t\rb  \rb ^{n-1} F_B\!\lb  \frac{\ln\R-\ln t +\nu - d\mu^2}{|\mu|}\rb  \dx t \label{equ: expression_v_exponential_decay}\\
    &\leq  \lb  1- \lb 1-a^2 \rb F_{\hatr}\! \lb \epsilon\rb  \rb ^{n-1} \underbrace{\int_0^\infty t^{d+1} F_B\!\lb  \frac{\ln\R-\ln t +\nu - d\mu^2}{|\mu|}\rb  \dx t}_{=\frac{1}{C_1d}\IE\!\lsb \hatr_1^2\rsb<\infty} \nonumber\\
    & \to0\nonumber
\end{align}
exponentially fast as $n\to\infty$.

Therefore, we have shown that there is an upper bound of $\mathscr{V}_{n,a}$ that decays at order $n^{-\frac{2}{d}}$. It is enough to further show that there is also a lower bound of $\mathscr{V}_{n,a}$ that decays at the same order.

We define
\[
\varpi_n:=\lb  \frac{1}{3(n-1)\lb 1-a^2 \rb C_1} \rb ^{\frac{1}{d}}.
\]
Then there exists $n_1\geq1$ such that for any $n>n_1$, we have $\varpi_n<\epsilon$, and 
\[
F_B\!\lb  \frac{\ln\R-\ln \varpi_n +\nu - d\mu^2}{|\mu|}\rb \geq\frac{1}{2}.
\]

For $n>n_1$, using the fact that $1-x\geq e^{-2x},\forall 0<x<\frac{1}{2}$, and notice that when $t<\varpi_n$,
\[
(n-1)\lb 1-a^2 \rb F_{\hatr}\! \lb t\rb \leq (n-1)\lb 1-a^2 \rb \frac{3}{2}C_1 \varpi_n^d = \frac{1}{2},
\]
we have

\begin{align*}
    \mathscr{V}_{n,a} 
    &\geq dC_1\frac{na^2}{A_n^2} \int_0^{\varpi_n} t^{d+1}\lb  1- \lb 1-a^2 \rb F_{\hatr}\! \lb t\rb  \rb ^{n-1} \frac{1}{2} \dx t\\
    &\geq dC_1\frac{na^2}{2A_n^2} \int_0^{\varpi_n} t^{d+1}\exp\!\lcb  - 2(n-1)\lb 1-a^2 \rb F_{\hatr}\! \lb t\rb  \rcb \dx t\\
    &\geq dC_1\frac{na^2}{2eA_n^2} \frac{\varpi_n^{d+2}}{d+2}\\
    &\asymp  n^{-\frac{2}{d}}.
\end{align*}

We now establish that \eqref{equ: asymptotic_u} holds. The joint density of $\lb  R_{\kappa(1)}, ...,R_{\kappa(n)}\rb $ is given by
\begin{equation}\label{equ: joint_density_R_kappa}
    f_{\lb  R_{\kappa(1)}, ...,R_{\kappa(n)}\rb }\lb r_1, ...,r_n\rb 
    =n! \int_{0<u_1<\cdots<u_n} \prod_{i=1}^n \lb  f_R(r_i) f_Y\!\lb  \frac{u_i}{r_i} \rb  \frac{1}{r_i}\dx u_i  \rb .
\end{equation}

To justify \eqref{equ: joint_density_R_kappa}, we first consider $n=2$. By symmetry, we have
\begin{align*}
    &\IP\!\lb  R_{\kappa(1)} \in \dx r_1, R_{\kappa(2)} \in \dx r_2  \rb \\
    &= \IP\!\lb  R_1 \in \dx r_1, R_2 \in \dx r_2, \hatr_1<\hatr_2  \rb  + \IP\!\lb  R_2 \in \dx r_1, R_1 \in \dx r_2, \hatr_2<\hatr_1  \rb \\
    &= 2\int_{0<u_1<u_2} \IP\!\lb  R_1 \in \dx r_1, R_2 \in \dx r_2, \hatr_1\in \dx u_1,\hatr_2 \in \dx u_2 \rb .
\end{align*}

The joint density of $\mathscr{R}:=\lcb \lb  R_i, \hatr_i\rb  \rcb_{i=1}^n$ is given by
\[
f_\mathscr{R}\lb r_1, \hat{r}_1,...,r_n, \hat{r}_n\rb =\prod_{i=1}^n f_{\lb  R, \hatr\rb } \!\lb r_i, \hat{r}_i\rb  ,
\]
where $f_{\lb  R, \hatr\rb }$ is the joint density of $\lb  R_i, \hatr_i\rb $, which can be computed by \eqref{equ: marginal_r} and \eqref{equ: density_Y}
\begin{equation}\label{equ: joint_r_rhat}
    f_{\lb  R, \hatr\rb } \lb r, \hat{r}\rb  = f_R(r)f_Y\!\lb  \frac{\hat{r}}{r} \rb  \frac{1}{r}.
\end{equation}

Thus,
\begin{align*}
    &\IP\!\lb  R_{\kappa(1)} \in \dx r_1, R_{\kappa(2)} \in \dx r_2  \rb \\
    &= 2!\int_{0<u_1<u_2} \prod_{i=1}^2 \lb  f_R(r_i) f_Y\!\lb  \frac{u_i}{r_i} \rb  \frac{1}{r_i}\dx u_i  \rb  \dx r_1 \dx r_2.
\end{align*}

These arguments can be generalized to arbitrary $n$.

The marginal density of $R_{\kappa(i)}$ can be computed by
\begin{align}
    &f_{R_{\kappa(i)}}\lb r_i\rb  \nonumber\\
    &= \int_{(0,\infty)^{n-1}} f_{\lb  R_{\kappa(1)}, ...,R_{\kappa(n)}\rb }\lb r_1, ...,r_n\rb  \prod_{j\neq i} \dx r_j \nonumber\\
    &= n! \int_{0<u_1<\cdots<u_n} f_R(r_i) f_Y\!\lb  \frac{u_i}{r_i} \rb  \frac{1}{r_i} \prod_{j\neq i}  \lb  \int_0^\infty f_R(r_j) f_Y\!\lb  \frac{u_j}{r_j} \rb  \frac{1}{r_j} \dx r_j   \rb  \prod_{j=1}^n\dx u_j \nonumber\\
    &= n! \int_{u_i=0}^{u_i=\infty} f_R(r_i) f_Y\!\lb  \frac{u_i}{r_i} \rb  \frac{1}{r_i} \lb \int_{0<u_1<\cdots<u_i} \prod_{j=1}^{i-1}\lb  f_{\hatr}(u_j) \dx u_j \rb  \rb  \lb \int_{u_i<\cdots<u_n} \prod_{j=i+1}^{n}\lb  f_{\hatr}(u_j) \dx u_j \rb  \rb  \dx u_i \nonumber\\
    &=n! \int_0^\infty f_R(r_i) f_Y\!\lb  \frac{u_i}{r_i} \rb  \frac{1}{r_i} \lb \frac{1}{(i-1)!}\int_{(0,u_i)^{i-1}} \prod_{j=1}^{i-1}\lb  f_{\hatr}(u_j) \dx u_j \rb  \rb  \lb  \frac{1}{(n-i)!}\int_{(u_i,\infty)^{n-i}} \prod_{j=i+1}^{n}\lb  f_{\hatr}(u_j) \dx u_j \rb  \rb  \dx u_i \nonumber\\
    &= n \binom{n-1}{i-1} \int_0^\infty f_R(r_i) f_Y\!\lb  \frac{u_i}{r_i} \rb  \frac{1}{r_i} F_{\hatr}(u_i)^{i-1}\lb  1 -  F_{\hatr}(u_i) \rb ^{n-i} \dx u_i,\label{eqi: marginal_R_kappa}
\end{align}
where the second last equality is due to symmetry.

Therefore, using the binomial theorem, we can write
\begin{align}
    \mathscr{U}_{n,a}
    &= \frac{na^2}{A_n^2}   \int_0^\R  \int_0^\infty t^2 f_R(t) f_Y\!\lb  \frac{u}{t} \rb  \frac{1}{t} \sum_{i=1}^n\binom{n-1}{i-1} \lb a^2 F_{\hatr}(u)\rb ^{i-1}\lb  1 -  F_{\hatr}(u) \rb ^{n-i} \dx u \dx t \label{equ: expression_u_computation}\\
    &=\frac{na^2}{A_n^2}   \int_0^\R  \int_0^\infty t^2 f_{\lb  R, \hatr\rb } \lb t, u\rb  \lb   1 - \lb  1 - a^2 \rb   F_{\hatr}(u)  \rb  ^{n-1} \dx u \dx t \nonumber\\
    &=\frac{na^2}{A_n^2} \IE\!\lsb  R_1^2 \lb   1 - \lb  1 - a^2 \rb   F_{\hatr}\!\lb \hatr_1\rb   \rb  ^{n-1} \rsb \nonumber\\
    &=\frac{na^2}{A_n^2} \IE\!\lsb  \IE\!\lsb  R_1^2 \given \hatr_1 \rsb \lb   1 - \lb  1 - a^2 \rb   F_{\hatr}\!\lb \hatr_1\rb   \rb  ^{n-1} \rsb. \label{equ: expression_U_conditional_Rhat}
\end{align}

The conditional density of $R_1$ given $\hatr_1$ can be directly computed by \eqref{equ: joint_r_rhat} and \eqref{equ: density_hatr}
\begin{align*}
    f_{R|\hatr} (r|\hat r) 
    &=\frac{r^{d-1} \frac{1}{|\mu|\sqrt{2\pi}}\exp\!\lcb  -\frac{1}{2\mu^2}\lb  \ln \hat r - \ln r -\nu\rb ^2 \rcb}{\hat r^{d-1} \exp\{ -\nu d+d^2\mu^2/2 \} F_B\lb  \frac{\ln\R-\ln \hat r +\nu - d\mu^2}{|\mu|}\rb } \I(0<r<\R).
\end{align*}

Using a change of variable, the conditional density of $\ln R_1$ given $\hatr_1$ is given by
\begin{equation}\label{equ: conditional_density_lnR_given_Rhat}
    f_{\ln R|\hatr} (s|\hat r) 
    =\frac{\frac{1}{|\mu|\sqrt{2\pi}}\exp\!\lcb  -\frac{1}{2\mu^2}\lb  s-(\ln \hat r -\nu +\mu^2d)\rb ^2 \rcb}{F_B\lb  \frac{\ln\R-\ln \hat r +\nu - d\mu^2}{|\mu|}\rb } \I(s<\ln\R).
\end{equation}
That is, $\lb \ln R_1 \given\hatr_1 \rb $ is a truncated normal with mean $\ln \hatr_1 -\nu +\mu^2d$ and variance $\mu^2$, truncated at $(-\infty,\ln \R]$.

From direct computation, we have
\begin{align}
    \IE\!\lsb  R_1^2 \given \hatr_1 = \hat r \rsb
    &=\int_{-\infty}^{\ln\R} e^{2s}f_{\ln R|\hatr} (s|\hat r)\dx s\nonumber\\
    &= \frac{F_B\!\lb  \frac{\ln\R-\ln \hat r +\nu - (d+2)\mu^2}{|\mu|}\rb }{F_B\!\lb  \frac{\ln\R-\ln \hat r +\nu - d\mu^2}{|\mu|}\rb } \hat r^2 e^{2\mu^2(d+1)-2\nu}\label{conditionxia1}\\
    &\leq C_2\hat r^2 ,\nonumber
\end{align}
where $C_2:= e^{2\mu^2(d+1)-2\nu}$. It follows from \eqref{equ: expression_U_conditional_Rhat} that
\begin{equation}\label{equ: expression_u_upper_bound}
    \mathscr{U}_{n,a}\leq C_2\mathscr{V}_{n,a}.
\end{equation}

For the lower bound, we use \eqref{equ: expression_U_conditional_Rhat} again to get
\[
\mathscr{U}_{n,a}
    \geq\frac{na^2}{A_n^2} \IE\!\lsb  \IE\!\lsb  R_1^2 \given \hatr_1 \rsb \lb   1 - \lb  1 - a^2 \rb   F_{\hatr}\!\lb \hatr_1\rb   \rb  ^{n-1} \I\!\lb  \hatr_1\leq \epsilon\rb  \rsb.
\]

Using the Mills ratio bound $(1-F_B(x))/f_B(x) < 1/x$ for $x>0$, together with the symmetry of the normal distribution, it follows that
\[
\frac{d}{dx}\left(\frac{f_B(x)}{F_B(x)}\right) < 0, \qquad \forall x \in \mathbb{R}.
\]
Consequently, for any $c>0$,
\[
\frac{d}{dx}\left(\frac{F_B(x-c)}{F_B(x)}\right) > 0, \qquad \forall x \in \mathbb{R},
\]
which implies that $F_B(x-c)/F_B(x)$ is increasing in $x$.

Applying this monotonicity property, we obtain from \eqref{conditionxia1} that, for $\hat r \leq \epsilon$,
\begin{align*}
    \IE\!\lsb  R_1^2 \given \hatr_1 = \hat r \rsb
    \geq C_3\hat r^2 ,
\end{align*}
where
\[
C_3:=\frac{F_B\!\lb  \frac{\ln\R-\ln \epsilon +\nu - (d+2)\mu^2}{|\mu|}\rb }{F_B\!\lb  \frac{\ln\R-\ln \epsilon +\nu - d\mu^2}{|\mu|}\rb }  e^{2\mu^2(d+1)-2\nu}.
\]

It follows from \eqref{equ: asymptotic_v} and \eqref{equ: expression_v_exponential_decay} that
\begin{align*}
    \mathscr{U}_{n,a}
    &\geq C_3 \mathscr{V}_{n,a}- \frac{na^2}{A_n^2} C_3  \underbrace{\IE\!\lsb  \hatr_1^2\lb   1 - \lb  1 - a^2 \rb   F_{\hatr}\!\lb \hatr_1\rb   \rb  ^{n-1} \I\!\lb  \hatr_1> \epsilon\rb  \rsb}_{\to 0 \text{ exponentially fast by \eqref{equ: expression_v_exponential_decay}}}.
\end{align*}
Together with \eqref{equ: expression_u_upper_bound} and \eqref{equ: asymptotic_v} yields the claim \eqref{equ: asymptotic_u}.

Finally, we establish that \eqref{equ: asymptotic_w} holds.  Following similar lines of reasoning as in \eqref{equ: joint_density_R_kappa} and \eqref{eqi: marginal_R_kappa}, we can compute the joint density of $\lb  R_{\kappa(1)}, ...,R_{\kappa(n)}, \hatr_{(i)}\rb $ and hence the joint density of $\lb  R_{\kappa(i)}, \hatr_{(i)}\rb $
\begin{equation*}
    f_{\lb  R_{\kappa(i)}, \hatr_{(i)}\rb }\lb r_i, \hat r_i\rb 
    =n \binom{n-1}{i-1}  f_R(r_i) f_Y\!\lb  \frac{\hat r_i}{r_i} \rb  \frac{1}{r_i} F_{\hatr}(\hat r_i)^{i-1}\lb  1 -  F_{\hatr}(\hat r_i) \rb ^{n-i} .
\end{equation*}

Then it follows from direct computation as in \eqref{equ: expression_u_computation} that
\begin{align*}
    \mathscr{X}_{n,a} 
    &= \frac{na^2}{A_n^2} \IE\!\lsb  \lb R_1-\hatr_1 \rb ^2 \lb   1 - \lb  1 - a^2 \rb   F_{\hatr}\!\lb \hatr_1\rb   \rb  ^{n-1} \rsb.
\end{align*}

Using the fact that $(A-B)^2\leq A^2+B^2$, we can bound
\begin{equation}\label{equ: expression_x_upper}
    \mathscr{X}_{n,a}\leq \mathscr{U}_{n,a} +\mathscr{V}_{n,a}\asymp n^{-\frac{2}{d}}.
\end{equation}

For the lower bound, we notice that
\begin{align*}
    \mathscr{X}_{n,a} 
    &=\frac{na^2}{A_n^2} \IE\!\lsb  \lcb \lb R_1- \IE\!\lsb  R_1 \given \hatr_1 \rsb\rb ^2 + \lb \IE\!\lsb  R_1 \given \hatr_1 \rsb - \hatr_1 \rb ^2 \rcb \lb   1 - \lb  1 - a^2 \rb   F_{\hatr}\!\lb \hatr_1\rb   \rb  ^{n-1} \rsb\\
    &\geq\frac{na^2}{A_n^2} \IE\!\lsb  \lb \IE\!\lsb  R_1 \given \hatr_1 \rsb - \hatr_1 \rb ^2  \lb   1 - \lb  1 - a^2 \rb   F_{\hatr}\!\lb \hatr_1\rb   \rb  ^{n-1} \rsb.
\end{align*}

From direct computation using \eqref{equ: conditional_density_lnR_given_Rhat}, we have
\begin{align*}
    \IE\!\lsb  R_1 \given \hatr_1 = \hat r \rsb
    &= \frac{F_B\!\lb  \frac{\ln\R-\ln \hat r +\nu - (d+1)\mu^2}{|\mu|}\rb }{F_B\!\lb  \frac{\ln\R-\ln \hat r +\nu - d\mu^2}{|\mu|}\rb } \hat r e^{\mu^2(d+\frac{1}{2})-\nu}.
\end{align*}
Notice that $e^{\mu^2(d+\frac{1}{2})-\nu} = e^{\sigma^2(d-\frac{1}{2})/\beta^2}>1$, so there exist a $\epsilon^* > 0$ such that
\[
C_4:= \frac{F_B\!\lb  \frac{\ln\R-\ln \epsilon^* +\nu - (d+1)\mu^2}{|\mu|}\rb }{F_B\!\lb  \frac{\ln\R-\ln \epsilon^* +\nu - d\mu^2}{|\mu|}\rb }  e^{\mu^2(d+\frac{1}{2})-\nu}>1.
\]
Then, for any $\hat r\leq \epsilon^*$, we have
\[
\IE\!\lsb  R_1 \given \hatr_1 = \hat r \rsb \geq C_4 \hat r > \hat r.
\]

It follows that
\begin{align*}
    \mathscr{X}_{n,a} 
    &\geq\frac{na^2}{A_n^2} \IE\!\lsb  \lb \IE\!\lsb  R_1 \given \hatr_1 \rsb - \hatr_1 \rb ^2  \lb   1 - \lb  1 - a^2 \rb   F_{\hatr}\!\lb \hatr_1\rb   \rb  ^{n-1} \I\!\lb  \hatr_1\leq \epsilon^*\rb  \rsb\\
    &\geq\frac{na^2}{A_n^2} (C_4-1)^2 \lb \mathscr{V}_{n,a} -  \underbrace{\IE\!\lsb  \hatr_1 ^2  \lb   1 - \lb  1 - a^2 \rb   F_{\hatr}\!\lb \hatr_1\rb   \rb  ^{n-1} \I\!\lb  \hatr_1 > \epsilon^*\rb  \rsb}_{\to 0 \text{ exponentially fast similar to \eqref{equ: expression_v_exponential_decay}}} \rb \\
    &\asymp n^{-\frac{2}{d}}.
\end{align*}

Together with \eqref{equ: expression_x_upper} proves that $\mathscr{X}_{n,a}$ decays at order $n^{-\frac{2}{d}}$, and the claim \eqref{equ: asymptotic_w} follows.

\end{proof}

\section*{Acknowledgments}

Work supported in part by Australian Research Council Grant No DP220102666 (AX).

\bibliographystyle{IEEEtran}
\bibliography{Bibliography}

@ARTICLE{WinEtAl2010a,
  author={Shen, Yuan and Win, Moe Z.},
  journal={IEEE Transactions on Information Theory}, 
  title={Fundamental Limits of Wideband Localization— Part I: A General Framework}, 
  year={2010},
  volume={56},
  number={10},
  pages={4956-4980},
  doi={10.1109/TIT.2010.2060110}}

@book{BaccelliBlaszczyszyn2009a,
  author  = {Baccelli, Fran{\c{c}}ois and B{\l}aszczyszyn, Bart{\l}omiej},
  title   = {Stochastic Geometry and Wireless Networks: Volume I Theory},
  publisher = {Now Publishers Inc.},
  year      = {2010},
  doi     = {10.1561/1300000006},
}

@book{Haenggi2012, place={Cambridge}, title={Stochastic Geometry for Wireless Networks}, publisher={Cambridge University Press}, author={Haenggi, Martin}, year={2012}}

@article{HaenggiEtAl2009,
author = {Haenggi, Martin and Andrews, Jeffrey and Baccelli, Fran{\c{c}}ois and Dousse, Olivier and Franceschetti, Massimo},
year = {2009},
month = {10},
pages = {1029 - 1046},
title = {Stochastic Geometry and Random Graphs for the Analysis and Design of Wireless Networks},
volume = {27},
journal = {Selected Areas in Communications, IEEE Journal on},
doi = {10.1109/JSAC.2009.090902}
}

@article{BergelNoam2018,
  author  = {Bergel, Itsik and Noam, Yair},
  title   = {Lower Bound on the Localization Error in Infinite Networks with Random Sensor Locations},
  journal = {IEEE Transactions on Signal Processing},
  year    = {2018},
  volume  = {66},
  number  = {5},
  pages   = {1228--1241},
  doi     = {10.1109/TSP.2017.2780040},
  url     = {https://ieeexplore.ieee.org/document/8186282}
}

@book{Kallenberg1983,
  author    = {Kallenberg, Olav},
  title     = {Random Measures},
  edition   = {3rd},
  publisher = {Academic Press},
  address   = {London},
  year      = {1983},
  note      = {MR0818219}
}

@article{KeelerRossXia2018,
  author  = {Keeler, Holger Paul and Ross, Nathan and Xia, Aihua},
  title   = {When Do Wireless Network Signals Appear {Poisson}?},
  journal = {Bernoulli},
  year    = {2018},
  volume  = {24},
  number  = {3},
  pages   = {1973--1994},
  doi     = {10.3150/16-BEJ917},
  url     = {https://projecteuclid.org/euclid.bj/1530087050}
}

@book{DaleyVereJones2008,
  author    = {Daley, D.J. and Vere-Jones, D.},
  title     = {An Introduction to the Theory of Point Processes: Volume II: General Theory and Structure},
  edition   = {2nd},
  publisher = {Springer},
  year      = {2008}
}

@book{sigma_values, place={Cambridge}, title={Wireless Communications}, publisher={Cambridge University Press}, author={Goldsmith, Andrea}, year={2005}}

@inproceedings{BlaszczyszynKarrayKeeler2013,
  author    = {Błaszczyszyn, Bartłomiej and Karray, Mohamed Kadhem and Keeler, Holger Paul},
  title     = {Using {Poisson} Processes to Model Lattice Cellular Networks},
  booktitle = {Proceedings of IEEE INFOCOM 2013},
  pages     = {773--781},
  year      = {2013},
  doi       = {10.1109/INFCOM.2013.6566864},
  url       = {https://ieeexplore.ieee.org/document/6566864}
}

@article{BlaszczyszynKarrayKeeler2015,
  author  = {Błaszczyszyn, Bartłomiej and Karray, Mohamed Kadhem and Keeler, Holger Paul},
  title   = {Wireless Networks Appear {Poissonian} Due to Strong Shadowing},
  journal = {IEEE Transactions on Wireless Communications},
  volume  = {14},
  number  = {8},
  pages   = {4379--4390},
  year    = {2015},
  doi     = {10.1109/TWC.2015.2421315},
  url     = {https://ieeexplore.ieee.org/document/7104150}
}

@ARTICLE{7745970,
  author={Ross, Nathan and Schuhmacher, Dominic},
  journal={IEEE Transactions on Information Theory}, 
  title={Wireless Network Signals With Moderately Correlated Shadowing Still Appear {Poisson}}, 
  year={2017},
  volume={63},
  number={2},
  pages={1177-1198},
  doi={10.1109/TIT.2016.2629482}}

@article{Blumenson1960,
 ISSN = {00029890, 19300972},
 URL = {http://www.jstor.org/stable/2308932},
 author = {Blumenson, Leslie E.},
 journal = {The American Mathematical Monthly},
 number = {1},
 pages = {63--66},
 publisher = {[Taylor & Francis, Ltd., Mathematical Association of America]},
 title = {A Derivation of n-Dimensional Spherical Coordinates},
 urldate = {2025-08-23},
 volume = {67},
 year = {1960}
}

@article{repulsive,
 ISSN = {00018678},
 URL = {http://www.jstor.org/stable/43563378},
 author = {Miyoshi, Naoto and Shirai, Tomoyuki},
 journal = {Advances in Applied Probability},
 number = {3},
 pages = {832--845},
 publisher = {Applied Probability Trust},
 title = {A CELLULAR NETWORK MODEL WITH {G}INIBRE CONFIGURED BASE STATIONS},
 urldate = {2025-08-31},
 volume = {46},
 year = {2014}
}

@INPROCEEDINGS{maternhardcore,
  author={Ibrahim, Abdelrahman M. and ElBatt, Tamer and El-Keyi, Amr},
  booktitle={2013 IEEE 24th Annual International Symposium on Personal, Indoor, and Mobile Radio Communications (PIMRC)}, 
  title={Coverage probability analysis for wireless networks using repulsive point processes}, 
  year={2013},
  volume={},
  number={},
  pages={1002-1007},
  doi={10.1109/PIMRC.2013.6666284}}

@article{empirical_Poisson,
author = {Lee, Chia-Han and Shih, Cheng-Yu and Chen, Yu-Sheng},
year = {2013},
month = {08},
pages = {},
title = {Stochastic geometry based models for modeling cellular networks in urban areas},
volume = {19},
journal = {Wireless Networks},
doi = {10.1007/s11276-012-0518-0}
}

@ARTICLE{Poisson_cluster,
  author={Ganti, Radha Krishna and Haenggi, Martin},
  journal={IEEE Transactions on Information Theory}, 
  title={Interference and Outage in Clustered Wireless Ad Hoc Networks}, 
  year={2009},
  volume={55},
  number={9},
  pages={4067-4086},
  doi={10.1109/TIT.2009.2025543}}

@article{correlated_shadowing1,
author = {Giancristofaro, Domenico},
year = {1996},
month = {06},
pages = {958 - 959},
title = {Correlation model for shadow fading in mobile radio channels},
volume = {32},
journal = {Electronics Letters},
doi = {10.1049/el:19960655}
}

@inproceedings{correlated_shadowing2,
	author = {Catrein, Daniel and Mathar, Rudolf},
	title = "{G}aussian Random Fields as a Model for Spatially Correlated Log-Normal Fading",
	booktitle = "Proceedings: Australasian Telecommunication Networks and Applications Conference (ATNAC)",
	address = {Adelaide, Australia},
	month = Dec,
	year = 2008,
	hsb = RWTH-CONV-223587,
	}

@article{MHPTypeI,
title = {Generalizations of {Matérn’s} hard-core point processes},
journal = {Spatial Statistics},
volume = {3},
pages = {33-53},
year = {2013},
issn = {2211-6753},
doi = {https://doi.org/10.1016/j.spasta.2013.02.001},
url = {https://www.sciencedirect.com/science/article/pii/S2211675313000043},
author = {J. Teichmann and F. Ballani and K.G. {van den Boogaart}}
}

@book{Statistical_inference_and_simulation_for_spatial,
  title={Statistical inference and simulation for spatial point processes},
  author={M{\o}ller, Jesper and Waagepetersen, Rasmus Plenge},
  year={2003},
  publisher={CRC press}
}

@book{Statistical_analysis_and_modelling_of_spatial,
  title={Statistical analysis and modelling of spatial point patterns},
  author={Illian, Janine and Penttinen, Antti and Stoyan, Helga and Stoyan, Dietrich},
  year={2008},
  publisher={John Wiley \& Sons}
}

@ARTICLE{MUSIC,
  author={Schmidt, R.},
  journal={IEEE Transactions on Antennas and Propagation}, 
  title={Multiple emitter location and signal parameter estimation}, 
  year={1986},
  volume={34},
  number={3},
  pages={276-280},
  doi={10.1109/TAP.1986.1143830}}

@ARTICLE{AoAGaussian,
  author={Vallet, Pascal and Mestre, Xavier and Loubaton, Philippe},
  journal={IEEE Transactions on Signal Processing}, 
  title={{Performance Analysis of an Improved MUSIC DoA Estimator}}, 
  year={2015},
  volume={63},
  number={23},
  pages={6407-6422},
  doi={10.1109/TSP.2015.2465302}}

@ARTICLE{MUSICML,
  author={Stoica, P. and Nehorai, Arye},
  journal={IEEE Transactions on Acoustics, Speech, and Signal Processing}, 
  title={{MUSIC, maximum likelihood, and Cramer-Rao bound}}, 
  year={1989},
  volume={37},
  number={5},
  pages={720-741},
  doi={10.1109/29.17564}}

@INPROCEEDINGS{circularGaussian3D,
  author={Nurminen, Henri and Suomalainen, Laura and Ali-Loytty, Simo and Piché, Robert},
  booktitle={2018 21st International Conference on Information Fusion (FUSION)}, 
  title={{3D Angle-of-Arrival Positioning Using von Mises-Fisher Distribution}}, 
  year={2018},
  volume={},
  number={},
  pages={2036-2041},
  doi={10.23919/ICIF.2018.8455205}}

@INPROCEEDINGS{circularGaussian3Dexpermental,
  author={Geng, Chunhua and Abrudan, Traian E. and Kolmonen, Veli-Matti and Huang, Howard},
  booktitle={ICC 2021 - IEEE International Conference on Communications}, 
  title={Experimental Study on Probabilistic {ToA and AoA} Joint Localization in Real Indoor Environments}, 
  year={2021},
  volume={},
  number={},
  pages={1-6},
  doi={10.1109/ICC42927.2021.9500283}}

@INPROCEEDINGS{circularGaussian2D,
  author={Marković, Ivan and Petrović, Ivan},
  booktitle={2012 IEEE/RSJ International Conference on Intelligent Robots and Systems}, 
  title={{Bearing-only tracking with a mixture of von Mises distributions}}, 
  year={2012},
  volume={},
  number={},
  pages={707-712},
  doi={10.1109/IROS.2012.6385600}}

@ARTICLE{ReviewCRLB,
  author={Caceres Najarro, Lismer Andres and Song, Iickho and Kim, Kiseon},
  journal={IEEE Sensors Journal}, 
  title={{Fundamental Limitations and State-of-the-Art Solutions for Target Node Localization in WSNs: A Review}}, 
  year={2022},
  volume={22},
  number={24},
  pages={23661-23682},
  doi={10.1109/JSEN.2022.3217335}}

@ARTICLE{WSNPPP1,
  author={O’Lone, Christopher E. and Dhillon, Harpreet S. and Buehrer, R. Michael},
  journal={IEEE Transactions on Wireless Communications}, 
  title={A Statistical Characterization of Localization Performance in Wireless Networks}, 
  year={2018},
  volume={17},
  number={9},
  pages={5841-5856},
  doi={10.1109/TWC.2018.2850310}}

@ARTICLE{WSNPPP3,
  author={Schloemann, Javier and Dhillon, Harpreet S. and Buehrer, R. Michael},
  journal={IEEE Transactions on Wireless Communications}, 
  title={Toward a Tractable Analysis of Localization Fundamentals in Cellular Networks}, 
  year={2016},
  volume={15},
  number={3},
  pages={1768-1782},
  doi={10.1109/TWC.2015.2496273}}

@article{LavancierEtAl2015,
author = {Lavancier, Frédéric and Møller, Jesper and Rubak, Ege},
year = {2015},
pages = {853--877},
title = {Determinantal Point Process Models and Statistical Inference},
volume = {77},
journal = {Journal of the Royal Statistical Society: Series B (Statistical Methodology)},
doi = {10.1111/rssb.12096}
}

@ARTICLE{GinibreTutorial,
  author={Naoto MIYOSHI, Tomoyuki SHIRAI },
  journal={IEICE TRANSACTIONS on Communications}, 
  title={Spatial Modeling and Analysis of Cellular Networks Using the {Ginibre} Point Process: A Tutorial}, 
  year={2016},
  volume={E99-B},
  number={11},
  pages={2247-2255},
  doi={10.1587/transcom.2016NEI0001},
  ISSN={1745-1345},
  month={November},}

@book{BaccelliBlaszczyszynKarray2024,
  author    = {Fran{\c{c}}ois Baccelli and Bart{\l}omiej B{\l}aszczyszyn and Mohamed Kadhem Karray},
  title     = {Random Measures, Point Processes, and Stochastic Geometry},
  year      = {2024},
  publisher = {Inria},
  note      = {HAL Id: hal-02460214v2},
  url       = {https://hal.inria.fr/hal-02460214v2}
}

@article{NakataMiyoshi2014,
title = {Spatial stochastic models for analysis of heterogeneous cellular networks with repulsively deployed base stations},
journal = {Performance Evaluation},
volume = {78},
pages = {7-17},
year = {2014},
issn = {0166-5316},
doi = {https://doi.org/10.1016/j.peva.2014.05.002},
url = {https://www.sciencedirect.com/science/article/pii/S0166531614000546},
author = {Itaru Nakata and Naoto Miyoshi}
}

@book{Jammalamadaka2001,
  title={Topics in Circular Statistics},
  author={Jammalamadaka, S. Rao and SenGupta, A.},
  year={2001},
  publisher={World Scientific}
}

@INPROCEEDINGS{Miranda2010,
  author={Miranda, J. and Abrishambaf, R. and Gomes, T. and Gonçalves, P. and Cabral, J. and Tavares, A. and Monteiro, J.},
  booktitle={2013 11th IEEE International Conference on Industrial Informatics (INDIN)}, 
  title={Path loss exponent analysis in Wireless Sensor Networks: Experimental evaluation}, 
  year={2013},
  volume={},
  number={},
  pages={54-58},
  doi={10.1109/INDIN.2013.6622857}}

@Article{Ding2025,
AUTHOR = {Ding, Weizhong and Li, Lincan and Chang, Shengming},
TITLE = {A Simple and Efficient Method for RSS-AOA-Based Localization with Heterogeneous Anchor Nodes},
JOURNAL = {Sensors},
VOLUME = {25},
YEAR = {2025},
NUMBER = {7},
ARTICLE-NUMBER = {2028},
URL = {https://www.mdpi.com/1424-8220/25/7/2028},
PubMedID = {40218541},
ISSN = {1424-8220},
DOI = {10.3390/s25072028}
}

@article{Diagne2020,
  author  = {Diagne, S. and Val, T. and Farota, A. and Diop, B. and Assogba, O.},
  title   = {Performances Analysis of a System of Localization by Angle of Arrival UWB Radio},
  journal = {International Journal of Communications, Network and System Sciences},
  volume  = {13},
  number  = {2},
  pages   = {15--27},
  year    = {2020},
  doi     = {10.4236/ijcns.2020.132002}
}

@book{BarbourHolstJanson1992,
  author    = {Barbour, A. D. and Holst, L. and Janson, S.},
  title     = {Poisson Approximation},
  series    = {Oxford Studies in Probability},
  volume    = {2},
  publisher = {Oxford University Press},
  address   = {London},
  year      = {1992},
  mrnumber  = {1163825},
}

@article{Kontorovich2025,
  author  = {Kontorovich, Aryeh},
  title   = {On the tensorization of the variational distance},
  journal = {Electronic Communications in Probability},
  volume  = {30},
  pages   = {1--10},
  year    = {2025},
  doi     = {10.1214/25-ECP680}
}

@article{BARBOUR19929,
title = {Stein's method and point process approximation},
journal = {Stochastic Processes and their Applications},
volume = {43},
number = {1},
pages = {9-31},
year = {1992},
issn = {0304-4149},
doi = {https://doi.org/10.1016/0304-4149(92)90073-Y},
url = {https://www.sciencedirect.com/science/article/pii/030441499290073Y},
author = {A.D. Barbour and T.C. Brown}
}

@inbook{Xia_survey,
author = {Aihua Xia},
title = {Stein's method and Poisson process approximation},
publisher = {Singapore: World Scientific Press.},
year = {2005},
booktitle = {An Introduction to Stein's Method},
chapter = {},
pages = {115-181},
doi = {10.1142/9789812567680_0003},
URL = {https://www.worldscientific.com/doi/abs/10.1142/9789812567680_0003},
eprint = {https://www.worldscientific.com/doi/pdf/10.1142/9789812567680_0003}
}

@article{Xia1995OnMI,
  title={On metrics in point process approximation},
  author={Aihua Xia},
  journal={Stochastics and Stochastics Reports},
  year={1995},
  volume={52},
  pages={247-263},
  url={https://api.semanticscholar.org/CorpusID:121435475}
}

@ARTICLE{DeterminantalGaussian,
  author={Li, Yingzhe and Baccelli, François and Dhillon, Harpreet S. and Andrews, Jeffrey G.},
  journal={IEEE Transactions on Communications}, 
  title={Statistical Modeling and Probabilistic Analysis of Cellular Networks With Determinantal Point Processes}, 
  year={2015},
  volume={63},
  number={9},
  pages={3405-3422},
  doi={10.1109/TCOMM.2015.2456016}}

@ARTICLE{DeterminantalGinibre,
  author={Deng, Na and Zhou, Wuyang and Haenggi, Martin},
  journal={IEEE Transactions on Wireless Communications}, 
  title={The Ginibre Point Process as a Model for Wireless Networks With Repulsion}, 
  year={2015},
  volume={14},
  number={1},
  pages={107-121},
  doi={10.1109/TWC.2014.2332335}}

@INPROCEEDINGS{DeterminantalGinibreSensor,
  author={Kong, Han-Bae and Wang, Ping and Niyato, Dusit},
  booktitle={2017 IEEE International Conference on Communications (ICC)}, 
  title={Performance analysis of wireless sensor networks with ginibre point process modeling}, 
  year={2017},
  volume={},
  number={},
  pages={1-6},
  doi={10.1109/ICC.2017.7996585}}

@book{Baker2002,
  author    = {Baker, Andrew},
  title     = {Matrix Groups: An Introduction to Lie Group Theory},
  year      = {2002},
  publisher = {Springer},
  address   = {New York}
}

@inbook{Data_Fusion_and_Filtering_Techniques,
author = {Figueiras, João},
publisher = {John Wiley \& Sons, Ltd},
isbn = {9781119068846},
title = {Data Fusion and Filtering Techniques},
booktitle = {Mobile Positioning and Tracking},
chapter = {5},
pages = {109-133},
doi = {https://doi.org/10.1002/9781119068846.ch5},
url = {https://onlinelibrary.wiley.com/doi/abs/10.1002/9781119068846.ch5},
eprint = {https://onlinelibrary.wiley.com/doi/pdf/10.1002/9781119068846.ch5},
year = {2017}
}

@INPROCEEDINGS{Wireless_Network_MCP_TCP,
  author={Chun, Young Jin and Hasna, Mazen Omar},
  booktitle={2014 International Conference on Information and Communication Technology Convergence (ICTC)}, 
  title={Analysis of heterogeneous cellular networks interference with biased cell association using Poisson cluster processes}, 
  year={2014},
  volume={},
  number={},
  pages={319-324},
  doi={10.1109/ICTC.2014.6983146}}

@ARTICLE{Wireless_Network_TCP,
  author={Yang, Lihua and Lim, Teng Joon and Zhao, Junhui and Motani, Mehul},
  journal={IEEE Transactions on Vehicular Technology}, 
  title={Modeling and Analysis of HetNets With Interference Management Using Poisson Cluster Process}, 
  year={2021},
  volume={70},
  number={11},
  pages={12039-12054},
  doi={10.1109/TVT.2021.3114739}}

\end{document}